\documentclass[aps,prx,twocolumn,english,balance,superscriptaddress,floats,showpacs,prb,footinbib]{revtex4-2}
\usepackage[latin9]{inputenc}
\usepackage{amsmath}
\usepackage{amssymb}
\usepackage{graphicx}
\usepackage{esint}
\usepackage{subfigure}
\usepackage{multirow}
\usepackage{mathtools}
\usepackage{xcolor}

\makeatletter

\newcommand{\beq}{\begin{equation}}
	\newcommand{\eeq}{\end{equation}}
\newcommand{\bea}{\begin{eqnarray}}
	\newcommand{\eea}{\end{eqnarray}}
\newcommand{\bwt}{\begin{widetext}}
	\newcommand{\ewt}{\end{widetext}}
\@ifundefined{textcolor}{}
{%
	\definecolor{BLACK}{gray}{0}
	\definecolor{WHITE}{gray}{1}
	\definecolor{RED}{rgb}{1,0,0}
	\definecolor{GREEN}{rgb}{0,1,0}
	\definecolor{BLUE}{rgb}{0,0,1}
	\definecolor{CYAN}{cmyk}{1,0,0,0}
	\definecolor{MAGENTA}{cmyk}{0,1,0,0}
	\definecolor{YELLOW}{cmyk}{0,0,1,0}
}

\newcommand{\bg}{\mathbf{g}}
\newcommand{\bu}{\mathbf{u}}
\newcommand{\br}{\mathbf{r}}

\newcommand{\bK}{\mathbf{K}}

\newcommand{\cK}{\mathcal{K}}
\newcommand{\cG}{\mathcal{G}}

\newcommand{\fvec}[1]{\boldsymbol{#1}}

\newcommand{\half}{\frac{1}{2}}

\newcommand{\rmd}{{\rm d}}
\newcommand{\diag}{{\rm diag}}

\makeatother

\begin{document}
	
	\title{Pseudo-magnetic fields, particle-hole asymmetry, and microscopic effective continuum Hamiltonians of twisted bilayer graphene}
	
	\author{Jian Kang}
	\email{kangjian@shanghaitech.edu.cn}
	\affiliation{School of Physical Science and Technology, ShanghaiTech University, Shanghai 200031, China}
	
	\author{Oskar Vafek}
	\email{vafek@magnet.fsu.edu}
	\affiliation{National High Magnetic Field Laboratory, Tallahassee, Florida, 32310, USA}
	\affiliation{Department of Physics, Florida State University, Tallahassee, Florida 32306, USA}

\begin{abstract}
	Using the method developed in the companion paper~\cite{}, we construct the effective continuum theories for two different microscopic tight binding models of the twisted bilayer graphene at the twist angle of $1.05^\circ$, one Slater-Koster based and the other ab-initio Wannier based. The energy spectra obtained from the continuum theory --either for rigid twist or including lattice relaxation-- are found to be in nearly perfect agreement with the spectra from the tight binding models when the gradient expansion is carried out to second order, demonstrating the validity of the method. We also analyze the properties of the Bloch states of the resulting narrow bands, finding non-negligible particle-hole symmetry breaking near the $\Gamma$ point in our continuum theory constructed for the ab-initio based microscopic model due to a term in the continuum theory that was previously overlooked. This reveals the difference with all existing continuum models where the particle-hole symmetry of the narrow band Hilbert space is nearly perfect. 
\end{abstract}

\maketitle

%\begin{abstract}
%	xx
%\end{abstract}

\section{Introduction}
Recent discoveries of electronic correlations in the twisted bilayer graphene (TBG), including correlated insulators~\cite{Pablo1}, superconductivity~\cite{Pablo2}, (quantum) anomalous Hall state~\cite{David, Young} and others~\cite{Cory1, Abhay19, Yazdani,Ashoori,Eva,Dmitry1,Stevan,Stevan19,Zeldov,Yazdani2,Shahal,Dmitry2,Young2,JiaSC,YuanCao2021, Young3,PabloNature2021,Shahal2,Young4,YazdaniSC,Yacoby2,Yacoby,JiaSOC,YoungCDW}, have generated enthusiasm among both experimentalists and theorists. It is becoming clear that the interplay between band topology and strong electronic interactions plays an essential role in understanding the remarkable phenomena~\cite{LiangPRX1,KangVafekPRX,Senthil1,GuineaPNAS,Balents19,BJYangPRX,Bernevig1,Leon2,Dai1,Grisha,KangVafekPRL,FengchengSC,Senthil2,MacDonald,Zaletel1,Zaletel2,Ashvin1,KangVafekPRB,ZaletelDMRG,Fengcheng,Sau,Dai2,YiZhang,Guinea2,EslamSoftMode,NickKekule,NickStrain,Fernandes2,BernevigTBG,SongHeavy,Xiaoyu,Parker21}. However, many of the key questions, such as the exact ground states and the mechanism of superconductivity, still remain open. 

The most common theoretical approach to studying the correlated states is to start with a continuum effective Hamiltonian, often referred to as the Bistritzer-MacDonald (BM) model~\cite{BMModel}, which gives isolated narrow bands for a range of near-magic twist angles, and then to project the Coulomb interaction onto the wavefunctions of the narrow bands (sometimes including few remote bands as well). 
The BM model~\cite{BMModel} has achieved success in many respects. It correctly predicts the first magic angle where the bands around the charge neutrality point (CNP) become extremely narrow and captures their band topology.
For relaxed structures, however, the BM model --which was originally derived for a rigid twist-- does not include terms which are nominally of the same order in gradient expansion as the ones which are kept, such as the pseudo-magnetic fields induced by the $C_3$ symmetric strain from lattice relaxation. Moreover, next order gradient terms are needed to accurately capture the narrow bands near the magic angle due to the anomalously small non-interacting bandwidth obtained without such terms~\cite{paper1}. In addition, the narrow band wavefunctions of the BM model are nearly particle-hole (p-h) symmetric~\cite{BernevigTBG}. The presence of the p-h symmetry is known to play important role in choosing the correlated ground states~\cite{Zaletel2,VafekKangPRL20,BernevigTBG,ZaletelDMRG}. Experimentally, it is also seen to be broken at low temperature in that various correlated states appear more stable on either the hole or the electron side of the CNP. This motivates development of a more accurate low energy effective continuum model for TBG.

The goal of this paper is to apply the general formulas developed in the previous companion paper~\cite{paper1} for an arbitrary smooth atomic displacement $\fvec u_j(\fvec r)$
to the specific case of TBG with the relative twist angle $\theta=1.05^\circ$. The atomic displacement fields' configurations are computed by first fixing $\theta$ and then minimizing the combination of the intra-layer elastic terms and the inter-layer adhesion terms computed using  generalized stacking fault energy (GSFE) functions. We do so for two sets of GSFE parameters found in the literature~\cite{KoshinoPRB17,KaxirasRelaxation}. In both cases the regions of AB stacking in the moire pattern grow at the expense of the AA regions compared to just the rigid twist configuration, although the quantitative differences between the two models lead to smoother deformation fields for the set of parameters in Ref.~\cite{KaxirasRelaxation}. For both models we perform the Helmholtz decomposition of the displacement field due to the atomic relaxation (see Eq.~\ref{Eqn:RelaxDecomp}) and find that in both models it is dominated by the curl of an out-of-plane field $\hat{\fvec z}\varepsilon^U(\fvec x)$. The scalar field $\varepsilon^U(\fvec x)$ is in turn spatially periodic with the triangular moire pattern and it is dominated by its first Fourier harmonic (see Eqs.(\ref{Eqn:FTofvarepsU}-\ref{Eqn:Varepssym}) and Table~\ref{Tab:ElasticParammeters}). The $\varepsilon^U(\fvec x)$ field for the set of relaxation parameters in Ref.~\cite{KoshinoPRB17} also obtains the contribution from  higher Fourier harmonics, leading to larger momentum transfer in the inter-layer tunneling. 
%XXX We may want to say what kind of an error it would induce if we would keep just the first harmonic for Koshino? 

\begin{widetext}
 \begin{figure*}[t]
	\centering
	\subfigure[\label{Fig:DispComp:Koshino1}]{\includegraphics[width=0.95\columnwidth]{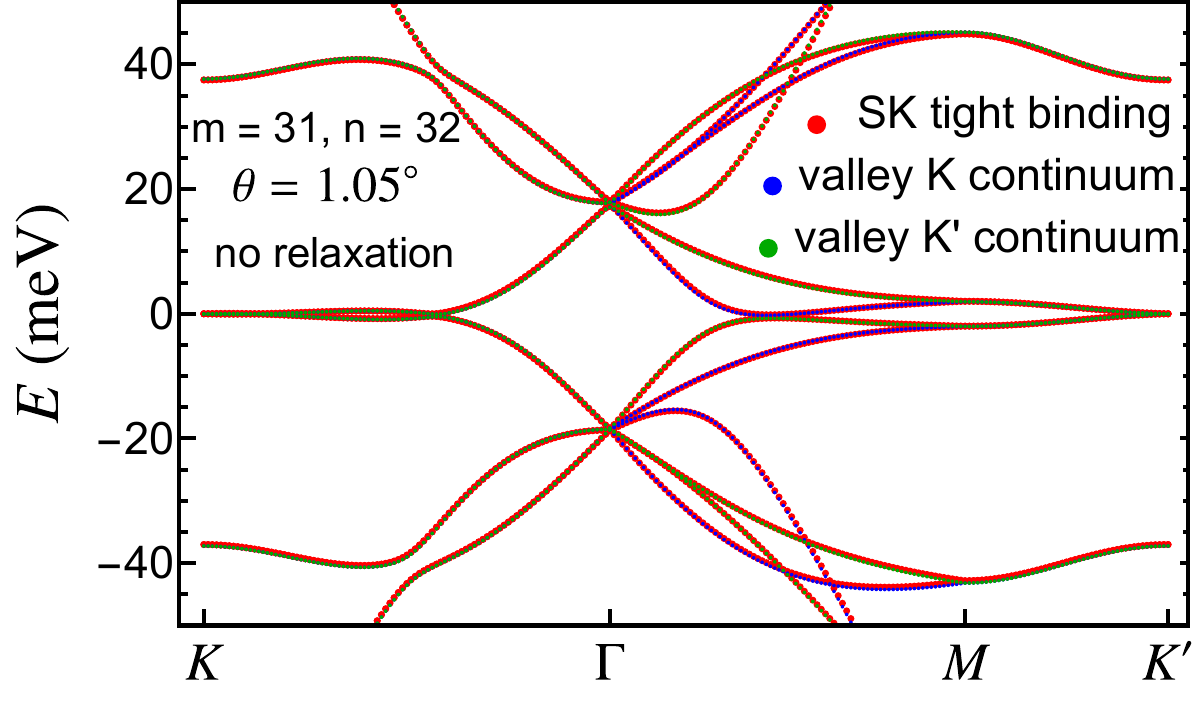}}
	\subfigure[\label{Fig:DispComp:Koshino2}]{\includegraphics[width=0.95\columnwidth]{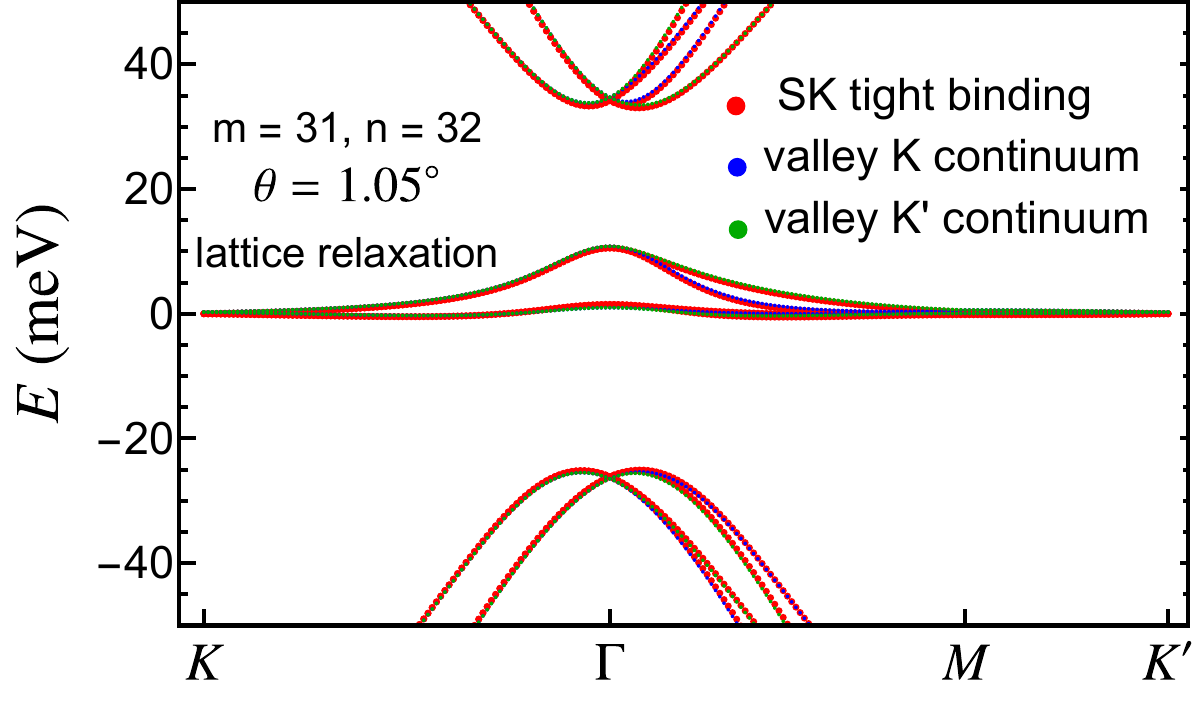}}	
	\subfigure[\label{Fig:DispComp:Shiang1}]{\includegraphics[width=0.95\columnwidth]{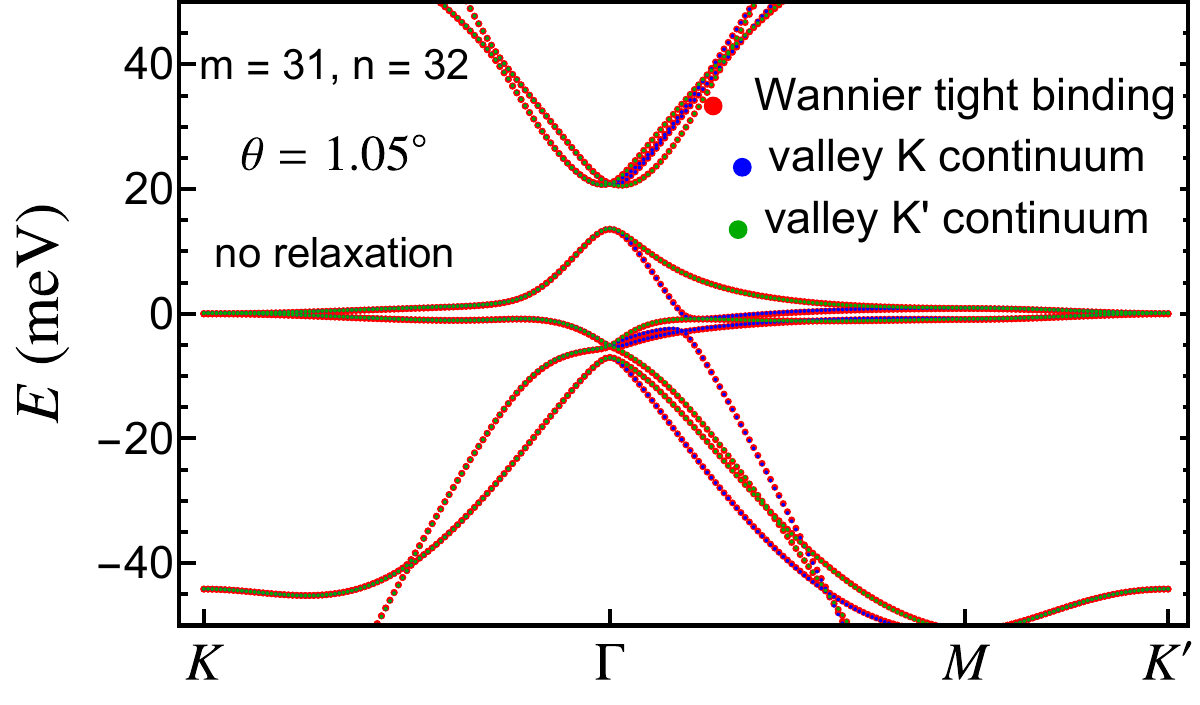}}	
	\subfigure[\label{Fig:DispComp:Shiang2}]{\includegraphics[width=0.95\columnwidth]{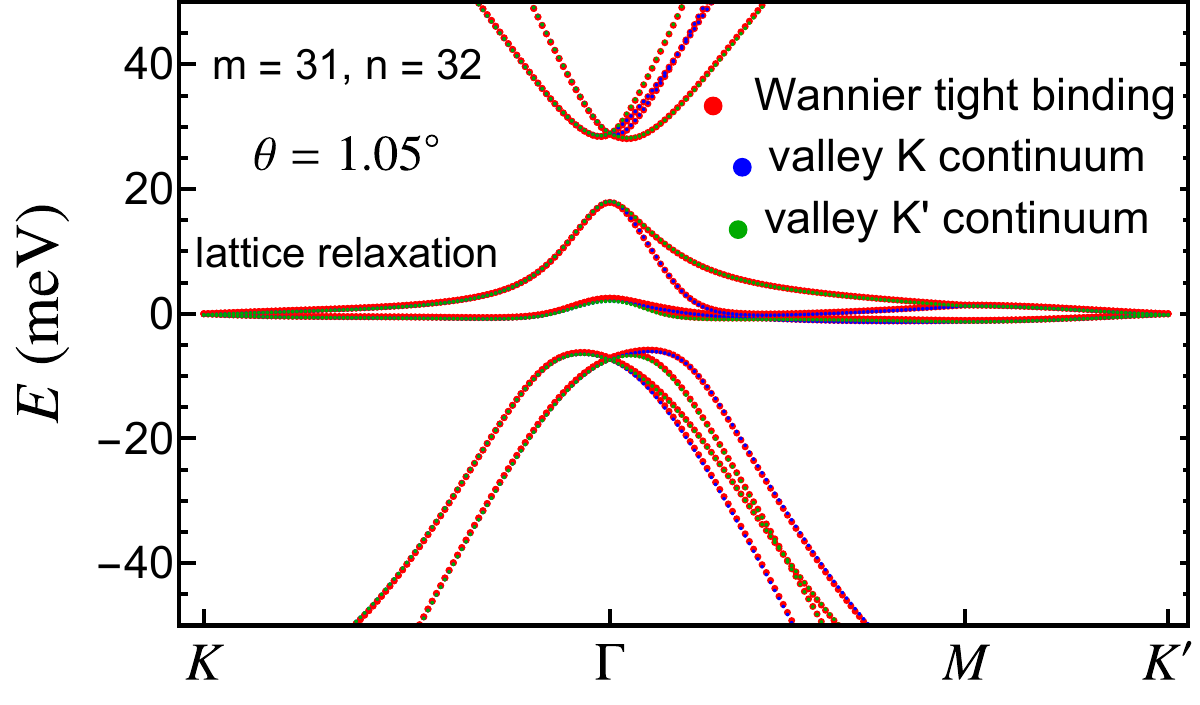}}				
	\caption{Comparison of the energy spectra near the CNP obtained using the microscopic tight binding model (red) and the continuum theory (blue for valley $\fvec K$ and green for valley $\fvec K'$) for the Slater-Koster (SK) based model in Ref.~\cite{KoshinoPRB12} (above) and Wannier based model of Ref.\cite{KaxirasPRB16} (below) in the absence (left) and presence (right) of the lattice relaxation.}
	\label{Fig:DispComp}
 \end{figure*}
\end{widetext}

We next input thus determined atomic displacement fields into the formulas for the continuum Hamiltonian developed in the previous paper, expanding up to second order in gradients in the intra-layer Hamiltonian and up to first order in gradients in the inter-layer Hamiltonian. 
For the intra-layer Hamiltonian, $H_{intra}$, we find an efficient way to compute the desired parameters of the continuum model from the microscopic tight binding functions of Ref.~\cite{KoshinoPRB12} and Ref.~\cite{KaxirasPRB16} by Poisson resumming the powerlaw decaying momentum space sums into real space where they fall off exponentially fast. Each moment of the position vector weighted with the intra-layer hopping function is accompanied by a gradient of either the fermion field or the atomic displacement field, and contributes a factor of $|\bg|a$, where $|\bg|\sim |\fvec K|\theta$ (with $\theta$ in radians), i.e. a factor of $\sim0.08$. 
For the inter-layer Hamiltonian $H_{inter}$, the tunneling falls off fast in the momentum space once the wavevector significantly exceeds the inverse of the inter-layer separation $1/d_0$, making the direct momentum summation efficient. At the same time, each moment of the position vector weighted with the inter-layer hopping function is also accompanied by a gradient of either the fermion field or the atomic displacement field, and thus contributes a factor of $\sim|\bg|d_0$. Since $d_0\simeq 1.36a$, the higher order gradient terms are suppressed by similar factors in the $H_{intra}$ and $H_{inter}$.

At $\theta=1.05^\circ$, the first order gradient of fermion fields or of the atomic displacement fields in $H_{intra}$ are of the same order as the contact terms in $H_{inter}$~\cite{Balents19}. The second order in gradients intra-layer terms are, in turn, of the same order as the first order in gradients inter-layer terms. This pattern continues for the higher order terms. As shown in the Fig.~\ref{FigS:SKGradient} and \ref{FigS:WannierGradient}, the disagreement between the first order continuum Hamiltonian spectrum and the exact tight binding spectrum is $\sim 10$meV, i.e. of the order of the narrow bandwidth. On the other hand, including the second order terms in the continuum Hamiltonian improves the agreement significantly as seen in the Figs.~\ref{Fig:DispComp}, with a nearly perfect agreement throughout the moire Brillouin zone; the largest disagreement is near the $\Gamma$ point where there is at most $0.7$meV difference for the model of Ref.~\cite{KoshinoPRB12} and at most $0.3$meV difference for the model of Ref.~\cite{KaxirasPRB16}.

Thus, the continuum Hamiltonian at the valley $\bK$ is
\begin{eqnarray}
    H^{\bK}_{eff}=H_{intra}+H_{inter}, \label{Eqn:EffH}
\end{eqnarray}
with $H_{intra}$ given in Eq.~(\ref{Eqn:HIntra}) and $H_{inter}$ in Eq.~(\ref{Eqn:HamInter}) together with the parameters in Table~\ref{Tab:ParammeterRelax} and \ref{Tab:ParammeterInter} constitute a highly accurate continuum model for the twisted bilayer graphene at $\theta=1.05^\circ$ obtained directly from the {\em ab initio} microscopic tight binding models, with or without relaxation, using the systematic gradient expansion. The effective Hamiltonian for the valley $\bK'$ is readily obtained from $H^{\bK}_{eff}$ by the spinless time reversal symmetry.

\begin{figure}[t]
	\centering
    \subfigure[\label{Fig:Schematic:mono}]{\includegraphics[width=0.45\columnwidth]{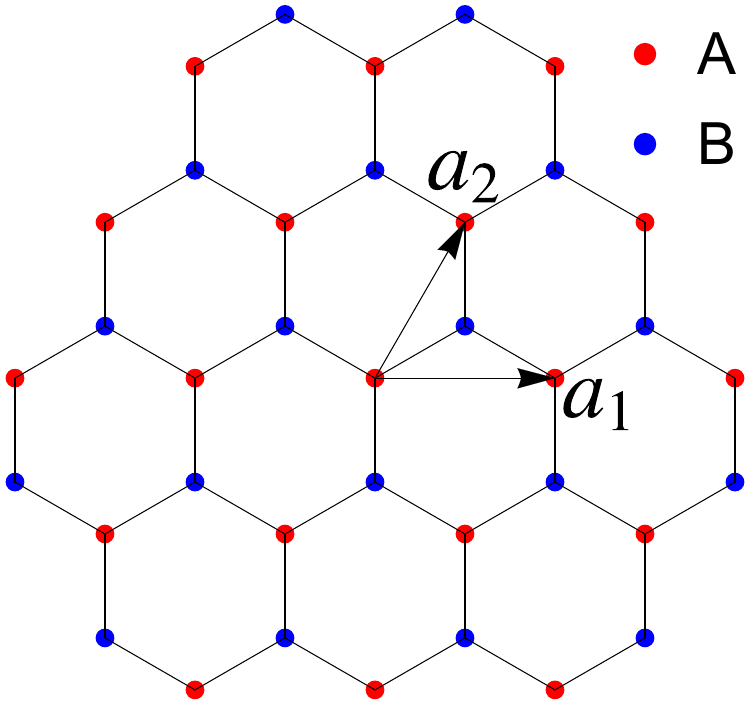}}			
	\subfigure[\label{Fig:Schematic:twist}]{\includegraphics[width=0.5\columnwidth]{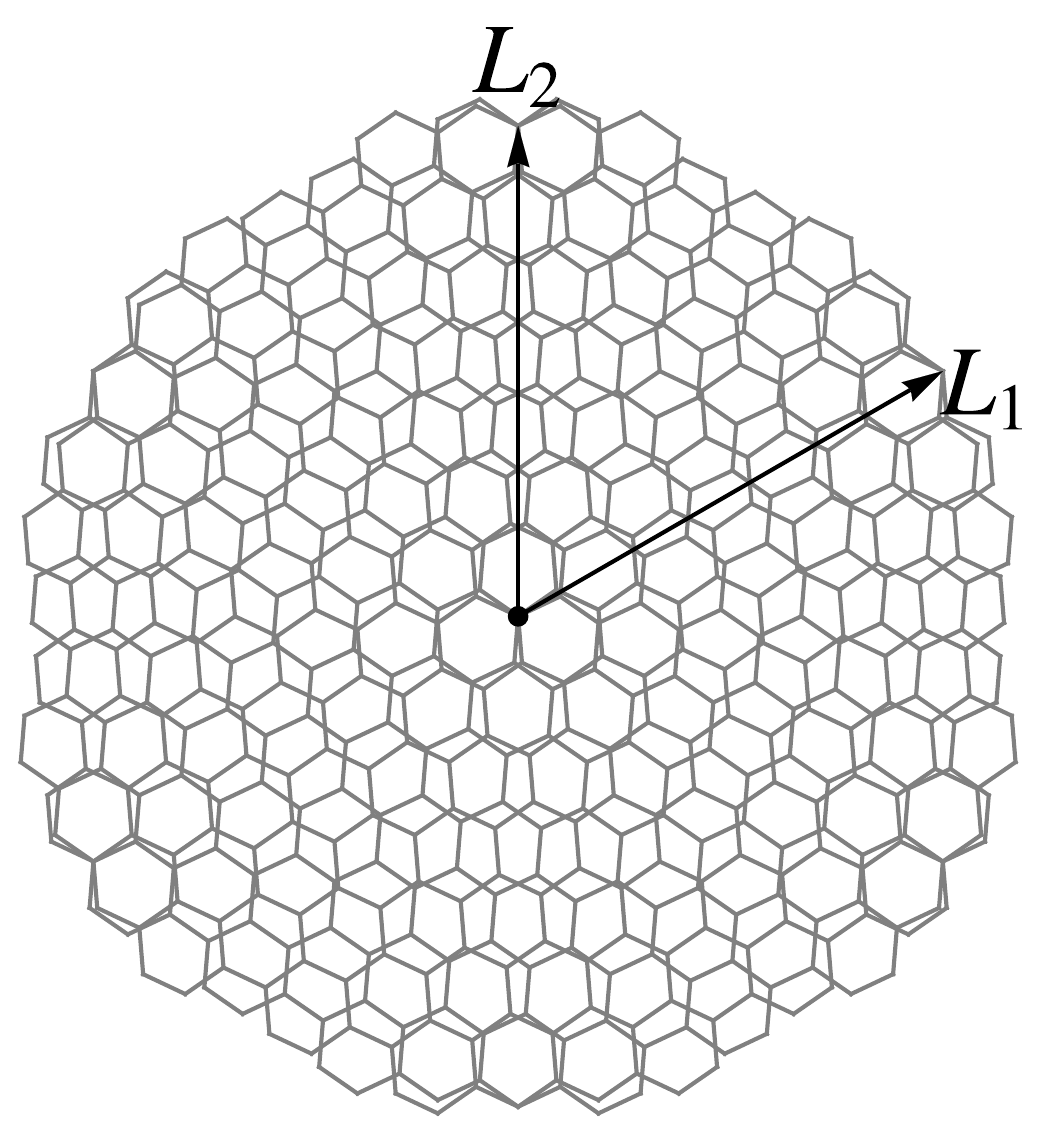}}	
    \subfigure[\label{Fig:Schematic:mom}]{\includegraphics[width=0.75\columnwidth]{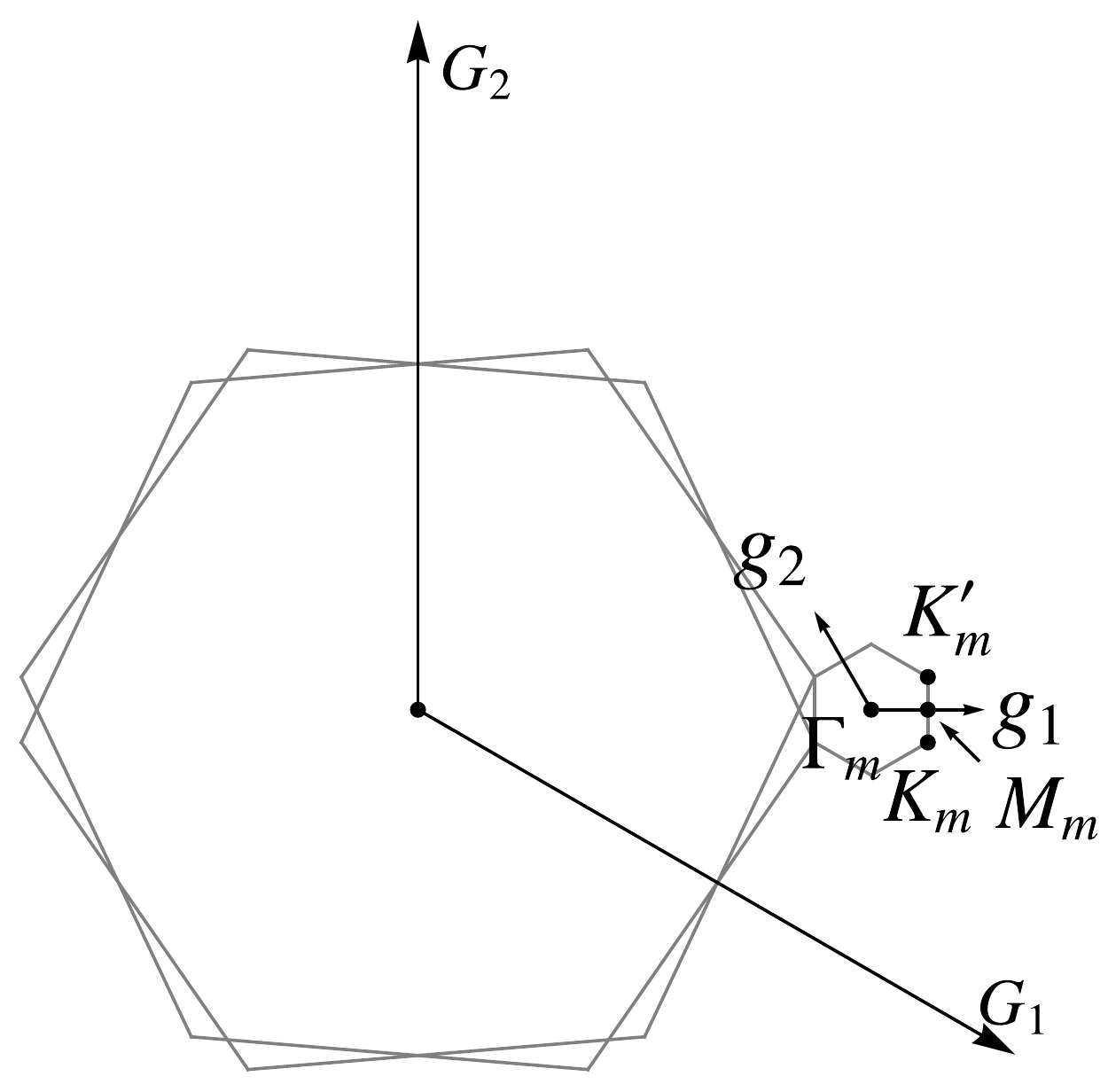}}	
	\caption{Schematic plot for (a) the monolayer lattice vectors $\fvec a_{1, 2}$, (b) the moire lattice vectors $\fvec L_{1,2}$, and (c) their associated reciprocal lattice vectors $\fvec G_{1, 2}$ and $\fvec g_{1, 2}$.}
	\label{Fig:Schematic}
\end{figure}

In addition to studying the energy spectra we also analyze the wavefunctions for the resulting isolated narrow bands. First, we do so by computing the sublattice polarization as well as the Wilson loops~\cite{Dai1,Bernevig1,KangVafekPRB}.
Second, we quantify the degree of the p-h asymmetry in our continuum models for different momenta in the moire BZ by computing the deviation from unitarity of the momentum resolved projected p-h operator. Momentum averaged version of this operator was analyzed for the BM model in Ref.~\cite{BernevigTBG}, where p-h symmetry was found to be nearly perfect. We define the deviation from unitarity as the difference of the smallest singular value from unity, confirming the finding of Ref.~\cite{BernevigTBG} of nearly perfect p-h symmetry in the BM model where the p-h is broken by at most $1\%$ near the $\Gamma$
point. Further, we do this analysis for the continuum model obtained from the tight binding model of Ref.~\cite{KoshinoPRB12}, with relaxed lattice configuration, where we find at most $1.8\%$ deviation from unitarity near the $\Gamma$ point (see Fig.~\ref{Fig:PHSVD:1}). On the other hand, for the continuum model obtained from the tight binding model of the Ref.~\cite{KaxirasPRB16}, interestingly, the p-h symmetry is broken by $\sim 16\%$ near the $\Gamma$ point (see Fig.~\ref{Fig:PHSVD:2}). The stronger p-h symmetry breaking in the model of Ref.~\cite{KaxirasPRB16} is due to the angle dependence of the microscopic inter-layer hopping, resulting in a larger p-h symmetry breaking contact inter-layer tunneling term in the continuum theory, which we dubbed $w_3$ previously (see the supplementary material of Ref.~\cite{VafekKangPRL20}).
The significance of such sizable p-h symmetry breaking for the correlated states will be presented in a separate paper.

Finally, we analyze the effect of the atomic relaxation induced pseudo-vector potential terms on the narrow bandwidth by studying the first order model. Such terms appear already at the first order in gradient expansion, so there is no justification for dropping them in the BM model with the relaxation. Because the pseudo-vector potential terms are of the same order as the contact inter-layer tunneling terms $w_{0,1}$, one may naively conclude that their effect is to broaden the bandwidth by a similar order and to prevent the magic angle phenomenon. While they do increase the bandwidth at the ``old'' magic angle (i.e. without the periodic relaxation induced vector potential), we find that their effect can be compensated by a change of the twist angle, recovering the narrow band at a new (smaller) magic angle. We were able to demonstrate this by solving the problem analytically in the chiral limit including the pseudo-vector potential terms absent in Ref.~\cite{Grisha}. We highlight the importance of $C_3$ symmetry for this compensation. 

This paper is organized as follows: in section II we calculate the lattice relaxation for two sets of GSFE parameters in Ref.~\cite{KoshinoPRB17} and Ref.~\cite{KaxirasRelaxation}. In section III, we present our effective continuum theory of the TBG for two microscopic tight binding models in Ref.~\cite{KoshinoPRB12} and \cite{KaxirasPRB16}, with the corresponding parameter values listed in Table~\ref{Tab:ParammeterRelax}, \ref{Tab:ParammeterInter}, and \ref{TabS:ExtraParammeterInter}. We also plot the energy spectra of the continuum effective theories including the remote bands up $\sim 200$meV. Their nearly perfect agreement with the spectra from the tight binding models demonstrates the validity of the constructed continuum theories. In section IV, we investigate the properties of the Bloch states of the narrow bands, including the sublattice polarization, Wilson loops, and the p-h asymmetry. As shown in Fig.~\ref{FigS:KaxirasPH}, the p-h asymmetry is dominated by $w_3$, a previously overlooked inter-layer contact coupling. Section V studies the exactly flat band limit when including the lattice relaxation induced pseudo magnetic field. Finally, the section VI is devoted to the summary.

\section{Relaxed lattice deformation in the vicinity of the first magic angle}
\label{Sec:LatRelax}

In this section, we follow the approach presented in Ref.~\cite{KoshinoPRB17} to obtain the lattice distortion when the twist angle is near the first magic angle. We assume that the lattice distortion is independent of the sublattice labeled by $S=A$ or $S=B$, i.e.~$\fvec U^{\parallel,\perp}_{j, S}(\fvec x) = \fvec U^{\parallel,\perp}_j(\fvec x)$, where $j=t$ refers to the top layer and $j=b$ refers to the bottom layer. We further neglect the lattice corrugation, so that $\fvec U^{\perp}_b = 0$ and $\fvec U^{\perp}_t = d_0 \hat z$.  Under these assumptions, the intra-layer elastic energy of the graphene system can be written as %(XXX what happened to the $\parallel$ symbol carried by $U$'s? I don't think the notation makes sense after this point)
\begin{align}
	U_E = & \half \sum_{j = t, b} \int\rmd^2 \fvec x \left[ \mathcal{K} (\partial_x U^{\parallel}_{j, x} + \partial_y U^{\parallel}_{j, y})^2 + \right. \nonumber \\
	& \left. \mathcal{G} \left( (\partial_x U^{\parallel}_{j, x} - \partial_y U^{\parallel}_{j, y})^2 + (\partial_y U^{\parallel}_{j, x} + \partial_x U^{\parallel}_{j, y})^2 \right) \right] 
\end{align}
where $\mathcal{K}$ and $\mathcal{G}$ are the bulk and shear modulus of the monolayer graphene; their values for two different models are given in the Table.~\ref{Tab:ElasticParammeters}. It is more convenient to introduce symmetric and anti-symmetric combinations
\begin{eqnarray}
\fvec U^+ &=& \half(\fvec U^{\parallel}_t + \fvec U^{\parallel}_b),\\
\label{Eqn:UminusDef}
\fvec U^- &=& \fvec U^{\parallel}_t - \fvec U^{\parallel}_b.
\end{eqnarray}
\begin{widetext}
	\centering
	\begin{table*}[t]
		\centering
		\begin{tabular}{|c|c|c|c|c|c|} \hline
			Elastic/Adhesion Parameter  & $\cK$ & $\cG$ & $c_1$ 	& $c_2$ & $c_3$ \\		\hline
			Ref.~\cite{KoshinoPRB17} &  $12.82$eV/\AA$^2$    & $9.57$eV/\AA$^2$ &  $3.206$meV/\AA$^2$ &  $0$   & $0$  \\ \hline
			Ref.~\cite{KaxirasRelaxation}    &    $13.265$eV/\AA$^2$     & $9.035$eV/\AA$^2$  &  $0.7755$meV/\AA$^2$ &    $-0.071$meV/\AA$^2$  &    $-0.018$meV/\AA$^2$  \\ \hline 
%			Ref.~xxx & $13.265$eV/\AA$^2$    & $9.035$eV/\AA$^2$  & $0.615$meV/\AA$^2$  & $-0.015$meV/\AA$^2$  &  $0.146$meV/\AA$^2$  \\ \hline \hline
			Lattice Relaxation & $ \tilde{\varepsilon}_1/a^2$ & $ \tilde{\varepsilon}_2/a^2$ & $ \tilde{\varepsilon}_3/a^2$ 	& $ \tilde{\varepsilon}_4/a^2$ & $ \tilde{\varepsilon}_5/a^2$  \\	\hline 
			Ref.~\cite{KoshinoPRB17} &   $0.4243$     & $0.0222$ &  $0.0354$ &  $0.0039$     & $0.0047$  \\ \hline
			Ref.~\cite{KaxirasRelaxation}    &    $0.2270$     & $0.0014$  &  $0.0064$  &    $-0.0002$     & $0.0002$  \\ \hline
%			Ref.xxx &  $0.165$  & $-0.003$ & $0.014$ & $0.0004$ & $0.0014$   \\ \hline
		\end{tabular}
		\caption{Parameters of the elastic theory and the lattice relaxation obtained from Ref.~\cite{KoshinoPRB17} and Ref.~\cite{KaxirasRelaxation}, where $a$ is the lattice constant of the undistorted monolayer graphene. }
		\label{Tab:ElasticParammeters}
	\end{table*}
\end{widetext}
 The intra-layer elastic energy then can be expressed as
\begin{align}
	U_E = & \int\rmd^2\fvec x\  \left[ \mathcal{K} (\partial_x U^+_x + \partial_y U^+_y)^2 + \right. \nonumber \\
	& \left.  \mathcal{G} \left(  ( \partial_x U^+_{x} - \partial_y U^+_{y})^2  +  ( \partial_y U^+_{x} + \partial_x U^+_{y})^2 \right) \right] + \nonumber \\
	&  \frac14 \int\rmd^2\fvec x\   \left[ \mathcal{K} ( \partial_x U^-_x + \partial_y U^-_y )^2 + \right. \nonumber \\
	& \left.  \mathcal{G} \left(  ( \partial_x U^-_{x} - \partial_y U^-_{y})^2  +  ( \partial_y U^-_{x} + \partial_x U^-_{y})^2 \right) \right]
\label{Eqn:UE}
\end{align}
with $\fvec U^+$ and $\fvec U^-$ decoupled.

In addition to the intra-layer elastic energy, we also include the inter-layer adhesion energy
\begin{equation}
U_B = \int \rmd^2\fvec x\ V[\fvec U^-(\fvec x)]
\label{Eqn:UB}
\end{equation} 
where $V[\fvec U^-(\fvec x)]$  is a periodic and even function of the relative displacement $\fvec U^-$ i.e.~$V[\fvec U^-] = V[-\fvec U^-]$ and $V[\fvec U^-] = V[\fvec U^- + \fvec a_i]$ ($i = 1$, $2$), where $\fvec a_1$ and $\fvec a_2$ are the primitive lattice vectors. As shown in the Fig.~\ref{Fig:Schematic}, they are defined as
\begin{align}
	\fvec a_1 = a(1, 0) \ , \quad \fvec a_2 = a\left( \half, \frac{\sqrt{3}}2 \right) \ .
\end{align}
Therefore, the Fourier transform of $V[\fvec U^-]$ can be expressed as
\begin{align}
	V[\fvec U^-(\fvec x)] & =  \sum_{\fvec G} V_{\fvec G} \cos\left( \fvec G \cdot \fvec  U^-(\fvec x) \right), \label{Eqn:LayerInt}
\end{align}
where $\fvec G = m \fvec G_1 + n \fvec G_2$ is a reciprocal lattice vector of the undistorted monolayer graphene, with integer $m,n$ and $\fvec G_1 = \frac{2\pi}a(1, -\frac1{\sqrt{3}})$ and $\fvec G_2 = \frac{2\pi}a(0, \frac2{\sqrt{3}})$. 
The Fourier coefficients $V_{\fvec G}$ fall off with large $\fvec G$, so the sum can be truncated after a few shells.
Furthermore, different $V_{\fvec G}$s are related by symmetries. As a consequence, the adhesion potential has the form~\cite{KaxirasRelaxation}
\begin{align}
	& V[\fvec U^-(\fvec x)]  = c_0  + c_1 \left( \cos(\fvec G_1 \cdot \fvec U^-) + \cos(\fvec G_2 \cdot \fvec U^-) \right. \nonumber \\ 
	& \left. + \cos((\fvec G_1 + \fvec G_2) \cdot \fvec U^-) \right)  + c_2 \left[ \cos((\fvec G_1 - \fvec G_2) \cdot \fvec U^-)  \right. \nonumber \\
	& \left. + \cos((2\fvec G_1 + \fvec G_2)\cdot \fvec U^-)  + \cos((\fvec G_1 + 2\fvec G_2) \cdot \fvec U^-) \right] \nonumber \\
	& +  c_3 \left[ \cos(2\fvec G_1 \cdot \fvec U^-) + \cos(2\fvec G_2 \cdot \fvec U^-) \right. \nonumber \\
	& \left.  + \cos(2(\fvec G_1 + \fvec G_2) \cdot \fvec U^-) \right].   \label{Eqn:InterBinding}
\end{align}
The values of $c_j$'s are given in the Table \ref{Tab:ElasticParammeters}.

For twisted bilayer graphene, the displacement vector field $\fvec U^-$ contains two parts, the relative twist between the two layers, and the relative displacement due to the lattice relaxation or the heterostrain~\cite{paper1}, 
\begin{equation} 
\label{Eqn:Uparametrization}
\fvec U^-(\fvec x) = \theta \hat z \times \fvec x + \delta \fvec U(\fvec x) \  . 
\end{equation}
It follows that
\begin{align}
	& \partial_x U^-_x \pm \partial_y U^-_y = \partial_x \delta U_x \pm \partial_y \delta U_y \nonumber \\
	&  \partial_x U^-_y + \partial_y U^-_x = \partial_x \delta U^-_y + \partial_y \delta U^-_x.
\end{align}
Therefore, as physically expected, the rigid twist term does not contribute to the intra-layer elastic energy. 

Introducing 
\begin{align}
   \fvec g_{\fvec G} = - \theta \hat z \times \fvec G, \label{Eqn:gGMap}
\end{align}
if the twist angle $\theta$ is small, then $\fvec g_{\fvec G}$ becomes a reciprocal vector of the moire superlattice. Note that with this definition we have a one-to-one mapping between the set of $\fvec G$'s and the set of all moire reciprocal lattice vectors, $\{\fvec g\}$.
As seen in Fig.~\ref{Fig:Schematic}, the basis vectors of the set $\{ \fvec g\}$ are
\begin{align}
\fvec g_1 & = \frac{2\pi}L \left( \frac2{\sqrt{3}}, 0 \right), & \fvec g_2 & = \frac{2\pi}L \left( - \frac1{\sqrt{3}}, 1 \right),  \  \label{Eqn:MoiregVec}
\end{align}
where $L = a/(2\sin\frac{\theta}2)$ is the length of $\fvec L_i$, and the primitive moire lattice vectors are
\begin{align}
	\fvec L_1 & = L\left( \frac{\sqrt{3}}2, \half \right), & \fvec L_2 & = L(0, 1). \label{Eqn:MoireLVec}
\end{align}
Indeed, for $\fvec G = m_1 \fvec G_1 + m_2 \fvec G_2$, with integer $m$, $n$,
\begin{eqnarray}
\label{Eqn:gGs}\fvec g_{\fvec G} = (m_2 - m_1)\fvec g_1 - m_1 \fvec g_2, 
\end{eqnarray}
because $\fvec g_{\fvec G_1} = -(\fvec g_1 + \fvec g_2)$ and $\fvec g_{\fvec G_2} = \fvec g_1$. 
Using (\ref{Eqn:gGMap}) and (\ref{Eqn:Uparametrization}), we obtain 
\begin{align}
	\cos\left(\fvec G \cdot \fvec U^-(\fvec x)\right) & =  \cos\left( \fvec g_{\fvec G} \cdot \fvec x +   \fvec G \cdot \delta \fvec U(\fvec x) \right).
\end{align}

In addition to the moire lattice constant $L$ set by interatomic distance $a$ and the twist angle $\theta$, the combination of the intra-layer elastic energy $U_E$ and the inter-layer adhesion potential $U_B$ introduces another characteristic length scale $l = a\sqrt{\frac{\mathcal{G} + \mathcal{K}}{c_1}}$~\cite{KoshinoPRB17}. If $L \gg l$, the inter-layer adhesion dominates over the intra-layer elastic energy and the relaxation maximizes the AB/BA stacking region while minimizing the AA stacking regions. As a consequence, the system breaks up into triangular domains of AB/BA stacking separated by domain walls~\cite{BasovScience2018}. On the other hand, if $L \ll l$ then the lattice relaxation is weak and the structure is close to the one with rigid twist only. In this case the size of AB/BA and AA stacking regions is about the same.

In the rest of this section, we focus on the bilayer system with the commensurate twist angle, i.e.~the moire unit cell vectors $\fvec L_{1, 2}$ satisfying
\begin{align}
	\fvec L_1 & = m \fvec a_1 + n \fvec a_2 \\
	\fvec L_2 & = -n \fvec a_1 + (m + n) \fvec a_2 
\end{align}
where $m$ and $n$ are two integers, with the corresponding twist angle $\theta = \cos^{-1}\left( \frac{m^2 + 4m n + n^2}{2(m^2 + m n + n^2)}  \right)$.

At the first magic angle $\theta=1.05^{\circ}$ (with $m = 31$ and $n = 32$)  and the parameters listed in Table.~\ref{Tab:ElasticParammeters}, $L \approx 0.65l$ for Ref.~\cite{KoshinoPRB17} and $L \approx 0.32l$ for Ref.~\cite{KaxirasRelaxation}. Therefore, the lattice relaxation is expected to be stronger for Ref.\cite{KoshinoPRB17}, with a larger increase of AB/BA stacking regions and a larger decrease of AA stacking regions. As such, the contribution of higher Fourier harmonics to the relaxation is larger, as confirmed by the value of $\tilde{\epsilon}_3$ in Table.~\ref{Tab:ElasticParammeters} defined via Eqs.\ref{Eqn:RelaxDecomp}-\ref{Eqn:Varepssym} and obtained from minimizing $U_E+U_B$ defined in Eqs.(\ref{Eqn:UE}) and (\ref{Eqn:UB}). For Ref.\cite{KaxirasRelaxation}, the lattice relaxation is weaker and smoother; it is dominated by the lowest harmonic terms, i.e.~$\tilde{\epsilon}_1$. As we show in the later sections, an important consequence of this lattice relaxation for the electronic structure is that one must go beyond the Bistritzer-MacDonald model~\cite{BMModel} and include the inter-layer tunneling terms with a larger momentum transfer than just the first shell in order to obtain an accurate description of the magic angle narrow bands. 

As mentioned, the lattice relaxation is obtained by minimizing $U_E + U_B$  with respect to $\delta \fvec U(\fvec x)$.
This leads to the differential equation 
\begin{align}
	& -\frac{1}{2}\begin{pmatrix}
		\left(\mathcal{G} + \mathcal{K}\right) \partial_x^2 + \cG \partial_y^2 &   \cK \partial_x \partial_y \\
		\cK\partial_x \partial_y  & \left( \cG + \cK\right) \partial_y^2 + \cG \partial_x^2 
	\end{pmatrix}
    \begin{pmatrix}
    	\delta U_x \\ \delta U_y
    \end{pmatrix} \nonumber \\
=  & \sum_{\fvec G} V_{\fvec G} \sin\left( \fvec g_{\fvec G} \cdot \fvec x + \fvec G \cdot \delta \fvec U(\fvec x) \right) \begin{pmatrix}
  G_x  \\   G_y
\end{pmatrix}.
	\label{Eqn:RelaxDiffEqn}
\end{align}
Because the lattice relaxation field $\delta \fvec U(\fvec x)$ is a periodic function of $\fvec x$ that satisfies $\delta \fvec U(\fvec x) = \delta \fvec U(\fvec x + \fvec L)$ where $\fvec L= n_1 \fvec L_1 + n_2 \fvec L_2$ is any moire superlattice vector, its Fourier transform can be written as 
\begin{align}  
	\delta \fvec  U(\fvec x) = \sum_{\fvec g} \delta  \tilde{\fvec U}(\fvec g) e^{i \fvec g \cdot \fvec x}  \label{Eqn:DeltaUComp}  \ .  
\end{align}
Here $\sum_{\fvec g}$ sums over all the reciprocal vectors of the moire superlattice, i.e. over the same set as in (\ref{Eqn:gGs}). Introducing the Fourier sum of $\sin(\fvec g_{\fvec G} \cdot \fvec x + \fvec G \cdot \delta \fvec U)$, we obtain
\begin{align}
	& \half \begin{pmatrix}
		(\cG + \cK) g_x^2 + \cG g_y^2 & \cK g_x g_y \\ \cK g_x g_y  & (\cG + \cK) g_y^2 + \cG g_x^2
	\end{pmatrix} \begin{pmatrix} \delta \tilde{U}_x(\fvec g) \\ \delta \tilde{U}_y(\fvec g) \end{pmatrix} \nonumber \\
	= & \sum_{\fvec G} V_{\fvec G} f^{\delta U}_{\fvec g}(\fvec G) \begin{pmatrix}  G_x \\   G_y \end{pmatrix} , \label{Eqn:RelaxMatrixEqn}
	\
	\end{align}
	where 
	\begin{align} & \sin\left(\fvec g_{\fvec G} \cdot \fvec x + \fvec G \cdot \delta \fvec U(\fvec x)\right) = \sum_{\fvec g} f^{\delta U}_{\fvec g}(\fvec G) e^{i \fvec g \cdot \fvec x}  \label{Eqn:InterBindingFourier}\ .
\end{align}
The term $\delta  \tilde{\fvec U}(\fvec g = 0)$ corresponds to a uniform relative translation between two twisted and deformed layers. In order to show that we can set it to zero, we decompose $\delta \fvec U$ into two parts: $\delta \fvec U (\fvec x) = \delta \fvec U_0 + \delta \fvec U_1(\fvec x)$, where  $\delta \fvec U_0 = \delta \tilde{\fvec U}(\fvec g = 0)$, and $\langle \delta \fvec U_1(\fvec x) \rangle_{\fvec x} = 0$, or equivalently, $\delta \tilde{\fvec U}_1(\fvec g = 0) = 0$. Applying Eq.~\ref{Eqn:gGMap}, 
\begin{align}
	 & \cos(\fvec g_{\fvec G} \cdot \fvec x + \fvec G \cdot \delta \fvec U) \nonumber \\
	= & \cos( \fvec g_{\fvec G} \cdot (\fvec x - \theta^{-1} \hat z \times \delta \fvec U_0) + \fvec G \cdot \delta \fvec U_1(\fvec x)).
\end{align} 
The inter-layer adhesion energy can now be written as a function that depends only on $\delta \fvec U_1$:
\begin{align}
	& U_B[\delta \fvec U]  = \int\rmd^2 \fvec x\ \sum_{\fvec G} V_{\fvec G} \cos\left( \fvec g \cdot(\fvec x - \theta^{-1} \hat z \times \delta \fvec U_0) + \fvec G \cdot \delta \fvec U_1(\fvec x) \right) \nonumber \\
	& = \int\rmd^2 \fvec x'\ \sum_{\fvec G} V_{\fvec G} \cos\left( \fvec g \cdot \fvec x' + \fvec G \cdot \delta \fvec U_1(\fvec x' + \theta^{-1} \hat z \times \delta \fvec U_0) ) \right) \nonumber \\
	& = U_B[\delta \fvec U_1']
\end{align}
where $\delta \fvec U_1'(\fvec x) = \delta \fvec U_1(\fvec x + \theta^{-1} \hat z \times \fvec \delta \fvec U_0)$. Because $\langle \delta \fvec U_1' \rangle_{\fvec x} = \langle \delta \fvec U_1 \rangle_{\fvec x} = 0$, the inter-layer adhesion energy of the configuration $\delta \fvec U$ is the same as the adhesion energy of $\delta \fvec U_1'$ whose spatial average vanishes. 
Additionally, the elastic energies of these two configurations are also the same since the energy depends only on the gradient of the lattice relaxation $\delta \fvec U$. 
Therefore, we can set $\delta \tilde{\fvec U}(\fvec g = 0) = 0$. In addition, the parity of  $\delta \fvec U$ is odd, i.e.~$\delta \fvec U(- \fvec x) = - \delta \fvec U(\fvec x)$, leading to the odd parity of (the purely imaginary) $\delta \tilde{\fvec U}(\fvec g)$.

\begin{figure}[t]	
	\subfigure[\label{Fig:LatRelaxEpsilon:Koshino}]{\includegraphics[width=0.58\columnwidth]{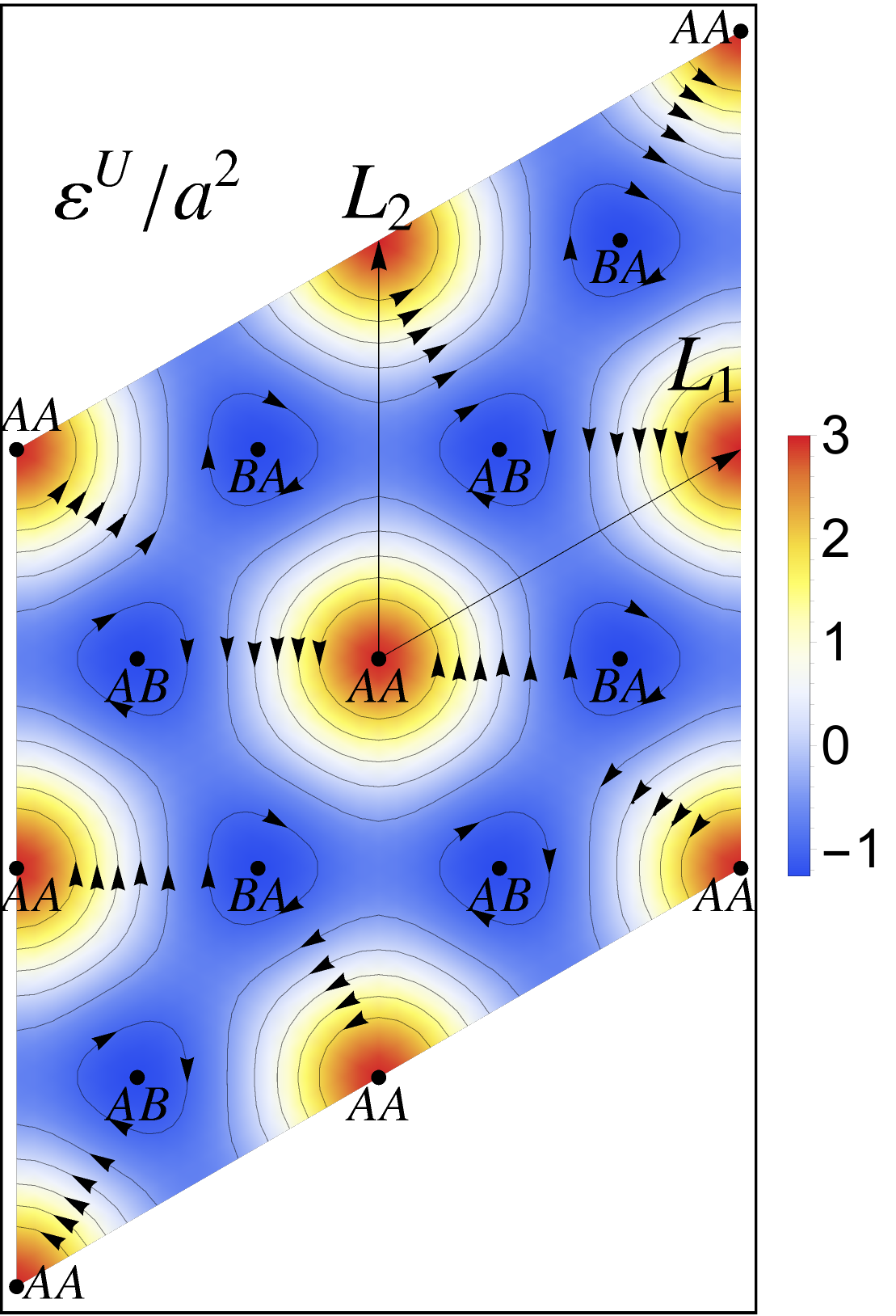}}	
	\subfigure[\label{Fig:LatRelaxEpsilon:Carr}]{\includegraphics[width=0.58\columnwidth]{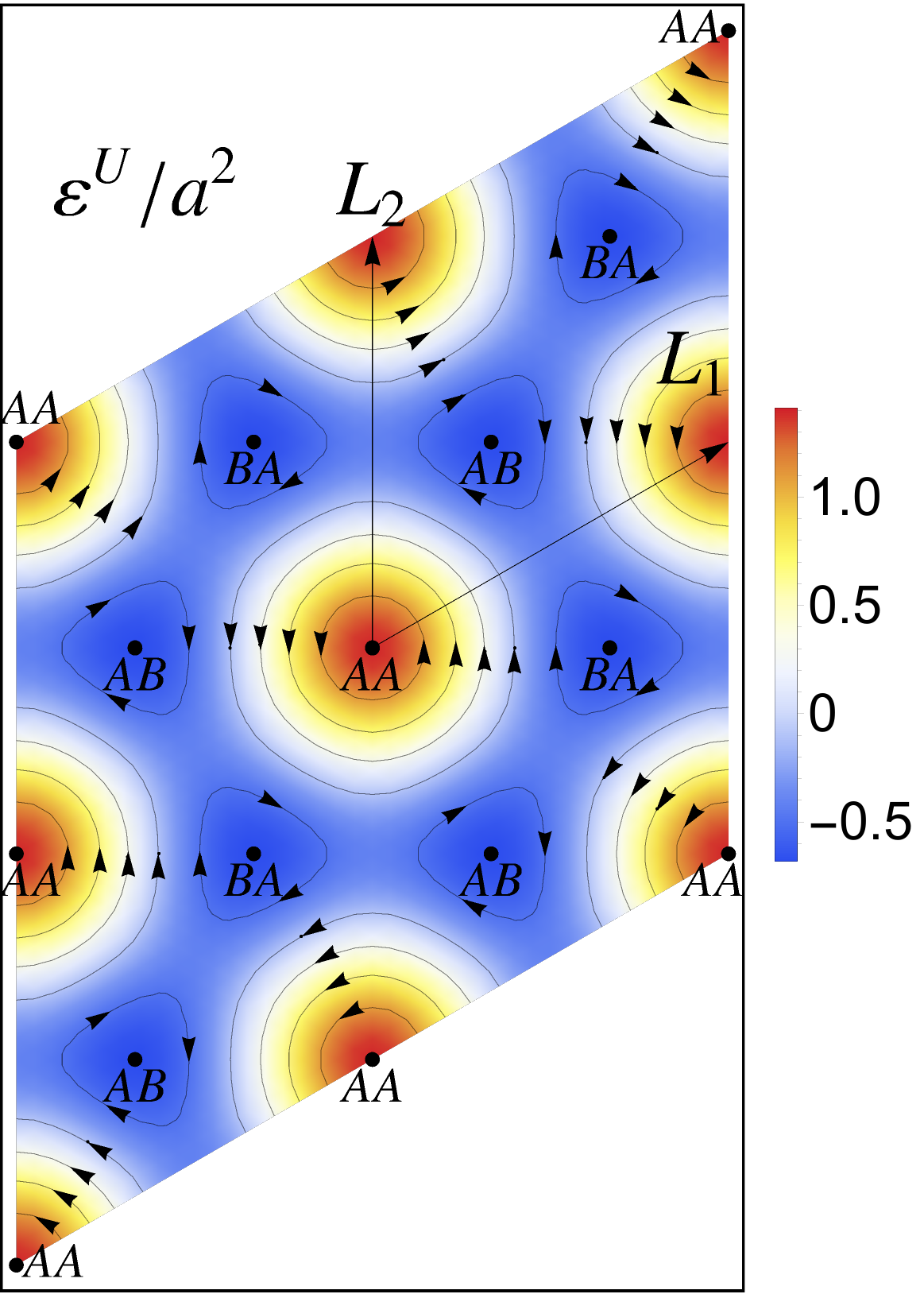}}		
	\caption{The contour plot of the scalar field $\varepsilon^U$  defining the solenoidal component of the atomic displacement field $\delta \fvec U$ in Eq.~\ref{Eqn:RelaxDecomp} for the two models of Ref.~\cite{KoshinoPRB17} (a) and Ref.\cite{KaxirasRelaxation} (b), where $a$ is the monolayer graphene lattice constant. The arrows point along the divergence-free part of $\delta \fvec U$. Near AA, it is in the same direction of the rigid twist of the uniform AA stacked configuration, leading to shrinking of AA stacking moire region, while around AB/BA, it is in the opposite direction of the rigid twist, resulting in the increase of the AB/BA stacking region. }
	\label{Fig:LatRelaxEpsilon}
\end{figure}

Although the sums in the Eq.~\ref{Eqn:LayerInt}, as well as Eqs.~\ref{Eqn:RelaxDiffEqn} and \ref{Eqn:RelaxMatrixEqn} formally include all the reciprocal lattice vectors $\fvec G$, only the terms with small magnitude of $\fvec G$ contribute significantly. This is because we are not in the limit of $L$ much larger than $l$ at the first magic angle, which would cause variations of the displacements over a length scale much shorter than $L$ (i.e. large gradients and therefore many $\fvec g$'s across the domain wall separating AB and BA regions).    
Correspondingly, to numerically solve Eq.~\ref{Eqn:RelaxMatrixEqn},  the Fourier sum in Eq.~\ref{Eqn:InterBindingFourier} can be truncated by including $\fvec g$s from just the first 5 shells as detailed below (see Eq.~\ref{Eqn:Varepssym}).

\begin{figure}[t]
	\centering
	\subfigure[\label{Fig:LatRelaxVarPhi:Koshino}]{\includegraphics[width=0.45\columnwidth]{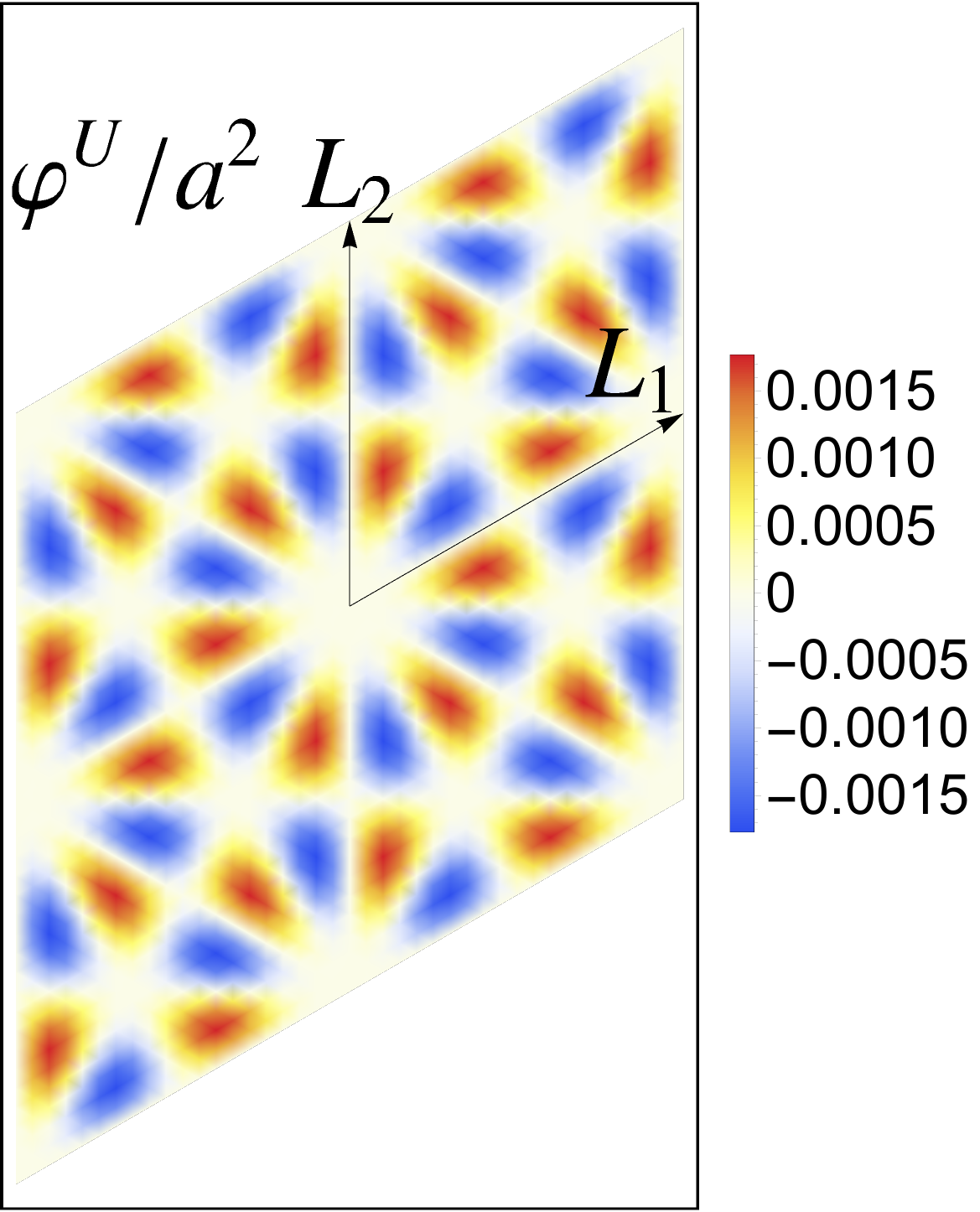}}
	\subfigure[\label{Fig:LatRelaxVarPhi:Carr}]{\includegraphics[width=0.45\columnwidth]{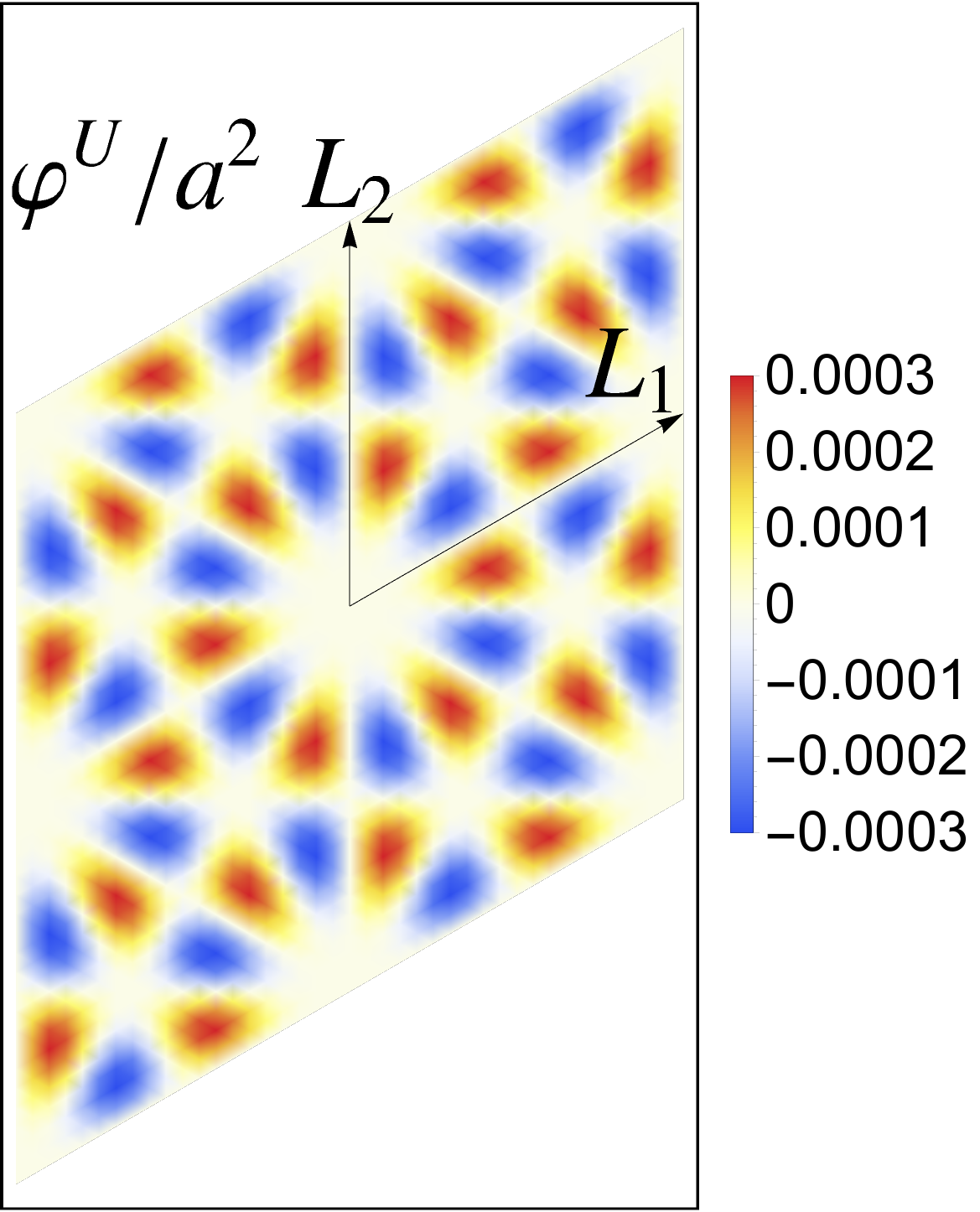}}
	\caption{The density plot of the scalar field $\varphi^U$ defining the irrotational component of the displacement field $\delta \fvec U$  in Eq.~\ref{Eqn:RelaxDecomp} for the model used in Ref.~\cite{KoshinoPRB17} (a) and in Ref.\cite{KaxirasRelaxation} (b); $a$ is the mononlayer graphene lattice constant. Note that the irrotational component is negligibly small compared to the solenoidal component shown in Fig.~\ref{Fig:LatRelaxEpsilon}}.
	\label{Fig:LatRelaxVarPhi}
\end{figure}

The nonlinear Eqs.~\ref{Eqn:RelaxMatrixEqn} and \ref{Eqn:InterBindingFourier}  can be efficiently solved by the iteration method. It starts with a trial solution with $\delta \fvec U(\fvec x) = 0$, feeding into Eqn.~\ref{Eqn:InterBindingFourier} to obtain $f_{\fvec g}^{\delta U_g}$, and then updating $\delta \fvec U(\fvec x)$ by solving its Fourier components $\delta \tilde{U}(\fvec g)$ from Eqn.~\ref{Eqn:RelaxMatrixEqn}. The iteration continues until $\delta \fvec U$ converges. Clearly, the solution is independent of the parameter $c_0$ of the potential in Eq.\ref{Eqn:InterBinding}.  Since the out-of-plane corrugation is not included in this model, $\delta \fvec U(\fvec x)$ is a two-dimensional vector field. As such, by Helmholtz theorem it can be decomposed into a sum of a curl-free part (irrotational) and a divergence-free part (solenoidal) as \begin{align}
	\delta \fvec U(\fvec x) = \fvec \nabla \varphi^U(\fvec x) + \fvec \nabla \times\left(\hat{\fvec z}\varepsilon^U(\fvec x)  \right).\label{Eqn:RelaxDecomp}
\end{align}
As shown in the Fig.~\ref{Fig:LatRelaxEpsilon} and \ref{Fig:LatRelaxVarPhi}, the numerically obtained lattice relaxation is dominated by the solenoidal part, $\fvec \nabla \times \left( \hat{\fvec z} \varepsilon^U(\fvec x) \right)$, for the lattice relaxation models of Ref.~\cite{KoshinoPRB17} and \cite{KaxirasRelaxation}, with their choice for parameters listed in Table~\ref{Tab:ElasticParammeters}. The resulting displacement field ~\cite{Ochoa22} is in qualitative agreement with the Bragg interferometry imaging of the strain fields in twisted bilayer graphene~\cite{Bediako21}. In the following calculations, we will neglect the small irrotational part, and include only the solenoidal  part.

Because the vector field $\delta \fvec U(\fvec x)$ is spatially periodic and vanishes on average, the scalar field $\varepsilon^U(\fvec x)$ is also periodic. It can therefore be can written as
\begin{align}
	\varepsilon^U(\fvec x) = \sum_{\fvec g} \tilde{\varepsilon}^U_{\fvec g} e^{i \fvec g \cdot \fvec x}
	\label{Eqn:FTofvarepsU}
\end{align}
with the $\fvec g$-sum truncated to 5 shells as for $\delta \fvec U(\fvec x)$. Numerically, we found that in both models $\varepsilon^U(\fvec x)$ is dominated by the following components:
\begin{align}
	& \tilde{\varepsilon}^U_{\fvec g = 0} = 0 \nonumber \\
	&  \tilde{\varepsilon}^U_{\pm \fvec g_1} =  \tilde{\varepsilon}^U_{\pm \fvec g_2} =   \tilde{\varepsilon}^U_{\pm (\fvec g_1 + \fvec g_2)} =  \tilde{\varepsilon}_1   \nonumber \\
	&  \tilde{\varepsilon}^U_{\pm (\fvec g_1 - \fvec g_2)} =  \tilde{\varepsilon}^U_{\pm (2 \fvec g_1 + \fvec g_2)} =   \tilde{\varepsilon}^U_{\pm (\fvec g_1 + 2\fvec g_2)} =  \tilde{\varepsilon}_2 \nonumber \\
	&  \tilde{\varepsilon}^U_{\pm 2\fvec g_1 } =  \tilde{\varepsilon}^U_{\pm 2 \fvec g_2} =   \tilde{\varepsilon}^U_{\pm 2(\fvec g_1 + \fvec g_2)} =  \tilde{\varepsilon}_3 \nonumber \\
	&  \tilde{\varepsilon}^U_{\pm (3\fvec g_1 + 2\fvec g_2) } =  \tilde{\varepsilon}^U_{\pm (3 \fvec g_1 +  \fvec g_2)} =   \tilde{\varepsilon}^U_{\pm (2\fvec g_1 + 3\fvec g_2)}  \nonumber \\
	& =   \tilde{\varepsilon}^U_{\pm (2\fvec g_1 - \fvec g_2)}  =   \tilde{\varepsilon}^U_{\pm (\fvec g_1 + 3 \fvec g_2)} =   \tilde{\varepsilon}^U_{\pm (\fvec g_1 - 2\fvec g_2)}  =  \tilde{\varepsilon}_4  \nonumber \\
	&  \tilde{\varepsilon}^U_{\pm 3\fvec g_1 } =  \tilde{\varepsilon}^U_{\pm 3\fvec g_2} =   \tilde{\varepsilon}^U_{\pm 3(\fvec g_1 + \fvec g_2)} =  \tilde{\varepsilon}_5.
\label{Eqn:Varepssym}
\end{align} 
The values of $ \tilde{\varepsilon}_{i}$($i = 1, \cdots, 5$) are listed in Table.~\ref{Tab:ElasticParammeters}. 

%\textcolor{red}{Eva Andrei?}

\section{Accurate Effective model near the first magic angle}
\label{Sec:EffModelTwist}
In the previous paper~\cite{paper1}, we have derived the effective continuum Hamiltonian using the gradient expansion of the slow envelope function of the fermions from $\fvec K$ and $\fvec K'$ points, and the slowly varying atomic displacement fields $\fvec U$ to be
\begin{widetext}
	\begin{eqnarray}
	&&	H^{\bK}_{eff} \simeq \frac{1}{A_{mlg}} \sum_{S,S'} \sum_{jj'} \sum_{\fvec G} e^{i\fvec G \cdot(\fvec{\tau}_S-\fvec{\tau}_{S'})} \int \rmd^2 \fvec x\	\mathcal{J}_{j}(\fvec x)
	\mathcal{J}_{j'}(\fvec x)	e^{i(\fvec G + \fvec K)\cdot \left(\fvec U^\parallel_{j}(\fvec x) - \fvec U^\parallel_{j'}(\fvec x) \right)}
	\int \rmd^2\fvec y e^{-i(\fvec G + \fvec K) \cdot \fvec y}
	\nonumber\\
	&& \times e^{i\frac{\fvec y}{2} \cdot \nabla_{\fvec x} \left( \fvec U^\parallel_{j}(\fvec x) + \fvec U^\parallel_{j'}(\fvec x) \right) \cdot (\fvec G + \fvec K)}	
	\left(t^{j j'}_{sym}\left[ \fvec y + \fvec U^\perp_{j}\left(\fvec x\right) - \fvec U^\perp_{j'}(\fvec x),
	\fvec \delta_S, \fvec \delta_{S'} \right]	
	+ t^{(1)}_{j j', S}(\fvec y) \frac13 \sum_{\alpha = 1}^3  \delta \theta_{j,S}^{(\alpha)} + t^{(2)}_{j j', S'}(\fvec y) \frac13\sum_{\alpha' = 1}^3 \delta \theta_{j', S'}^{(\alpha')}
	\right)\nonumber\\
	&&  \times	\left[ \Psi^\dagger_{j,S}(\fvec x) \Psi_{j',S'}(\fvec x) + \frac{\fvec y}2 \cdot\left( \left(\nabla_{\fvec x} \Psi^\dagger_{j,S}(\fvec x) \right) \Psi_{j',S'}(\fvec x) - \Psi^\dagger_{j,S}(\fvec x) \nabla_{\fvec x} \Psi_{j',S'}(\fvec x)\right)
	\right] \nonumber \\
	&& + \sum_{j, S} \int \rmd^2\fvec x\ \left( \epsilon_0 + \kappa \fvec \nabla \cdot \fvec U^{\parallel}_j(\fvec x) \right) \Psi_{j, S}^{\dagger}(\fvec x) \Psi_{j, S}(\fvec x) \  . 
	\label{Eqn:EffContH}
\end{eqnarray}
\end{widetext}
In the above $S$ and $S'$ sum over the sublattices $A$ and $B$, and $j$ sums over the layers top and bottom. We also consider the possibility that the hopping constant depends not only on the displacement $\fvec y + \fvec U^{\perp}_j(\fvec x) - \fvec U^{\perp}_{j'}(\fvec x)$, but also on the orientation of the nearest neighbor bonds~\cite{KaxirasPRB16}. Without the lattice distortion, the hoppings are given by $t^{j j'}_{sym}\left[ \fvec y + \fvec U^\perp_{j}\left(\fvec x\right) - \fvec U^\perp_{j'}(\fvec x),
\fvec \delta_S, \fvec \delta_{S'} \right]$. In the presence of the lattice distortion, the hoppings can be expanded to the first order of the change of bond angles $\delta \theta_{j, S}^{(\alpha)}$ and $\delta \theta_{j', S'}^{(\alpha')}$ where $\alpha$ and $\alpha'$ are the index of nearest neighbor bonds, ranging from $1$ to $3$.  Thus, the correction is $ t^{(1)}_{j j', S}(\fvec y) \frac13 \sum_{\alpha = 1}^3  \delta \theta_{j,S}^{(\alpha)} + t^{(2)}_{j j', S'}(\fvec y) \frac13\sum_{\alpha' = 1}^3 \delta \theta_{j', S'}^{(\alpha')}$, where $t^{(1)}_{j\neq j',S}(\fvec y) = \frac{\partial t^{j\neq j'}_{sym}}{\partial \theta_{j, S}}$ and $t^{(2)}_{j\neq j',S'}(\fvec y) =  \frac{\partial t^{j\neq j'}_{sym}}{\partial \theta_{j',S'}}$.

It is worth emphasizing that the electron-phonon coupling can be readily obtained from the formula of the effective continuum theory in Eqn.~\ref{Eqn:EffContH}. For this purpose, the lattice deformation is decomposed into the static and dynamic parts, $\fvec U_{j, S} = \fvec U_{j,S}^{(0)} + \fvec U_{j,S}^{(1)}$. $\fvec U_{j,S}^{(0)}$ is the static configuration of the lattice deformation obtained by minimizing the sum of the intra-layer elastic energy $U_E$ in Eqn.~\ref{Eqn:UE} and the interlayer adhesion energy $U_B$ in Eqn.~\ref{Eqn:UB}. As argued in Sec.~\ref{Sec:LatRelax}, without external strains, $\fvec U_{j,S}^{(0)} = \pm \half \delta\fvec U$ (the sign $+$ and $-$ is for the top (bottom) layer), with $\delta\fvec U$ given by Eqns.~\ref{Eqn:RelaxDecomp}, \ref{Eqn:FTofvarepsU}, and \ref{Eqn:Varepssym}. For the two models proposed in Ref.~\cite{KoshinoPRB17} and \cite{KaxirasRelaxation},  the numerical values of $\delta \fvec U$ are presented in Table.~\ref{Tab:ElasticParammeters}. $\fvec U_{j,S}^{(1)}$ is the oscillation part of $\fvec U_{j,S}$, i.~e., the phonon in the bilayer system. Therefore, the expansion of Eqn.~\ref{Eqn:EffContH} to a desired order of $\fvec U_{j,S}^{(1)}$ naturally leads to the coupling between phonons and electrons in such a system. 

In what follows, we will apply this formula to derive the effective Hamiltonian of the magic angle twisted bilayer graphene with and without the lattice relaxation which we obtained in the Sec.\ref{Sec:LatRelax}, and compare the energy spectra with the tight binding models of Ref.~\cite{KoshinoPRB12} and \cite{KaxirasPRB16}.

As the first step, we consider the Jacobian factor 
\begin{equation}\label{Eqn:calJ}
 \mathcal{J}_{j,S}(\fvec x)  = \left| \det\left( \frac{\partial (\fvec x - \fvec U^{\parallel}_j(\fvec x))}{\partial \fvec x} \right)  \right|^{1/2} .
 \end{equation}
Since $\fvec U^{\parallel}_j(\fvec x)$ varies smoothly in the real space, its gradient $|\fvec \nabla \fvec U_j^{\parallel}| \ll 1$, and the determinant can be approximated as
\begin{align}
	\left( \mathcal{J}_{j,S}(\fvec x) \right)^2  \approx 1 - \partial_{\mu} U^{\parallel}_{j, \mu}(\fvec x) \ .
\end{align}
Using $\fvec U_j^{\parallel}(\fvec x) = \pm\frac{1}{2} \left( \theta \hat z \times \fvec x + \delta \fvec U(\fvec x) \right)$, with the sign $+$ ($-$) for the top(bottom) layer respectively, and applying Eq.~\ref{Eqn:RelaxDecomp}, we obtain 
\begin{align}  
  & \left( \mathcal{J}_{j,S}(\fvec x) \right)^2 \approx  1 \mp \half \nabla^2  \varphi^U(\fvec x), \nonumber \\
 \Longrightarrow & \mathcal{J}_{j,S}(\fvec x) \approx 1  \mp \frac14  \nabla^2  \varphi^U(\fvec x).
\end{align}
So to the linear order of $\delta \fvec U$, the deviation of $\mathcal{J}$ from $1$ depends only on the curl-free part $\varphi^U$. As shown in Fig.~\ref{Fig:LatRelaxVarPhi}, for both models of the lattice relaxation, $|\varphi^{U}| \sim 10^{-3} a^2$ and it varies over the length scale much larger than $a$, leading to $|\nabla^2  \varphi^U| \lesssim 10^{-5}$ over the whole real space. As a consequence, the factor $\mathcal{J}$ is very close to $1$. We have checked this numerically by directly computing $\mathcal{J}_{j,S}(\fvec x)$ from the Eq.\ref{Eqn:calJ} and confirmed that its value does not deviate from $1$ by more than $\sim10^{-5}$. Thus, any deviation from $1$ in this factor can be safely neglected in the following calculations. 

It is worth emphasizing here that $\mathcal{J}\approx 1$ relies on the particular form of the lattice relaxation $\delta \fvec U$ that is dominated by its solenoidal part. In the more general case (not considered explicitly here), the presence of the position dependent $\mathcal{J}$ in the intra-layer terms can be interpreted as a spatial variation of the Fermi velocity of the massless Dirac fermion~\cite{Vozmediano2012}.

\begin{widetext}
	\centering
	\begin{table*}[htb]
		\centering
		\begin{tabular}{|c|c|c|c|c|c|c|c|c|} \hline
			intra-layer & $\mu$ (eV) & $v_F/a$ (eV) & $\beta_0/a^2$ (eV) &  $\beta_1/a^2$ (eV) & $v_F \gamma$ (eV)  & $C_0/a$ (eV) &  $D_0/a$ (eV)\\ \hline
			Ref.~\cite{KoshinoPRB12} & $0.7878$ & $2.1256$ & $-0.1846$ & $-0.3714$ & $-3.3644$ & 0.9426 & $-0.7491$ \\
			Ref.~\cite{KaxirasPRB16}    & $-0.3460$  & $2.1790$ &  $-0.1305$ &  $-0.5673$   & $-4.3195$ & $-2.3724$ & $-1.9308$     \\ \hline 
		\end{tabular}
		\caption{Numerical values of the parameters entering in the intra-layer Hamiltonian $H_{intra}$ in Eq.(\ref{Eqn:HIntra}) for two different microscopic models.}
		\label{Tab:ParammeterRelax}
	\end{table*}
\end{widetext}

\subsection{Intra-layer Couplings}
The effective continuum Hamiltonian $H_{eff}^{\fvec K}$ can be decomposed into two parts: the intra-layer $H_{intra}$ and the inter-layer tunneling $H_{inter}$ as in Eq.\ref{Eqn:EffH}. Expanding to the second order gradients we obtain
\begin{widetext}
	\begin{align}
	H_{intra} & = H^{(0)}_{intra} + \delta H_{intra} \label{Eqn:HIntra} \\
	H^{(0)}_{intra} & = \int \rmd^2\fvec x\ \sum_{j=t,b} \sum_{S S'}\Psi_{j, S}^{\dagger}(\fvec x) \left\{ \mu \delta_{SS'} + v_F \bar{\fvec \sigma}_{S S'} \cdot \left( \fvec p^{(j)} + \gamma \mathcal{A}^{(j)}(\fvec x) \right) + \beta_0 \fvec p^2 \delta_{S S'} +  \frac{C_0}2 \left( \fvec p \cdot \mathcal{A}(\fvec x) + \mathcal{A}(\fvec x) \cdot \fvec p \right) \delta_{S S'} \right. \nonumber \\
	& \left.  + \beta_1 \left( (  p_x^2 -  p_y^2) \sigma_1 + 2   p_x   p_y \sigma_2 \right)_{S S'} + \half\sum_{\mu} \left(  p_{\mu}  \xi_{\mu,SS'}(\fvec x) + \xi_{\mu,SS'}(\fvec x)  p_{\mu}  \right)   \right\} \Psi_{j, S}(\fvec x),  \label{Eqn:H0Intra}
	\end{align}
\end{widetext}
where $j$ is summed over the top $(t)$ and bottom $(b)$ layers, $S$,$S'$ are summed over the the $A$ and $B$ sublattices, and $\mu$ over $x$ and $y$ components.
In the above, we split $H_{intra}$ into two terms: $H^{(0)}_{intra}$ and $\delta H_{intra}$. The first term, $H^{(0)}_{intra}$, contains all the contributions up to the second order in gradients whose energy scale is above $1$meV and dominates the second term $\delta H_{intra}$. The numerical values of the coefficients appearing in $H^{(0)}_{intra}$ for the two different microscopic models~\cite{KoshinoPRB12,KaxirasPRB16} can be found in the Table.~\ref{Tab:ParammeterRelax}. The definition of the lattice distortion induced pseudo vector fields $\fvec{\mathcal A}(\fvec x)$ and the fields $\fvec{\xi}_{SS'}(\fvec x)$ are discussed below. The second term, $\delta H_{intra}$, contains other second order gradient contributions that are smaller than $1$meV. We have checked numerically that inclusion of $\delta H_{intra}$ does not improve the agreement between the spectra of the tight binding Hamiltonians and $H_{eff}$ in any significant way, as demonstrated in the Fig.~\ref{Fig:DispComp} where $\delta H_{intra}$ is omitted. Therefore, we will focus on $H^{(0)}_{intra}$ in the main text. For completeness, we spell out the details of $\delta H_{intra}$  in the appendix Sec.~\ref{SecS:Intra}.

In Eq.~\ref{Eqn:H0Intra}, $\bar{\fvec \sigma} = (\sigma_x, - \sigma_y)$, $\fvec p = -i \hbar \fvec \nabla=-i\hbar \frac{\partial}{\partial \fvec x}$ is the momentum operator, and $\fvec p^{(j)}$ is the operator $\fvec p$ rotated by $\mp \theta/2$. When the angle $|\theta| \ll 1$,
\begin{align}
	\fvec p^{(j)} \approx \left( p_x + \frac{\theta_j}2 p_y\ ,\ p_y - \frac{\theta_j}2 p_x \right)  \ , 
\end{align}
where we introduced the notation $\theta_t = - \theta_b = \theta$.  The pseudo-vector field $\mathcal{A}(\fvec x)$ is induced by the lattice distortion, having the form of
\begin{align}
	\mathcal A(\fvec x) &= (\partial_x \delta   U_x - \partial_y \delta   U_y\ , \ -(\partial_x \delta   U_y + \partial_y \delta  U_x))   \nonumber \\
	& \approx (2 \partial_x \partial_y \varepsilon^U(\fvec x)\ , \ (\partial_x^2 - \partial_y^2)\varepsilon^U(\fvec x) ),  \label{Eqn:AField}
\end{align}
and $\mathcal{A}^{(j)}(\fvec x)$ is defined as
\begin{align}
	\mathcal{A}^{(j)}_{\mu}(\fvec x) & = \pm R\left( \frac{\theta_j}2 \right)_{\mu\nu} \mathcal A_{\nu} \nonumber \\
	& \approx \pm \left( \mathcal A_x - \frac{\theta_j}2 \mathcal A_y\ , \ \mathcal A_y + \frac{\theta_j}2 \mathcal A_x \right)_{\mu}  \ , 
\end{align}
where $R(\theta) = \cos(\theta) I_{2\time 2} - i \sin(\theta) \sigma_2$ is the $2\times 2$ matrix corresponding to the counterclockwise rotation along $\hat z$ by the angle of $\theta$. When $|\theta| \ll 1$, $R(\theta)_{\mu\nu} \approx \delta_{\mu\nu} - \theta \epsilon_{\mu\nu}$. The sign $+$ and $-$ are for $j$ corresponding the top and bottom layers respectively, reflecting the fact that the lattice distortions on two layers are opposite.

The field $\fvec \xi(\fvec x)$ is also induced by the lattice distortion. It is given by
\begin{align}
	 \xi_{x, SS'}(\fvec x) & =   \left[ \left(\frac{v_F}2 + 2D_0\right)(\partial_x \delta  U_x)  \sigma_1 - \right. \nonumber \\
	& \ \left. \left[ \left(\frac{v_F}2 + D_0\right)\partial_y \delta  U_x + D_0  \partial_x \delta  U_y  \right] \sigma_2 \right]_{S S'} \ , \nonumber \\
   \xi_{y, SS'}(\fvec x) & =  \left( \left[ \left(\frac{v_F}2 + D_0\right)\partial_x \delta  U_y + D_0 \partial_y \delta   U_x  \right] \sigma_1 \right. \nonumber \\
 & \ \left. - \left(\frac{v_F}2 + 2D_0\right)(\partial_y \delta   U_y ) \sigma_2  \right)_{S S'}  \ .
\end{align}
The values of the constants $\mu$, $v_F$, $\gamma$, $C_0$, and $D_0$ are listed in Table.~\ref{Tab:ParammeterRelax} for two different microscopic tight binding models. All of these constants can be expressed via the microscopic hopping function. Detailed formulas for their efficient evaluation are derived in the appendix Sec.~\ref{SecS:Intra}. 

%Note that 

%%%%%%%%%%%%%%%%%%%%%%%%%%%%%%%%%%%%%%%%%%%%%%%%%%%%%%%%%%%%%%%%%%%%%%%%%%%%%%%%%%%

In Eq.~\ref{Eqn:HIntra}, the term $\mu \Psi_{j,S}^{\dagger} \Psi_{j, S}$ leads to an overall shift of the energy spectrum and thus is irrelevant in most calculations. Among other terms, the leading one is $v_F  \bar{\fvec \sigma} \cdot \fvec p$, that produces the Dirac cone of the monolayer graphene. At the first magic angle $\theta=1.05^\circ$, this term in both models has the energy scale of $v_F k_{\theta} \sim160-170$meV, where $k_{\theta} = 4\pi \theta/(3  a)$. Using the values listed in Table~\ref{Tab:ParammeterRelax}, we can estimate the coupling between the fermion and the pseudo-vector field $\mathcal{A}(\fvec x)$. We found $|v_F \gamma \mathcal{A}^{(j)}(\fvec x)| \lesssim 100$meV, the same order as $v_F k_{\theta}$, showing the necessity of including this term in the effective continuum Hamiltonian, even if we were to only keep the first order gradients. 

\begin{figure}[t]
	\centering
	\includegraphics[width=0.95\columnwidth]{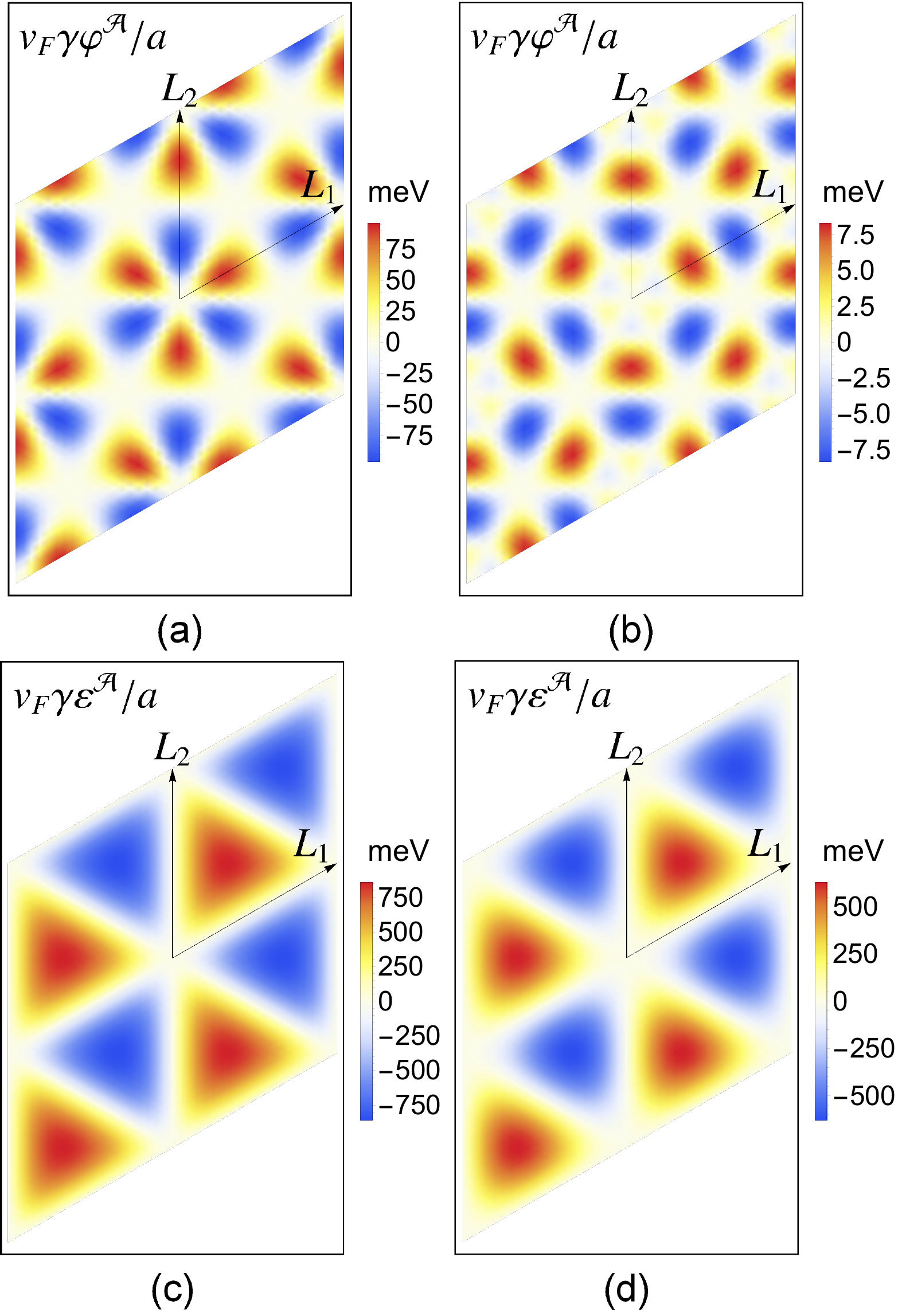}
	%	\subfigure[\label{Fig:AField:2}]{\includegraphics[width=0.95\columnwidth]{PseudoMagField.pdf}}					
	\caption{$v_F \gamma \varphi^{\mathcal{A}}$ (above) and $v_F \gamma \varepsilon^{\mathcal{A}}$ (below) defined in Eq.~\ref{Eqn:ADecomp} for the two models developed in Ref.~\cite{KoshinoPRB17} (left) and \cite{KaxirasRelaxation} (right), where $v_F$ is the Fermi velocity of the undistorted monolayer Dirac cone.}
	\label{Fig:AField}
\end{figure}

\begin{figure}[htb]
	\centering
	\includegraphics[width=0.99\columnwidth]{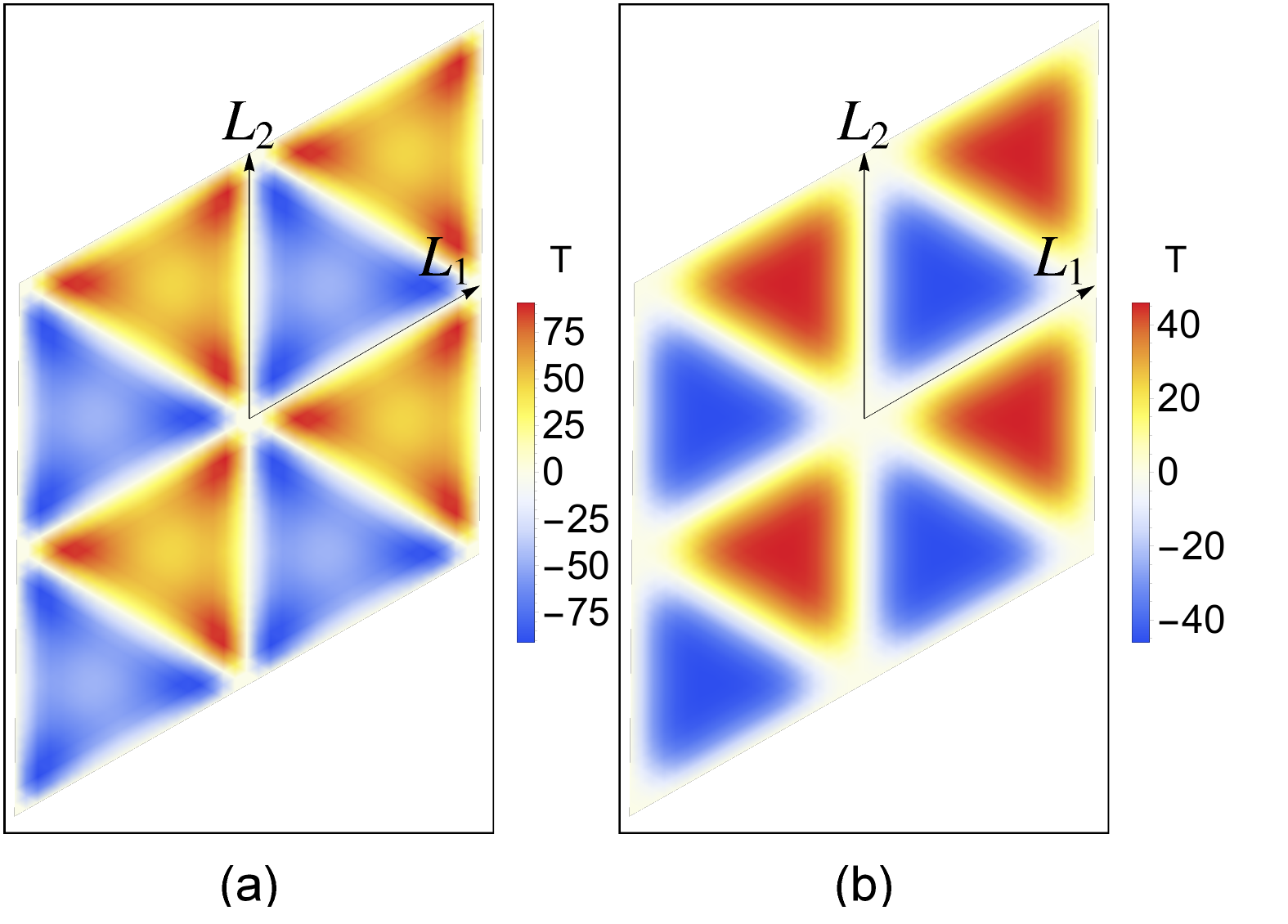}				
	\caption{The pseudo magnetic field $\mathcal{B}$ induced by the lattice relaxation as defined in Eq.~\ref{Eqn:MagField} for the two models developed in Ref.~\cite{KoshinoPRB17} (left) and \cite{KaxirasRelaxation} (right).}
	\label{Fig:BField}
\end{figure}

Since the $\mathcal{A}$ is also a two-dimensional vector field and $\langle \mathcal{A} \rangle = 0$ averaged over the whole space, it can also be decomposed into the irrotational and the solenoidal parts
\begin{align}  
	\mathcal{A} = \fvec \nabla \varphi^{\mathcal{A}} + \fvec \nabla \times \left(\hat{\fvec z} \varepsilon^{\mathcal{A}} \right) \ .  \label{Eqn:ADecomp}
\end{align}

As shown in Fig.~\ref{Fig:AField}, the solenoidal part of the pseudo vector field $\mathcal{A}$ is larger than its irrotational. Interestingly, the induced pseudo-magnetic field resulting from $\mathcal{A}$, defined as
\begin{align}
	\mathcal{B} \hat z = \frac{c}e \fvec \nabla \times (\gamma \mathcal{A}) \label{Eqn:MagField} \,
\end{align} 
is about $30\mathrm{T}$ around the AB/BA stacked regions, and can be as high as $75$T near the AA stacked region, as shown in Fig.~\ref{Fig:BField}.

\subsection{Inter-layer Tunnelings}
\begin{figure}[t]
	\centering
	\includegraphics[width=0.7\columnwidth]{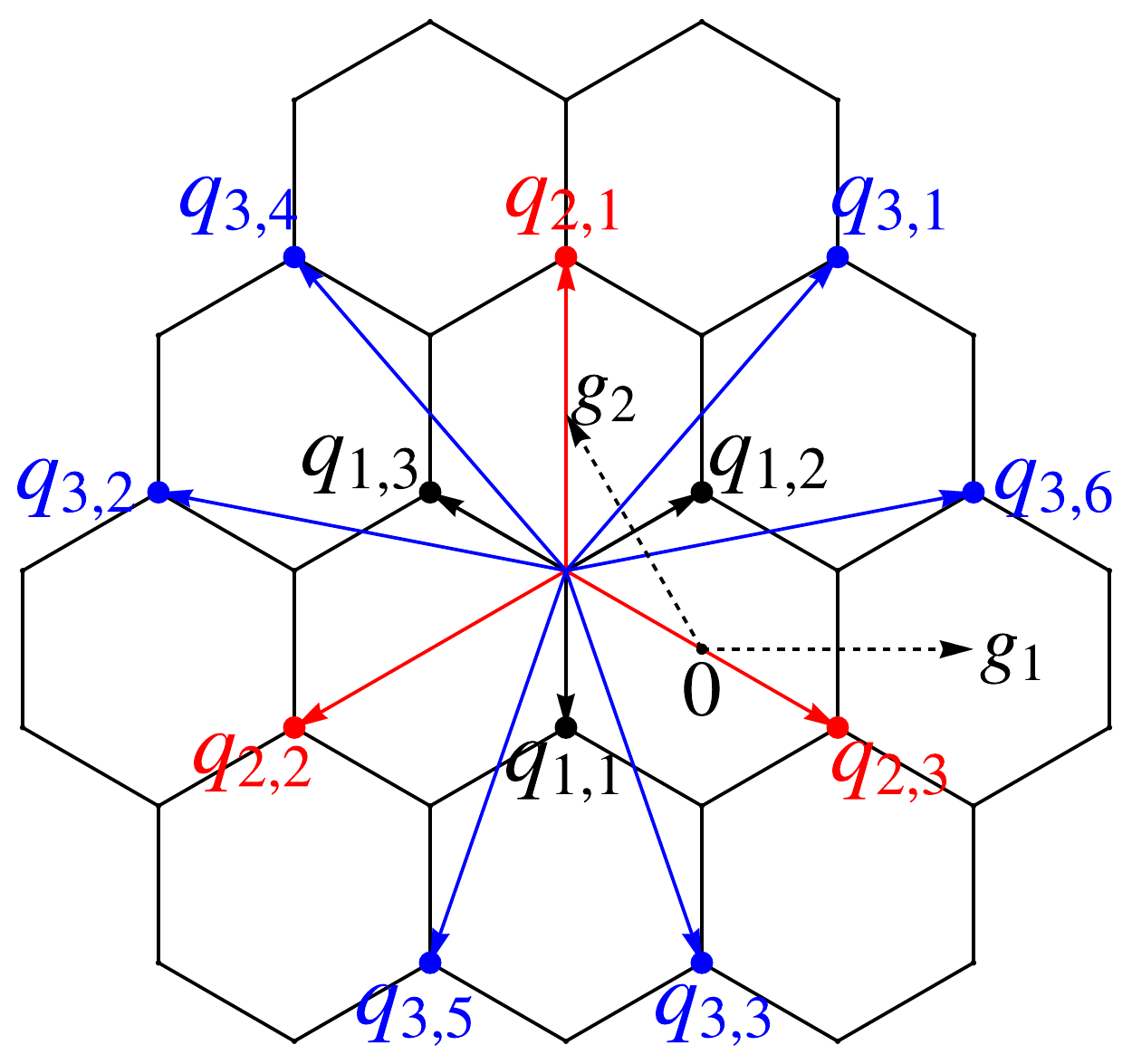}
	%	\subfigure[\label{Fig:AField:2}]{\includegraphics[width=0.95\columnwidth]{PseudoMagField.pdf}}				
	\caption{The first and extended BZ of the moire superlattice. The vectors $\fvec q_{\mu,l}$ in the first three shells  are also plotted here.}
	\label{Fig:QShell}
\end{figure}

Up to the first order gradients, the inter-layer tunneling part of the effective continuum Hamiltonian in Eq.~\ref{Eqn:EffH} can be written as
\begin{widetext}
\begin{align}
	 H_{inter}  =  \sum_{S S'}\int\rmd^2 \fvec x\Psi^{\dagger}_{t,S}(\fvec x) \left( T_{S S'}(\fvec x) + \frac{1}{2}\left\{\fvec p,\fvec \Lambda_{S S'}(\fvec x)  \right\} \right) \Psi_{b, S'}(\fvec x)  + h.c. .\label{Eqn:HamInter}
\end{align}
\end{widetext}
The first and the second terms in the above parenthesis describe the contact and gradient inter-layer couplings, respectively. As before, $\fvec p = -i \hbar \fvec \nabla$ is the momentum operator.
The scalar field $T_{S S'}$ and the vector field $\fvec \Lambda_{S S'}$ can be expanded as
\begin{align}
	T_{S S'}(\fvec x) = \sum_{\mu, j} T_{S S'}^{(\mu, l)} e^{i \fvec q_{\mu, l} \cdot \fvec x},  \\
	\fvec \Lambda_{S S'}(\fvec x) = \sum_{\mu, l} \fvec \Lambda_{S S'}^{(\mu, l)} e^{i \fvec q_{\mu, l} \cdot \fvec x}, \label{Eqn:WLambdaFT}
\end{align} 
where the vectors $\fvec q_{\mu, l}$ form shells in the extended moire BZ as illustrated in the Fig.\ref{Fig:QShell}. Any vector $\fvec q_{\mu, l}$ can be decomposed as $\fvec q_{\mu, l} = \fvec q_1 + \fvec g$ with $\fvec q_1 = -4\pi \hat {\fvec y}/(3 |L_1|) $ and $\fvec g$ being a reciprocal lattice vector of the moire superlattice, defined in Eq.~\ref{Eqn:MoiregVec}.  Different $\fvec q$ vectors are distinguished by their subscript indices $(\mu, l)$, with $\mu$ denoting the shell ordered by its radius $|\fvec q|$ from small to large, and $l$ labeling different $\fvec q$ vectors inside the same shell. 

%For the systems with the twist angle near the first magic angle, the Fourier summation in Eq.~\ref{Eqn:WLambdaFT} can be accurately truncated by included only the $\fvec q$ vectors shown in  Fig.~\ref{Fig:QShell}, i.e.~all the $\fvec q$s in the first three shells. To be more specific, they are
%\begin{align}
%	\fvec q_{1, 1} & = \fvec q_1 \ , & \fvec q_{1, i} = \left( C_3 \right)^{i-1} \fvec q_{1,1} \nonumber \\
%	\fvec q_{2, 1} & = \fvec q_1 + \fvec g_1 + 2\fvec g_2  & \fvec q_{2, i} = \left( C_3 \right)^{i-1} \fvec q_{2,1} \nonumber \\
%	\fvec q_{3, 1} & = \fvec q_1 + 2(\fvec g_1 + \fvec g_2) &  \fvec q_{3, i} = \left( C_3 \right)^{i-1} \fvec q_{3,1} \nonumber \\
%	\fvec q_{3, 4} & = \fvec q_1 + 2\fvec g_2  & \fvec q_{3, 3 + i} = \left( C_3 \right)^{i-1} \fvec q_{3,4} 
%\end{align} 
%where $i = 1$, $2$, and $3$, and $C_3$ are the in-plane counterclockwise rotation by $2\pi/3$. 

%

Symmetries further constrain the form of $T_{SS'}$ and $\fvec \Lambda_{S S'}$. The lattice distortion $\fvec U(\fvec x)$ considered in this paper is invariant under  $C_2 \mathcal{T}$, $C_{2x}$, and $C_3$ transformations, and so is the effective continuum Hamiltonian. For example, under $C_2\mathcal{T}$ transformation, the fermion fields $\Psi_{j, S}(\fvec x) \longrightarrow \mathcal{K} \Psi_{j, \bar{S}}(-\fvec x)$, where $\bar{S}$ is the sublattice index different from $S$, and $\mathcal{K}$ is the complex conjugation. Therefore, the inter-layer tunneling matrices must satisfy the constraints $T(\fvec x) = \sigma_x T^{\ast}(- \fvec x) \sigma_x$ and $\fvec \Lambda(\fvec x) = \sigma_x \fvec \Lambda^{\ast}(- \fvec x) \sigma_x$. Correspondingly, their Fourier components must satisfy
\begin{align}
	T_{S S'}^{(\mu, l)} = \left( \sigma_x \big(T^{(\mu, l)} \big)^{\ast} \sigma_x \right)_{S S'}, \nonumber \\
	\fvec \Lambda_{S S'}^{(\mu, l)} = \left( \sigma_x \big(\fvec \Lambda^{(\mu, l)} \big)^{\ast} \sigma_x \right)_{S S'} .
\end{align}
This implies that the above Fourier components can be written as
\begin{align}
	T_{S S'}^{(\mu, l)} & = \left( w_0^{(\mu, l)} \sigma_0 + w_1^{(\mu, l)} \sigma_1 + w_2^{(\mu, l)} \sigma_2 + i w_3^{(\mu, l)} \sigma_3  \right)_{S S'}, \nonumber \\
	\fvec \Lambda_{S S'}^{(\mu, l)} & = \left( \fvec \lambda_0^{(\mu, l)} \sigma_0 + \fvec \lambda_1^{(\mu, l)} \sigma_1 + \fvec \lambda_2^{(\mu, l)} \sigma_2 + i \fvec \lambda_3^{(\mu, l)} \sigma_3  \right)_{S S'},
\end{align}
where $w^{(\mu, l)}_{i}$ and the vectors $\fvec \lambda^{(\mu, l)}_{i}$ are all real. The more detailed symmetry analysis, including $C_{2x}$ and $C_3$, is presented in the appendix~\cite{appendix}. Here, we only list several constraints from which all $w_i^{(\mu, l)}$s and $\fvec \lambda_i^{(\mu, l)}$s with $\mu \leq 3$ can be obtained based on Tab.~\ref{Tab:ParammeterInter}. For $1\leq l \leq 3$, due to $C_3$ symmetry, the parameters $w_i^{(\mu, l)}$ need to satisfy
\begin{align}
	& w_i^{(\mu, l)} = w_i^{(\mu, 1)}, \quad w_i^{(3, l+3)} = w_i^{(3, 4)}, \quad \mbox{for } i = 0, 3 \label{Eqn:WC3Sym} \\
%	& w_3^{(\mu, l)} = w_3^{(\mu, 1)}  \quad  w_3^{(3, l+3)} = w_3^{(3, 4)}   \\
	& w_1^{(\mu, l)} - i w_2^{(\mu, l)} = e^{i \frac{2\pi}3 (l -1)} \left( w_1^{(\mu, 1)} - i w_2^{(\mu, 1)} \right),    \\
	& w_1^{(3, l+3)} - i w_2^{(3, l+3)} = e^{i \frac{2\pi}3 (l -1)} \left( w_1^{(3, 4)} - i w_2^{(3, 4)} \right).   
\end{align}
Similarly, for the vectors $\fvec \lambda_i^{(\mu, l)}$, if we restrict $1\leq l \leq 3$, the $C_3$ symmetry leads to the following constraints. For $i = 0$ and $3$, 
\begin{align}
	& \left( \fvec \lambda_i^{(\mu, l)} \right)_{\alpha} = R\left( \frac{2\pi}3(l - 1) \right)_{\alpha \beta} \left( \fvec \lambda_i^{(\mu, 1)} \right)_{\beta}  \\ 
	& \left( \fvec \lambda_i^{(3, l+3)} \right)_{\alpha} = R\left( \frac{2\pi}3(l - 1) \right)_{\alpha \beta} \left( \fvec \lambda_i^{(3, 4)} \right)_{\beta} 
\end{align}
In addition,
\begin{align}
	\left( \fvec \lambda_1^{(\mu, l)} - i \fvec \lambda_2^{(\mu, l)} \right)_{\alpha} & = e^{i \frac{2\pi}3 (l -1)} R\left( \frac{2\pi}3(l - 1) \right)_{\alpha \beta} \times \nonumber \\
	& \ \left( \fvec \lambda_1^{(\mu, 1)} - i \fvec \lambda_2^{(\mu, 1)} \right)_{\beta}  \\
	 \left( \fvec \lambda_1^{(3, l+3)} - i \fvec \lambda_2^{(3, l+3)} \right)_{\alpha} & = e^{i \frac{2\pi}3 (l -1)} R\left( \frac{2\pi}3(l - 1) \right)_{\alpha \beta}\times \nonumber \\ 
	 & \ \left( \fvec \lambda_1^{(3, 4)} - i \fvec \lambda_2^{(3, 4)} \right)_{\beta} 
%	&  \fvec \lambda_1^{(3, l+3)} - i \fvec \lambda_2^{(3, l+3)} = e^{i \frac{2\pi}3 (l-1)} \left( C_3 \right)^{l-1} \left( \fvec \lambda_1^{(3, 4)} - i \fvec \lambda_2^{(4, 4)} \right) \nonumber \\
%  	& w_{i}^{(3, 1)}  = w_i^{(3, 4)} \quad \fvec \lambda_i^{(3, 1)} = C_{2x} \fvec \lambda_i^{(3, 4)} \quad \mbox{for } i \neq 2 \quad   \nonumber \\
% 	& w_2^{(3, 1)}  = - w_2^{(3, 4)}  \quad    	\fvec \lambda_2^{(3, 1)} = -C_{2x} \fvec \lambda_2^{(3, 4)}
\end{align}
where $R(\theta) = \cos(\theta) I_{2 \times 2} -i \sin(\theta) \sigma_2$ is the $2\times 2$ matrix corresponding to the counterclockwise rotation along $\hat z$ by the angle of $\theta$. Furthermore, the $C_{2x}$ symmetry imposes the constraints on both $w_i^{(\mu, l)}$ and $\fvec \lambda_i^{(\mu, l)}$. For $i \neq 2$,
\begin{align}
	& w_{i}^{(3, 1)}  = w_i^{(3, 4)} \\ 
	& \left( \fvec \lambda_i^{(3, 1)} \right)_{\alpha} = \left( \tau_3 \right)_{\alpha \beta} \left( \fvec \lambda_i^{(3, 4)} \right)_{\beta} 
\end{align}
For $w_2^{(\mu, l)}$ and $\fvec \lambda_2^{(\mu, l)}$, we obtain 
\begin{align}
	& w_2^{(3, 1)}  = - w_2^{(3, 4)}\\
	& \left( \fvec \lambda_2^{(3, 1)} \right)_{\alpha} = - \left( \tau_3 \right)_{\alpha \beta} \left( \fvec \lambda_2^{(3, 4)} \right)_{\beta} \label{Eqn:LambdaC2xSym}
\end{align}
where the superscripts $\alpha$ and $\beta$ label the components of the vectors field $\fvec \lambda_i^{(\mu, l)}$. We should also emphasize that the constraints listed in Eq.~\ref{Eqn:WC3Sym}--\ref{Eqn:LambdaC2xSym} are not complete. For example, by $C_{2x}$ symmetry, we can also derive $w_2^{(1,1)} = 0$. The more detailed and complete discussion on symmetry constraints are presented in the appendix Sec.~\ref{SecS:Sym}.

If we keep only the innermost $\fvec q$ shell, then the contact term in the inter-layer Hamiltonian of our theory limits to,
\begin{align}
	T_{SS'}^{(1)}(\fvec x) & = \sum_{j = 1}^3 e^{i \fvec q_j \cdot \fvec x} \left(  w_0 I_{2\times 2} + iw_3 \sigma_3 +\right. \nonumber \\
	& \ \left. w_1 \left( \cos(\frac{2\pi(j - 1)}3) \sigma_1 - \sin(\frac{2\pi(j - 1)}3) \sigma_2    \right)   \right)_{SS'},
\end{align}
since $\fvec q_1 = \fvec q_{1,1}$, $\fvec q_2 = \fvec q_{1, 2}$, and $\fvec q_{1, 3}$ and 
\begin{align}
	w_0 = w_0^{(1,1)} \ , \quad w_1 = w_1^{(1,1)}\ , \quad w_3 = w_3^{(1,1)}.
\end{align} 
Therefore, our theory recovers the inter-layer term in the BM continuum model~\cite{BMModel,Senthil1,BernevigTBG}
if we set $w_3=0$ and keep only $\fvec q$s in the first shell and neglect the gradient couplings $\fvec \Lambda_{SS'}$. Furthermore, in the absence of the lattice relaxation, as derived in the appendix~\cite{appendix}, our theory gives $w_0 = w_1$~\cite{BMModel}. As shown in the next section, the $w_3$ term is responsible for non-negligible p-h asymmetry for the model of Ref.~\cite{KaxirasPRB16}. 

Fig.~\ref{FigS:SKQShellSpec} and \ref{FigS:WannierQShellSpec} show the comparison of the spectrum obtained from $H_{eff}^{\bK}$, truncating to a different number of $\fvec q$ shells in the inter-layer tunneling terms. For the rigid twist (i.e. when the lattice relaxation is absent), the approximation of including only the innermost $\fvec q$ shell gives the spectrum that is almost identical with the one produced by the tight binding model in most of the moire BZ, with the mismatch of only $\sim 2$meV around the center of the moire Brillouin zone point $\fvec \Gamma$; the bandwidth of the narrow bands for the rigid twist is about $40$meV and $20$meV for the models in Ref.~\cite{KoshinoPRB12} and Ref.~\cite{KaxirasPRB16} respectively, with at least one of the bandgaps to the remote bands vanishing. All these features have been well reproduced by including only one $\fvec q$ shell in $H^\bK_{eff}$. To further improve the agreement, we include the first two $\fvec q$ shells and achieve the accuracy presented in Fig.~\ref{Fig:DispComp:Koshino1} and \ref{Fig:DispComp:Shiang1}. The excellent agreement obtained with only $2$ shells reflects the fact that the Fourier transform of the inter-layer hopping quickly decays as the function of the momentum~\cite{BMModel,appendix}.  

On the other hand, in the presence of the lattice relaxation, we need to include more shells to achieve the comparable accuracy. This is demonstrated in the Fig.~\ref{FigS:SKQShellSpec} and \ref{FigS:WannierQShellSpec}. The increase of needed shells results from the factor $e^{i (\fvec G + \fvec K)\cdot (\fvec U_{j, S}^{\parallel}(\fvec x) - \fvec U_{j', S'}^{\parallel}(\fvec x))}$ in Eq.~\ref{Eqn:EffContH}. The Fourier transform of the inter-layer hopping (i.e. for $j \neq j'$) is the largest for $\fvec G$s satisfying $|\fvec G + \fvec K| = |\fvec K|$. Because of the spatial inhomogeneity of the lattice relaxation $\delta \fvec U$ (Eq.~\ref{Eqn:DeltaUComp}), the mentioned exponential factor induces the inter-layer scattering with the momentum transfer of all possible $\fvec q_{\mu, l}$; the strength of the scattering is proportional to the Fourier transform of the exponential factor. For the lattice relaxation in Ref.~\cite{KaxirasRelaxation}, $\delta \fvec U$ is dominated by the lowest wavevectors $\pm \fvec g_1$, $\pm \fvec g_2$ and $\pm (\fvec g_1 + \fvec g_2)$. As a consequence, $H_{inter}$ should include, at least, the terms with the momentum transfer of $\fvec q_{1, l} \pm \fvec g_1$, $\fvec q_{1, l} \pm \fvec g_2$, and $\fvec q_{1, l} \pm (\fvec g_1 + \fvec g_2)$, i.e.~all the $\fvec q$ vectors in the first three shells. While all the values of $w_i^{(\mu, l)}$ and $\fvec \lambda_i^{(\mu, l)}$ in the first three shells can be obtained from Table~\ref{Tab:ParammeterInter} and the formula listed in Eq.~\ref{Eqn:WC3Sym}--\ref{Eqn:LambdaC2xSym}, the values in the next three shells can be calculated in the same way based on Table~\ref{TabS:ExtraParammeterInter} and Eq.~\ref{EqnS:C3Constraints}--\ref{EqnS:C2xConstraints} in the appendix. Fig.~\ref{FigS:WannierQShellSpec}(b) shows that the first four shells are needed to achieve the accuracy of $0.3$meV at $\fvec \Gamma$ for the lattice configuration of Ref.~\cite{KaxirasRelaxation}. 

For Ref.~\cite{KoshinoPRB17}, both the first, $\tilde{\epsilon}_1$, and the third, $\tilde{\epsilon}_3$, harmonics of $\delta \fvec U$ are sizable (see Table~\ref{Tab:ElasticParammeters}). The wavevectors of the third harmonic are $\pm 2 \fvec g_1$, $\pm 2 \fvec g_2$, and $\pm 2(\fvec g_1 + \fvec g_2)$. Following the argument in the above paragraph, we therefore expect that $H_{inter}$ should include the terms with the momentum transfer of all $\fvec q$s in the first $6$ shells. Indeed, as demonstrated numerically in the Fig.~\ref{FigS:SKQShellSpec}(b), we achieve the accuracy of $0.8$meV around $\fvec \Gamma$ with six $\fvec q$ shells.

\begin{widetext}
	\centering
	\begin{table*}[htb]
		\centering
		\begin{tabular}{|c|c|c|c|c|c|c|c|c|c|c|c|c|} \hline
			   &  $w_0^{(1,1)}$ & $w_1^{(1,1)}$  & $w_2^{(1,1)}$  & $w_3^{(1,1)}$  &   $w_0^{(2,1)}$  & $w_1^{(2,1)}$  & $w_2^{(2,1)}$  & $w_3^{(2,1)}$ &  $w_0^{(3,1)}$  & $w_1^{(3,1)}$ & $w_2^{(3,1)}$  & $w_3^{(3,1)}$    \\ \hline
			\begin{tabular}{@{}c@{}}Ref.~\cite{KoshinoPRB12} \\ unrelaxed \end{tabular}   &  $110.9$ & $110.9$ & $0$ & $0$ & $1.6$  & $1.6$  & $0$ & $0$ & \multicolumn{4}{|c|}{negligible}  \\  \hline
			\begin{tabular}{@{}c@{}}Ref.~\cite{KoshinoPRB12} \\ relaxed \end{tabular} & $54.4$ & $124.9$ & $0$ & $0$ & $-6.9$ &  $9.0$ & $0$ & $0$ & $17.5$ & $-10.8$ & $-18.8$ & $0$ \\ \hline \hline
			\begin{tabular}{@{}c@{}}Ref.~\cite{KaxirasPRB16} \\ unrelaxed \end{tabular} &   $104.0$  & $104.0$ & $0$  &  $-2.9$ & $1.1$ & $1.1$ & $0$ & $0$  & \multicolumn{4}{|c|}{negligible}  \\ \hline
			\begin{tabular}{@{}c@{}}Ref.~\cite{KaxirasPRB16} \\ relaxed \end{tabular} & $78.6$ & $113.1$ & $0$ & $-2.8$  & $-0.3$ & $3.4$ & $0$ & $-0.5$  & $11.0$ & $-5.6$ & $-9.7$ & $-0.6$ \\ \hline						
%			\begin{tabular}{@{}c@{}}Ref.~XXX \\ relaxed \end{tabular} & $85.8$ & $111.2$ & $0$ & $-2.8$ & $0.8$ & $2.7$ & $0$ & $-0.4$  & $7.0$ & $-5.3$ & $-7.0$ & $-0.5$ \\ \hline
			 & $\fvec \lambda_{0}^{(1,1)}/a$   & $\fvec \lambda_{1}^{(1,1)}/a$   & $\fvec \lambda_2^{(1,1)}/a$ & $\fvec \lambda_3^{(1,1)}/a$  & $\fvec \lambda_{0}^{(2,1)}/a$   & $\fvec \lambda_{1}^{(2,1)}/a$   & $\fvec \lambda_2^{(2,1)}/a$ & $\fvec \lambda_3^{(2,1)}/a$ & $\fvec \lambda_{0}^{(3,1)}/a$   & $\fvec \lambda_{1}^{(3,1)}/a$   & $\fvec \lambda_2^{(3,1)}/a$ & $\fvec \lambda_3^{(3,1)}/a$ \\ \hline
			\begin{tabular}{@{}c@{}}Ref.~\cite{KoshinoPRB12} \\ unrelaxed \end{tabular} &  $(-91.9, 0)$ & $(-91.9, 0)$ & $(0, 0)$ & $(0, 0)$ & $(1.8, 0)$ & $(1.8, 0)$ & $(0, 0)$ & $(0,0)$ & \multicolumn{4}{|c|}{negligible}  \\ \hline
			\begin{tabular}{@{}c@{}}Ref.~\cite{KoshinoPRB12} \\ relaxed \end{tabular} & $(-102.0, 0)$ & $(-74.45, 0)$ & $(0, -27.6)$ & $(0, 0)$ & $(-5.4, 0)$ & $(1.3, 0)$ & $(0, -6.6)$ & $(0, 0)$ & $(8.5, -15.6)$ & $(-3.3, 7.8)$ & $(-7.8, 11.7)$ & $(0, 0)$ \\ \hline 
			\begin{tabular}{@{}c@{}}Ref.~\cite{KaxirasPRB16} \\ unrelaxed \end{tabular} &  $(-84.2, 0)$  & $(-84.2, 0)$ &  $(0, -76.1)$  & $(0.6, 0)$ & $(2.0, 0)$  & $(2.0, 0)$ & $(0, 0.3)$ & $(0, 0)$ & \multicolumn{4}{|c|}{negligible}  \\ \hline
			\begin{tabular}{@{}c@{}}Ref.~\cite{KaxirasPRB16} \\ relaxed \end{tabular} & $(-90.8, 0)$ & $(-90.0, 0)$ & $(0, -83.3)$  & $(0.7, 0)$ & $(-0.2,0)$ & $(0, 0)$ & $(0, -1.7)$ & $(0, 0)$ & $(4.5, -7.7)$ & $(3.8, 7.3)$ & $(-7.3, 4.7)$ & $(0, 0)$ \\ \hline
%			\begin{tabular}{@{}c@{}}Ref.xxx \\ relaxed \end{tabular} & $(-89.4, 0)$ & $(-88.9, 0)$ & $(0, -81.7)$  & $(0.7, 0)$ & $(0.6,0)$ & $(0.6, 0)$ & $(0, -1.0)$ & $(0, 0)$ & $(4.4, -7.8)$ & $(3.7, 7.4)$ & $(-7.4, 4.7)$ & $(0, 0)$ \\ \hline
		\end{tabular}
		\caption{Parameters of the inter-layer tunneling terms for two models, in the absence/presence of the lattice relaxation. $a$ is the magnitude of the primitive lattice vector, and all numbers are in the unit of meV. }
		\label{Tab:ParammeterInter}
	\end{table*}
%\end{widetext}
\end{widetext}

\begin{figure}[htb]
	\centering
	\subfigure[\label{Fig:KoshinoRemoteDispComp:1}]{\includegraphics[width=0.9\columnwidth]{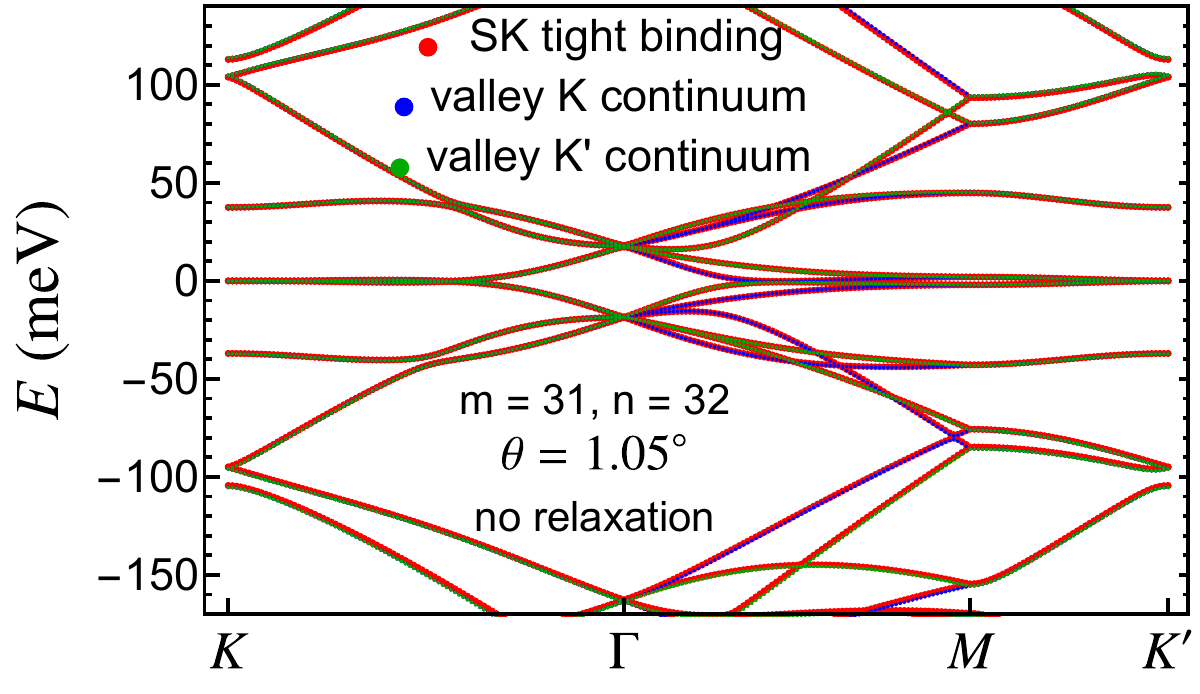}}
	\subfigure[\label{Fig:KoshinoRemoteDispComp:2}]{\includegraphics[width=0.9\columnwidth]{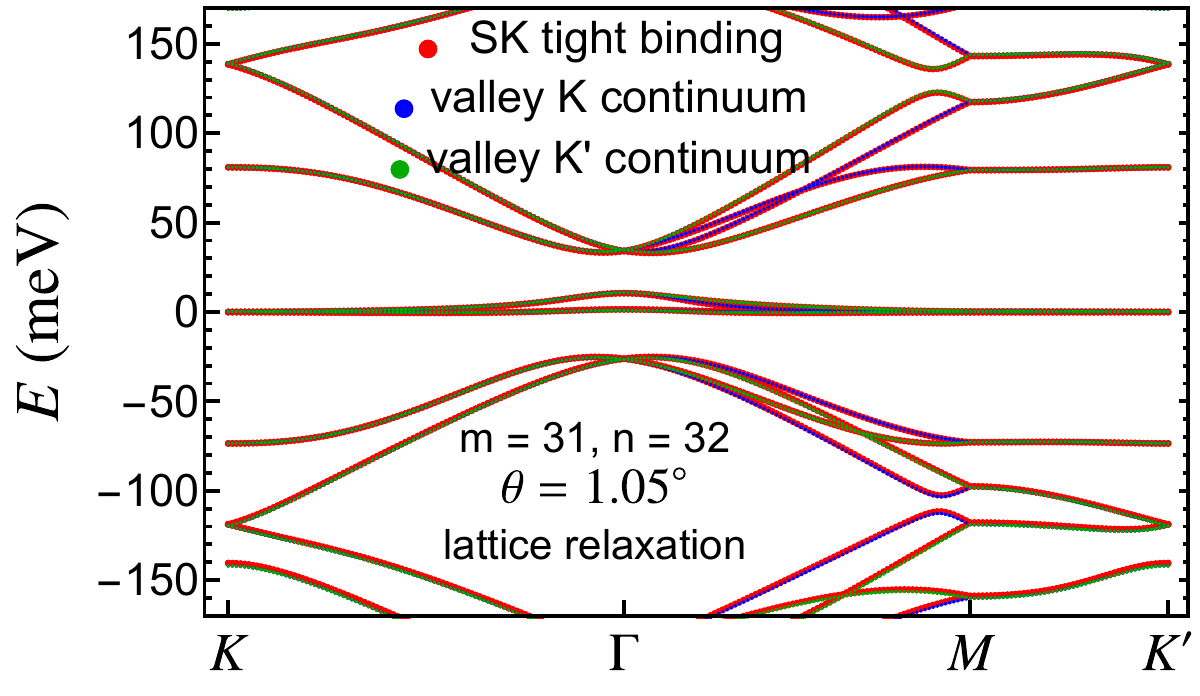}}			
	\caption{Comparison between the microscopic tight binding model (red) in Ref.~\cite{KoshinoPRB12} and the corresponding continuum model (blue for valley $\fvec K$ and green for valley $\fvec K'$)  in the absence (above) and presence (below) of the lattice relaxation.}
	\label{Fig:KoshinoRemoteDispComp}
\end{figure}
\begin{figure}[htb]
	\centering		
	\subfigure[\label{Fig:ShiangRemoteDispComp:1}]{\includegraphics[width=0.9\columnwidth]{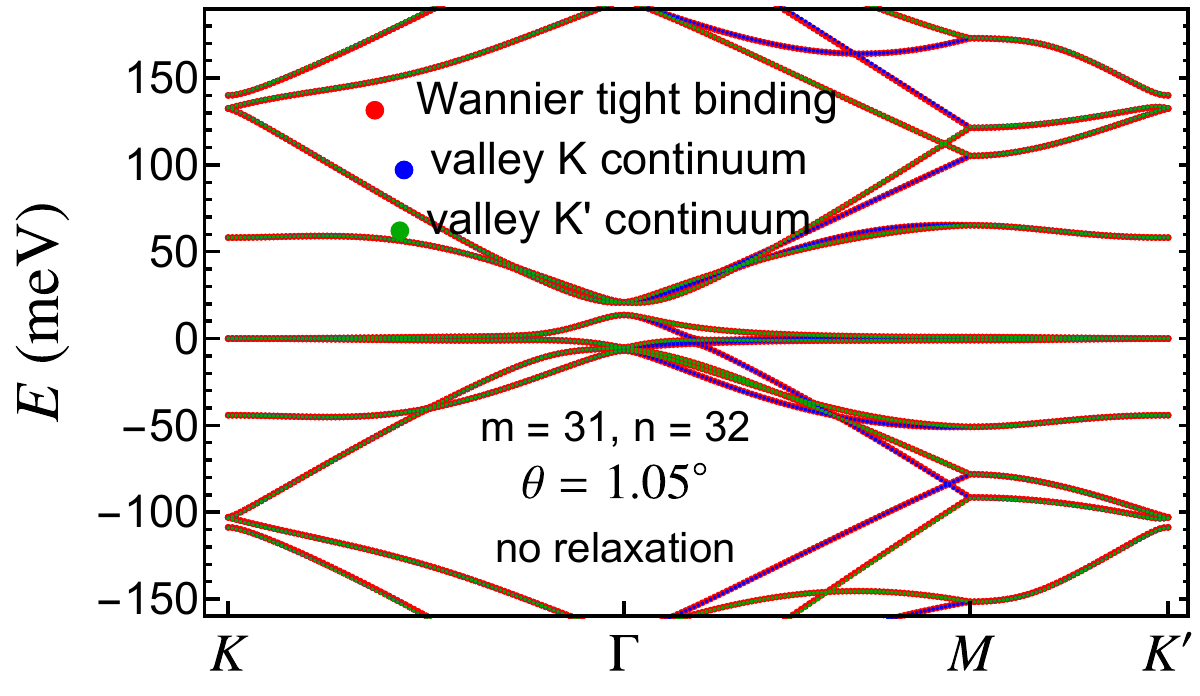}}	
	\subfigure[\label{Fig:ShiangRemoteDispComp:2}]{\includegraphics[width=0.9\columnwidth]{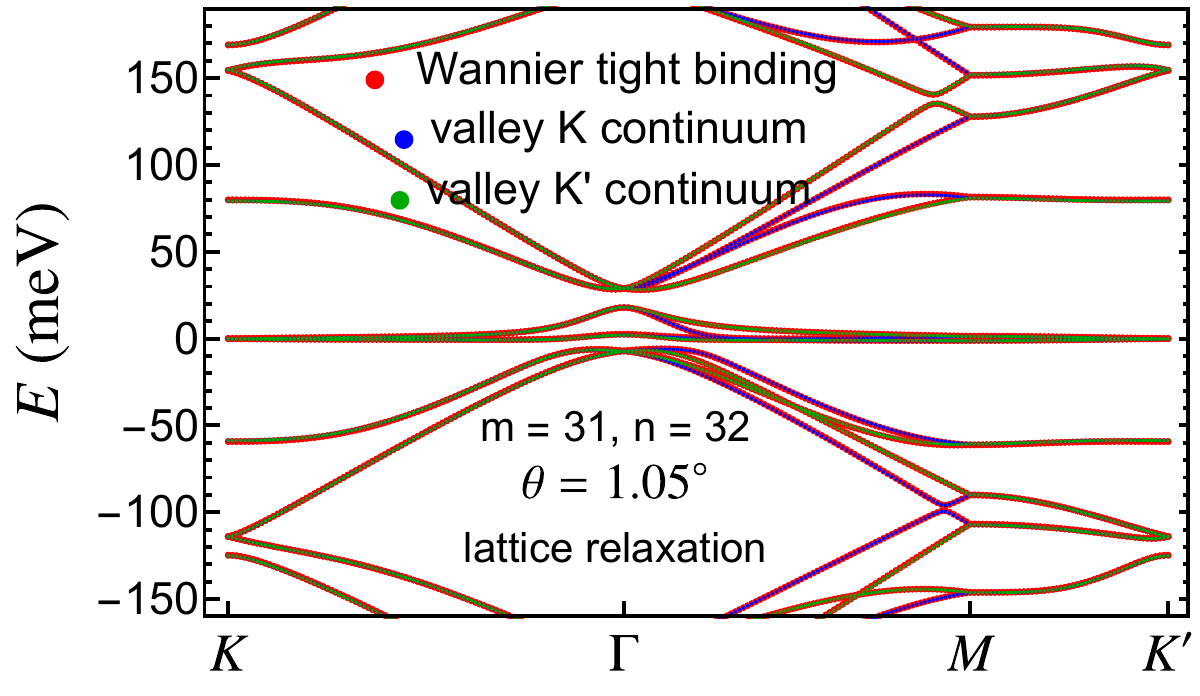}}		
	\caption{Comparison between the microscopic tight binding model (red) in Ref.~\cite{KaxirasPRB16} and the corresponding continuum model (green and blue)  in the absence (above) and presence (below) of the lattice relaxation.}
	\label{Fig:ShiangRemoteDispComp}
\end{figure}

Having obtained both the intralayer and interlayer parts of the continuum model for the moire periodic distortions, utilizing the Bloch theorem, we diagonalize $H_{eff}$ in the moire momentum space~\cite{KoshinoPRB20, GuineaModel}.
As shown in Fig.~\ref{Fig:KoshinoRemoteDispComp} and \ref{Fig:ShiangRemoteDispComp}, the spectra of $H_{eff}$ (for both valleys) and the microscopic tight binding model agree with each other beyond the narrow band regime. We have found that both spectra are consistent with each other until the energy reaches $\sim \pm 0.7$eV, where significant deviations start to rapidly grow. 

%\subsection{Wannier based hopping parameterization} 
%\label{Sec:TBMKaxiras}
%Shiang Fang and Kaxiras introduced a new set of Wannier functions into the graphene system. While the usual $p_z$ orbital is isotropic, the crystal field breaks the azimuthal symmetry into $C_{3z}$, and thus, the Wannier orbital hybridizes with higher angular momentum states. As a consequence, the interlayer tunneling is anisotropic  and given by the formula with ab initio based parameterization. (ref.)

\section{Analysis of the narrow band Hilbert space: sublattice polarization, p-h symmetry and Wilson loops}
\label{Sec:Analysis}
Having obtained the energy spectrum of $H^\bK_{eff}$ presented in the previous section, we now turn to the properties of the Hilbert space spanned by the narrow bands. While narrow bands appear in both models near the CNP when the twist angle is $1.05^{\circ}$, the corresponding states are found to be notably different. In this section, we consider three properties of the narrow band Hilbert space at the valley $\bK$: the sublattice polarization, the deviation from the p-h symmetry, and the Wilson loop.

\begin{figure}[htb]
	\centering		
	\subfigure[\label{Fig:Sublattice:1}]{\includegraphics[width=0.65\columnwidth]{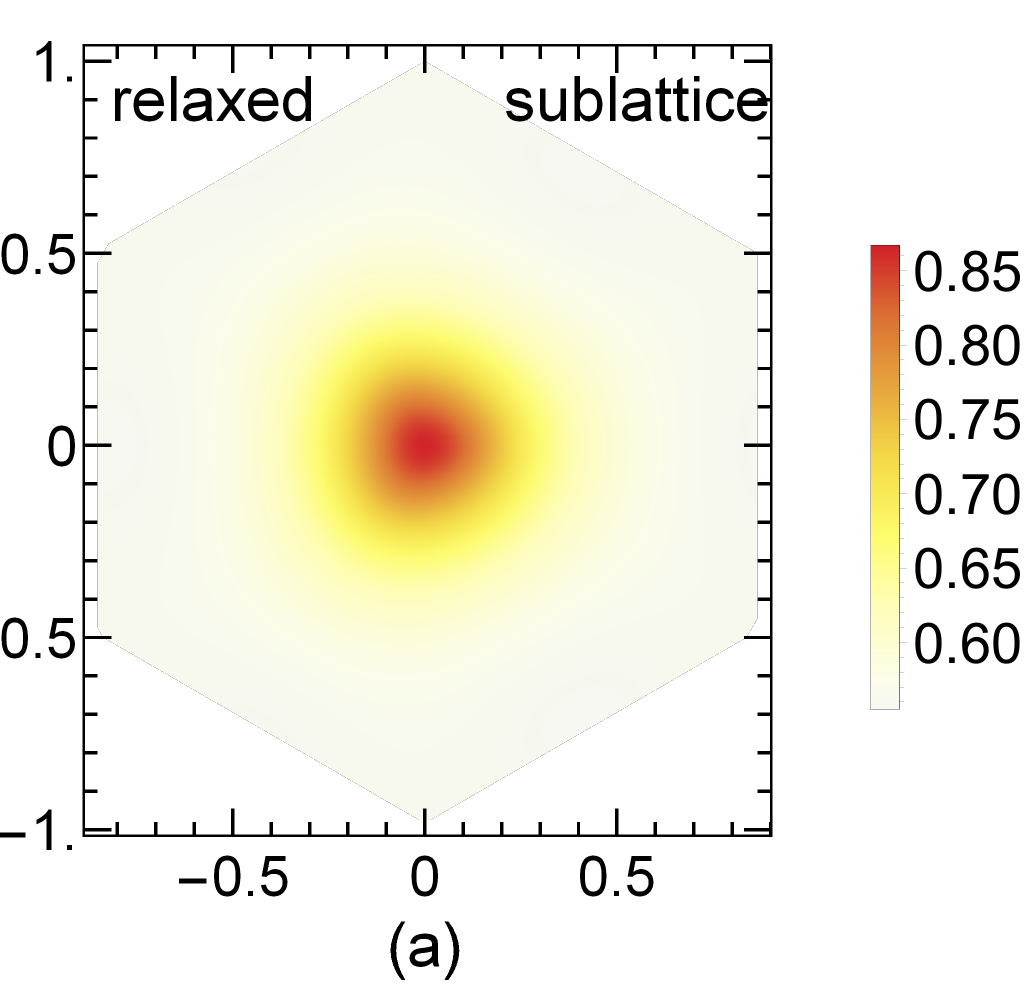}}		
	\subfigure[\label{Fig:Sublattice:2}]{\includegraphics[width=0.65\columnwidth]{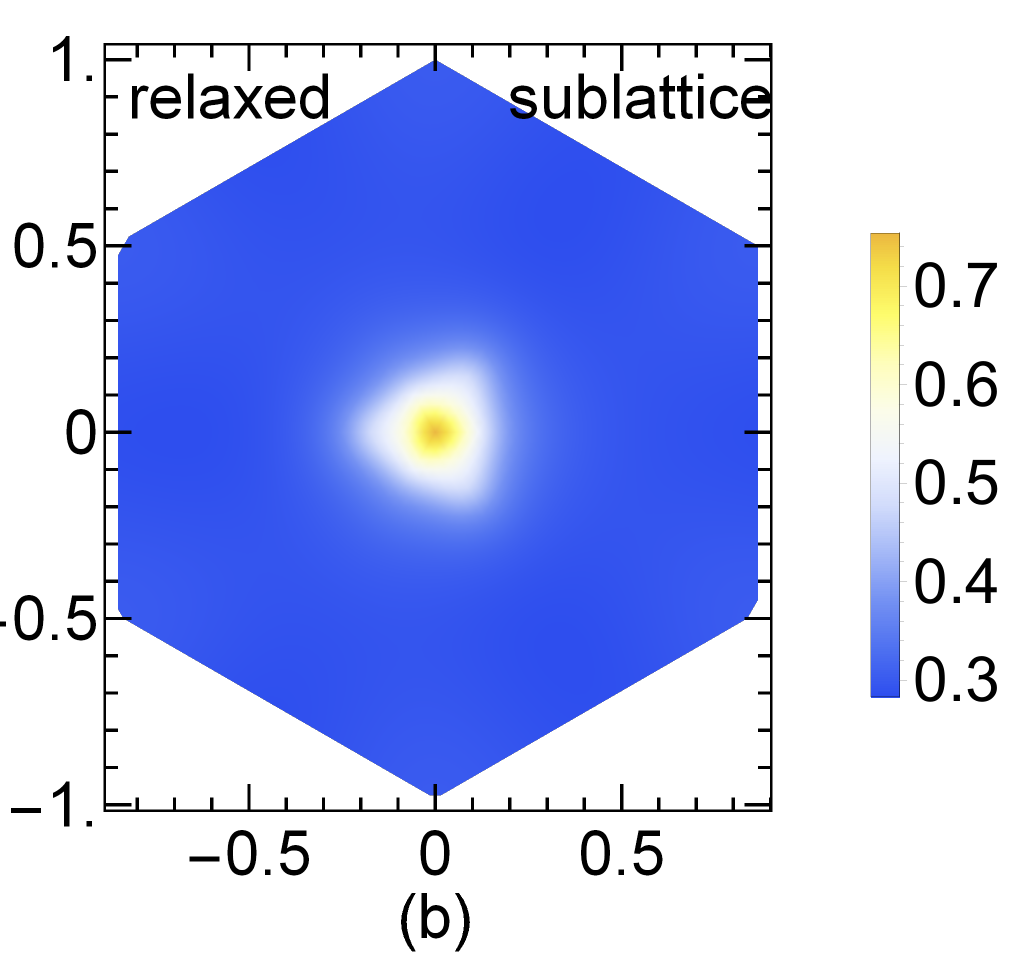}}	
	\caption{The sublattice polarization of the narrow bands for two microscopic models in (a) Ref.~\cite{KoshinoPRB12} and (b) Ref.~\cite{KaxirasPRB16}.}
	\label{Fig:Sublattice}
\end{figure}

The sublattice polarization of the narrow bands is defined via the eigenvalues of the $2\times 2$ projected sublattice matrix $\mathcal{S}_{ij}(\fvec k) = \langle \Psi_i(\fvec k) | \sigma_z | \Psi_j(\fvec k) \rangle$, where $\Psi_{i}(\fvec k)$ is the Bloch state with the momentum of $\fvec k$ in the band $i$, and $\sigma_z$ is the sublattice polarization operator. Because of the $C_2\mathcal{T}$ symmetry, the two eigenvalues of the projected $\sigma_z$ have the same magnitude with opposite signs, $\pm1$ corresponding to the perfect polarization obtained in the chiral limit~\cite{Grisha}. The sublattice polarization calculated based on the $H_{eff}^{\bK}$ is shown in the Fig.~\ref{Fig:Sublattice} for the two microscopic models. While the narrow band Hilbert space in either model is not perfectly polarized, the one proposed in Ref.~\cite{KoshinoPRB12} has larger sublattice polarization than the one in Ref.~\cite{KaxirasPRB16}, implying the former model is closer to the chiral limit than the latter. 

\begin{figure}[t]
	\centering		
	\subfigure[\label{Fig:PHSVD:1}]{\includegraphics[width=0.99\columnwidth]{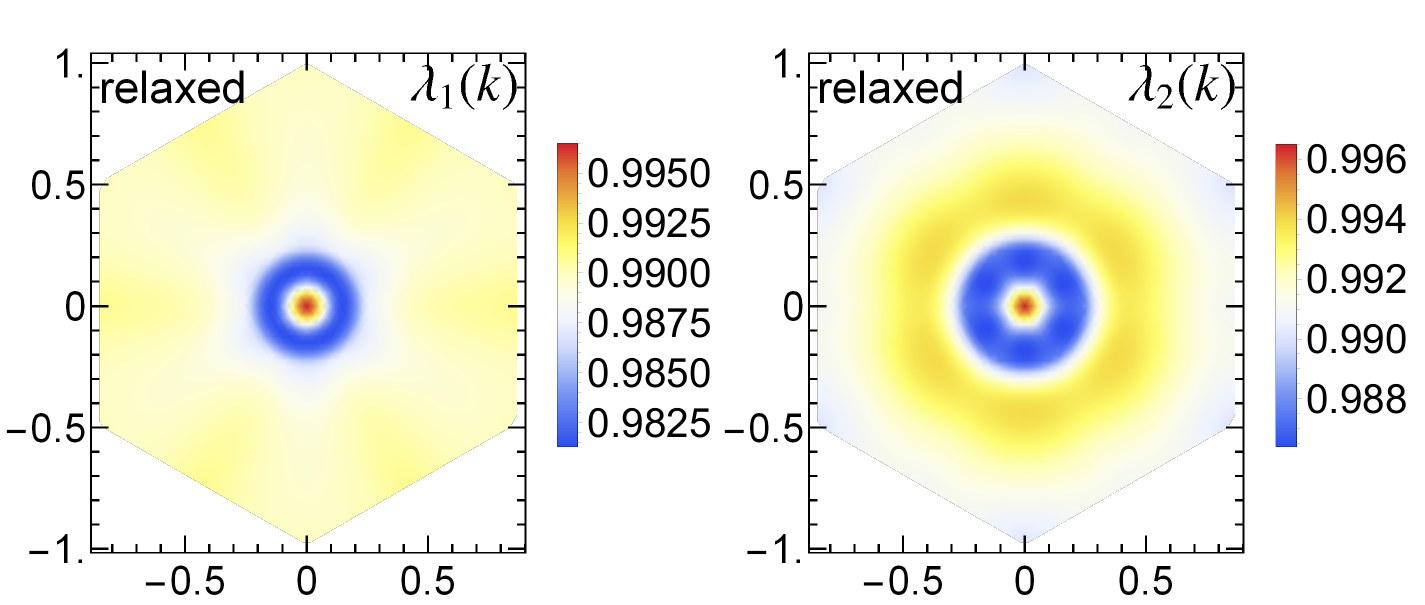}}		
	\subfigure[\label{Fig:PHSVD:2}]{\includegraphics[width=0.99\columnwidth]{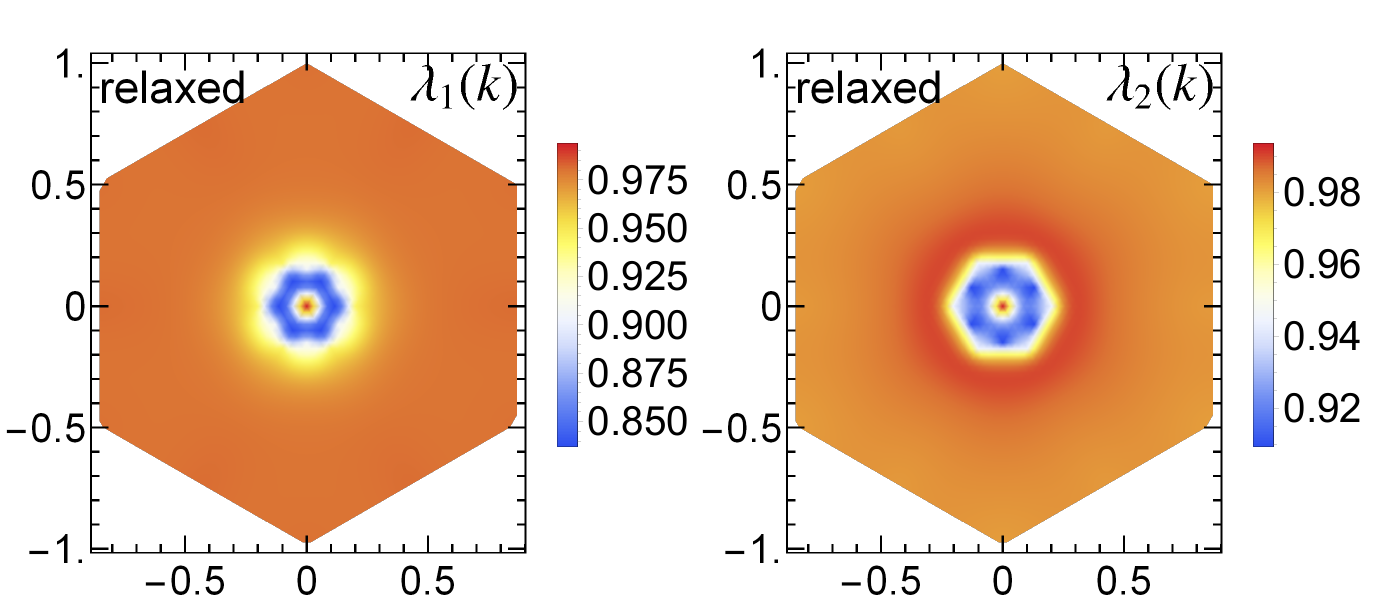}}	
	\caption{The two singular values (left) $\lambda_1(\fvec k)$ and (right) $\lambda_2(\fvec k)$  of the p-h matrix $\mathcal{P}(\fvec k)$ for two models in Ref.~\cite{KoshinoPRB12} and Ref.~\cite{KaxirasPRB16}. Their deviations from $1$ measure the p-h asymmetry of the narrow bands.}
	\label{Fig:PHSVD}
\end{figure}

The p-h symmetry~\cite{Bernevig1} plays an important role in that it leads to the $U(4)$ symmetry of the projected Coulomb interaction, and helps in identifying the ground state in the strong coupling limit. The p-h transformation $\hat{\mathcal{P}}$ acts within a valley and is defined as $i \mu_y$  --the interchange of the two layers and change the sign of the top layer--  followed by the in-plane inversion $\fvec r \rightarrow - \fvec r$. When keeping only the first order intra-layer gradient terms and only the contact inter-layer terms in the BM model, we have $\hat{\mathcal{P}}^{\dagger} H_{BM}(\fvec k) \hat{\mathcal{P}} = - H_{BM}(-\fvec k)$. However, $\hat{\mathcal{P}}$ is only an approximate symmetry as it is generally broken by the higher order gradient terms. For example, it is broken by the $O(k^2)$ and $O(k\partial U)$  terms in $H_{intra}$, as well as $w_3$ and the vector couplings $\fvec \lambda_{i \neq 3}$ in the inter-layer tunnelings. In order to quantify the degree of the p-h symmetry violation within the narrow bands, we define the $2 \times 2$ projected p-h matrix as $\mathcal{P}_{i j}(\fvec k) = \langle \Psi_{i}(-\fvec k) | \hat{\mathcal{P}} |  \Psi_{j, \fvec k} \rangle$.
%i.e.~$\hat P = i \mu_y$ followed by the inversion $\fvec r \rightarrow - \fvec r$, and $\mu_y$ is the Pauli matrix operated in the layer space. $\Psi_{i, \fvec k}$ is the state on the narrow band $i$ with the momentum of $\fvec k$.  
If the p-h symmetry is exact, it is expected that the matrix $\mathcal{P}_{ij}(\fvec k)$ is unitary for arbitrary $\fvec k$, and thus both the singular values, $\lambda_1(\fvec k)$ and $\lambda_2(\fvec k)$, are $1$. Otherwise, $\lambda_1(\fvec k)$ and $\lambda_2(\fvec k)$ are smaller than $1$, and therefore, the deviation of $\lambda_i$ ($i = 1$, $2$) from $1$ measures the p-h asymmetry of the Hilbert space. Fig.~\ref{Fig:PHSVD} illustrates the two singular values $\lambda_i$ for both models. While the narrow bands in Ref.~\cite{KoshinoPRB12} are almost perfectly p-h symmetric, those in Ref.~\cite{KaxirasPRB16} shows significant p-h asymmetry. As we demonstrate in the Fig.~\ref{FigS:KaxirasPH}, the dominant source of the p-h asymmetry in the model of Ref.~\cite{KaxirasPRB16} comes from the inter-layer contact coupling $w_3$ and the sub-dominant contribution comes from the gradient coupling $\fvec \lambda$.

\begin{figure}[htb]
	\centering		
	\subfigure[\label{Fig:WSLoop:1}]{\includegraphics[width=0.9\columnwidth]{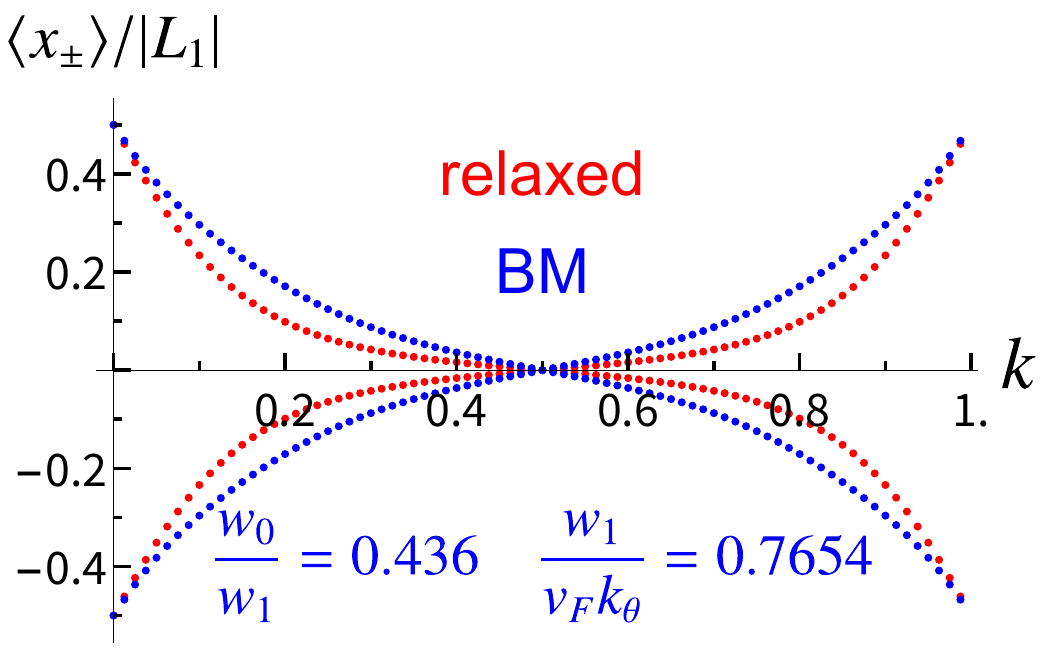}}		
	\subfigure[\label{Fig:WSLoop:2}]{\includegraphics[width=0.9\columnwidth]{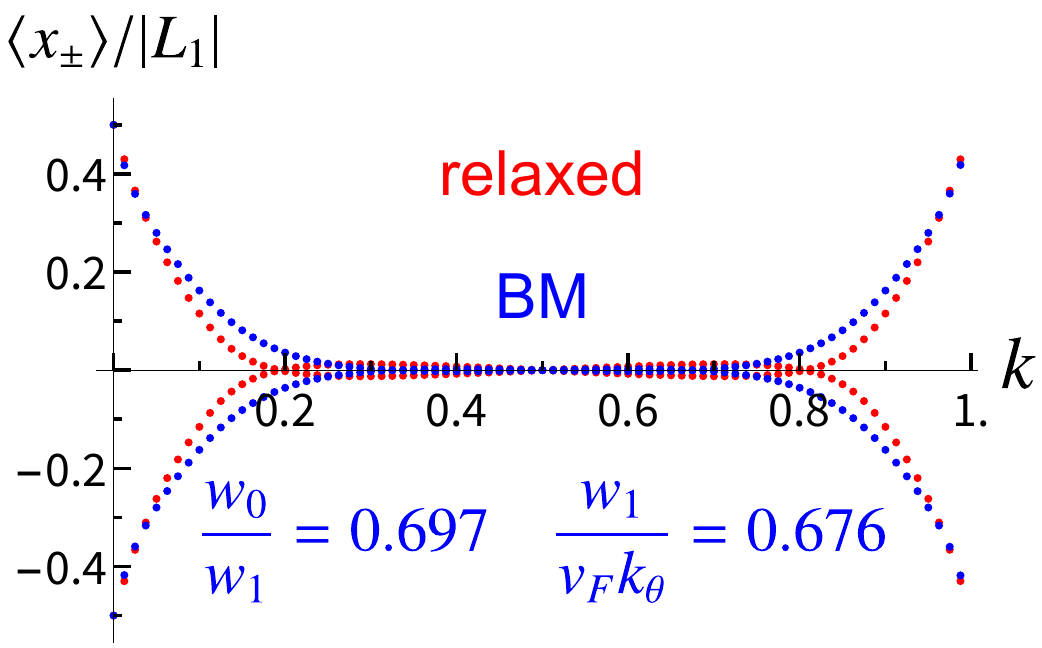}}	
	\caption{The phase of the eigenvalues of the Wilson loop operator of the two valley-polarized narrow bands (red) with different continuum model constructed for two microscopic tight binding models proposed by (a) Ref.~\cite{KoshinoPRB12} and (b) Ref.~\cite{KaxirasPRB16}. For comparison, the same phase of the eigenvalues has also been calculated for the BM model (blue) that contains only $v_F$, $w_0$ and $w_1$ with their numerical values taken from Table.~\ref{Tab:ParammeterRelax} and \ref{Tab:ParammeterInter}.}
	\label{Fig:WSLoop}
\end{figure}

Finally, we consider the eigenvalues of Wilson-loop operator as another property of the Hilbert space of the narrow bands. This operator $\hat O$ is defined as $P e^{-i \frac1N_1 \fvec g_1 \cdot \fvec r} P$ where $N_1$ is the number of unit cells along the direction of $\fvec L_1$ in the entire lattice with periodic boundary conditions, and $P$ is the projection operator onto the Bloch states of the narrow bands. Since this operator commutes with the momentum operator along $\fvec g_2$, its eigenstate is labeled by the momentum $k$ along $\fvec g_2$. In the BM model, the phase of the eigenvalues of $\hat O$, labeled as $\langle x_{\pm} \rangle$, has the winding number of $\pm 1$ as $k$ runs from $0$ to $1$, illustrating the nontrivial topological properties of the narrow band system~\cite{Bernevig1,Dai1,KangVafekPRB}.

In addition, if $w_0 = 0$, the system is in the chiral limit and $\langle x_{\pm} \rangle$ is almost a linear function of $\fvec k$, but becomes quite flat when $\fvec k \sim 0.5$ if the system is far away from the chiral limit ($w_0 \lesssim w_1$)~\cite{Bernevig1} (also see Fig.~1 in Ref.~\cite{KangVafekPRB}).  This behavior is also qualitatively reproduced in our constructed $H_{eff}^{\bK}$, as shown in Fig.~\ref{Fig:WSLoop}. For comparison, Fig.~\ref{Fig:WSLoop} also shows the winding of the phase $\langle x_{\pm} \rangle$ obtained from the BM model, with the values of the parameters $v_F$, $w_0$, and $w_1$ taken from Tab.~\ref{Tab:ParammeterRelax}. It is found that the curve for the BM model is straighter, suggesting that the terms neglected in the BM model but present in $H_{eff}^{\bK}$, drive the system further away from the chiral limit. Moreover, the curve for the model in Ref.~\cite{KoshinoPRB12} is straighter than the one for the model in Ref.~\cite{KaxirasPRB16}, and thus consistent with the former model being closer to the chiral limit than the latter.

\section{Exactly flat band Limit with relaxation induced pseudomagnetic fields}
\label{Sec:ChiralLimit}
The inclusion of the strain is believed to greatly increase the bandwidth at the magic angle~\cite{BiStrain, NickStrain,NickKekule}. Having seen that the relaxed atomic configuration of the twisted bilayer graphene obtained in Sec.~\ref{Sec:LatRelax} expands the AB/BA stacked regions and shrinks the AA stacked regions relative to just a rigid twist, 
thus intrinsically inducing strain, it is interesting to ask whether the bandwidth undergoes the increase as well. 
Motivated by this question, the goal of this section is to generalize the chiral limit introduced and analyzed in Ref.~\cite{Grisha} for the BM model including the relaxation induced pseudo-magnetic vector potential $\mathcal{A}$. 
While the relaxation induced $\mathcal{A}$ indeed increases the vanishing bandwidth at the magic angle found in Ref.~\cite{Grisha} without $\mathcal{A}$, we demonstrate below that decreasing the twist angle can compensate the effect of $\mathcal{A}$ on the bandwidth, resulting in exactly flat bands at the CNP at a new (smaller) magic angle. Throughout our analysis we pay particular attention to the importance of $C_3$ symmetry (preserved by $\mathcal{A}$) in making the compensation possible, noting that extrinsically induced strain generally breaks $C_3$.

For the purposes of this section, we start from the Hamiltonian 
\begin{align}
	H_{chiral} = \begin{pmatrix}
		v_F  \bar{\fvec \sigma}_{\frac{\theta}2} \cdot (\fvec p + \gamma \mathcal{A}) & T(\fvec x) \\ T^{\dagger}(\fvec x) & v_F \bar{\fvec \sigma}_{-\frac{\theta}2} \cdot ( \fvec p - \gamma \mathcal{A}) 
	\end{pmatrix}  \label{Eqn:HamExactFlat}
\end{align}
that acts on the four component spinor $\Psi(\fvec x) = (\psi_{t, A}(\fvec x), \psi_{t, B}(\fvec x), \psi_{b, A}(\fvec x), \psi_{b, B}(\fvec x))^T$, where the subscripts $t$/$b$ label the top/bottom layers and $A$/$B$ labels the sublattice.

In the equation above, $\bar{\fvec \sigma}_{\theta/2} = e^{-i\frac{\theta}4 \sigma_3} \bar{\fvec \sigma} e^{i \frac{\theta}4 \sigma_3}$ and $\mathcal{A}(\fvec x)$ is the real inhomogeneous pseudo-magnetic vector potential induced by the lattice relaxation, as calculated by Eq.~\ref{Eqn:AField}; just as before  $\bar{\fvec \sigma} = (\sigma_x, - \sigma_y)$. This $\mathcal{A}$ field is invariant under all the symmetry transformations discussed in Sec.~\ref{Sec:LatRelax}, such as $C_3$, $C_2 \mathcal{T}$, etc. Compared with the full $H_{intra}$ in Eq.~\ref{Eqn:HIntra}, all the second order terms have been neglected in Eq.~\ref{Eqn:HamExactFlat}. In addition, the inter-layer tunnelings also neglect the gradient coupling $\fvec \Lambda$, as well as the $w_0$ and $w_3$ terms in the contact coupling. Thus, the inter-layer tunneling $T(\fvec x)$ in $H_{chiral}$ can be written as
\begin{align}
    T(\fvec x) = \sum_{\mu, l} \left( w_1^{(\mu, l)} \sigma_1 + w_2^{(\mu, l)} \sigma_2 \right) e^{i \fvec q_{\mu, l} \cdot \fvec x}. \label{Eqn:ChiralT}
\end{align}
Since $T(x)$ contains only $\sigma_1$ and $\sigma_2$, $e^{- i \frac{\theta}4 \sigma_3} T(\fvec x ) e^{-i \frac{\theta}4 \sigma_3} = T(\fvec x)$. Introducing the unitary diagonal matrix: $\mathcal{U} =  \diag( e^{-i \frac{\theta}4}, e^{i \frac{\theta}4}, e^{i \frac{\theta}4}, e^{-i \frac{\theta}4})$,  $H_{chiral}$ can be simplified by applying the unitary transformation 
\begin{align}
    & H'_{chiral} = \mathcal{U}^{\dagger} H_{chiral} \mathcal{U} \nonumber \\
    =  & \begin{pmatrix}
    	v_F \bar{\fvec \sigma} \cdot (\fvec p + \gamma \mathcal{A}) & T(\fvec x) \\ T^{\dagger}(\fvec x) & v_F \bar{\fvec \sigma} \cdot (\fvec p - \gamma \mathcal{A}) 
    \end{pmatrix} 
\end{align}
while the transformed spinor is labelled as $\Phi(\fvec x) = \mathcal{U}^{\dagger} \Psi(\fvec x) = (\phi_{t, A}(\fvec x), \phi_{t, B}(\fvec x), \phi_{b, A}(\fvec x), \phi_{b, B}(\fvec x))^T$. Again, note that each $2 \times 2$ blocks of $H'_{chiral}$ contains only $\sigma_1$ and $\sigma_2$, thus $\{ \sigma_3 \otimes I, H'_{chiral}\} = 0$, i.e.~the chiral Hamiltonian is anti-symmetric under the chiral p-h transformation $\sigma_3 \otimes I$.  

First, we consider the states near the CNP at the corner of the moire BZ, i.e.~$\fvec K_m$ or $\fvec K_m'$. Turning off the inter-layer tunneling $T(\fvec x)$, we will show that two zero modes still exist even in the presence of the $\mathcal{A}$ field. To prove it, consider the equation for the zero modes at $\fvec K_m$,
\begin{align}
    & \bar{\fvec \sigma} \cdot (\fvec p + \gamma \mathcal{A}) \begin{pmatrix}
        \phi_{t,A}(\fvec x) \\ \phi_{t,B}(\fvec x)  
    \end{pmatrix}= 0, \\
    \longrightarrow & \left\{ \begin{array}{l}
       ((-i \partial_1 + \partial_2) + \gamma(\mathcal{A}_1 + i \mathcal{A}_2)) \phi_{t,B} = 0,   \\
 (-i \partial_1 - \partial_2) + \gamma(\mathcal{A}_1 - i \mathcal{A}_2) \phi_{t,A} = 0.
    \end{array} \right.   \label{Eqn:ZeroMode}
\end{align}
Using the Helmholtz decomposition in Eq.~\ref{Eqn:ADecomp}, $\mathcal{A} = (\partial_1 \varphi^{\mathcal{A}} + \partial_2 \varepsilon^{\mathcal{A}}, \ \partial_2 \varphi^{\mathcal{A}} - \partial_1 \varepsilon^{\mathcal{A}})$, we immediately obtain the two independent solutions to Eq.~\ref{Eqn:ZeroMode}, 
\begin{align}
    \left\{ \begin{array}{l}
         \phi_{t,A} = e^{-i \gamma \varphi^A} e^{\gamma \varepsilon^A}  \\
         \phi_{t,B} = 0
    \end{array} \right. \quad \left\{ \begin{array}{l}
         \phi_{t,A} = 0  \\
         \phi_{t,B} = e^{-i \gamma \varphi^A} e^{-\gamma \varepsilon^A} 
    \end{array} \right. \label{Eqn:ZeroModeKSol}
\end{align}
The pseudo vector field $\mathcal{A}$ is periodic and its average over space $\langle \mathcal{A} \rangle = 0$, and so are $\varphi^{\mathcal{A}}$ and $\varepsilon^{\mathcal{A}}$. Therefore, the two solutions in Eq.~\ref{Eqn:ZeroModeKSol} are also bounded and periodic, giving the two zero modes at $\fvec K_m$. 

Under $C_3$ transformation, the spinor $\Phi(\fvec x) \rightarrow e^{i \frac{2\pi}3 \sigma_z} \Phi(\fvec x')$, where $\fvec x'$ is the position $\fvec x$ rotated clockwise by $2\pi/3$~\cite{Grisha}. Therefore, the two zero modes in Eq.~\ref{Eqn:ZeroModeKSol} carry the extra phases of $e^{i2\pi/3}$ and $e^{-i2\pi/3}$ respectively, and thus transform differently under $C_3$. Furthermore, the chiral p-h transformation $\sigma_z \otimes I$ commutes with $C_3$. As the inter-layer tunneling $T(\fvec x)$ is gradually turned on~\cite{Grisha}, each of these two modes at $\fvec K_m$ must transform to itself under $\sigma_z \otimes I$, and therefore each still has zero energy. As a consequence, the two bands around the CNP touch at the Dirac cone at $\fvec K_m$ even when $\mathcal{A}$ is included. 

Following the arguments presented in Ref.~\cite{Grisha}, we can also express the Fermi velocity of the Dirac cone in terms of the wavefunction at $\fvec K_m$. For this purpose, we choose the basis  $\Phi' = (\phi_{t,A}, \phi_{b, A}, \phi_{t, B}, \phi_{b, B})$, and the zero modes at $\fvec K_m$ satisfy the equation
\begin{align}
    & \begin{pmatrix}
        0 & \mathcal{D}(\fvec x) \\ \mathcal{D}^{\dagger}(\fvec x) & 0
    \end{pmatrix} \begin{pmatrix}
        \Phi_{\fvec K_m, A}(\fvec x) \\ \Phi_{\fvec K_m, B}(\fvec x)
    \end{pmatrix} = 0,
\end{align}
where $\Phi_{\fvec K_m, A} = (\phi_{\fvec K_m, t, A},\ \phi_{\fvec K_m, b, A})$ and $\Phi_{\fvec K_m, B} = (\phi_{\fvec K_m, t, B},\ \phi_{\fvec K_m, b, B})$ are two component spinors. $\mathcal{D}(\fvec x)$ is a $2 \times 2$ matrix differential operator of the form
\begin{align}
   \mathcal{D}(\fvec x) = \begin{pmatrix}
        v_F \pi_+(\fvec x) & U(\fvec x) \\
        U(-\fvec x) & v_F \pi_-(\fvec x) 
   \end{pmatrix},
\end{align}
where 
\begin{align}
    U(\fvec x) & = \sum_{\mu, l} \left( w_1^{(\mu, l)} - i w_2^{(\mu, l)}\right) e^{i \fvec q_{\mu, l}\cdot \fvec x}, \\
    \pi_{\pm} & = p_1 + i p_2 \pm \gamma (\mathcal{A}_1 + i \mathcal{A}_2).
\end{align}
Since $\mathcal{A}(\fvec x) = \mathcal{A}(- \fvec x)$ due to the $C_2 \mathcal{T}$ symmetry, $\mathcal{D}^{\dagger}(\fvec x) = \mathcal{D}^*(- \fvec x)$. Thus, if the two component spinor $\varPsi(\fvec x)$ satisfies $\mathcal{D}(\fvec x) \varPsi(\fvec x) = 0$ so that $(0,\ \varPsi(\fvec x))$ is a zero mode at $\fvec K_m$, the spinor $\varPsi^*(-\fvec x)$ satisfies $\mathcal{D}^{\dagger}(\fvec x) \varPsi^*(-\fvec x) = 0$, and therefore, $(\varPsi^*(-\fvec x),\ 0)$ is another zero mode. As the momentum $\fvec p$ slightly deviates from $\fvec K_m$, $\bar{\fvec \sigma} \cdot \fvec p$ can be treated as perturbation and thus the Fermi velocity of the Dirac cone at $\fvec K_m$ is
\begin{align}
    v_{Dirac} = v_F \frac{\left|\langle \varPsi^*(-\fvec x) | \varPsi(\fvec x)\rangle \right|}{\langle \varPsi | \varPsi \rangle}  \ . \label{Eqn:vDirac}
\end{align}
Because $v_{Dirac}$ is a real number, in principle, it can vanish by tuning the inter-layer coupling constants. 

For the BM model in the chiral limit, Ref.~\cite{Grisha} showed that the bands around the CNP become exactly flat as long as $v_{Dirac} = 0$. Their argument is still valid  when the relaxation induced pseudo-vector field $\mathcal{A}$ is present. To prove this statement, we consider the equation 
\begin{align}
	\mathcal{D}(\fvec x) \varPsi_{\fvec k, B}(\fvec x) = 0 \ , \label{Eqn:DxEqn}
\end{align}
where $\fvec k$ is the Bloch momentum. Introducing the complex coordinates $z = x_1 + i x_2$ and $\bar{z} = x_1 - i x_2$, we find 
\begin{align}
	\mathcal{D}(\fvec x) = v_F \begin{pmatrix}
		 -2i \bar{\partial} + \gamma (\mathcal{A}_1 + i \mathcal{A}_2) &  v_F^{-1} U(\fvec x) \\ v_F^{-1} U(-\fvec x) & -2i \bar{\partial} - \gamma (\mathcal{A}_1 + i \mathcal{A}_2) 
	\end{pmatrix},
\end{align}
where $\bar{\partial} = \partial_{\bar{z}} = \half(\partial_1 + i \partial_2)$. Note that the differential operators in $\mathcal{D}(\fvec x)$ contains only $\bar{\partial}$, i.e. $\partial_z$ is absent. Since we already showed that $\Phi_{\fvec K_m, B}$ is the solution of Eq.~\ref{Eqn:DxEqn},  $f(z) \Phi_{\fvec K_m, B}$ is also a solution as long as $f(z)$ is a holomorphic function because then $\bar{\partial} f(z) = 0$. To construct the solution with the Bloch boundary conditions at an arbitrary momentum $\fvec k$, introduce\cite{Grisha}
\begin{align}
	\eta(z = x_1 + i x_2) = \frac{\vartheta_{\frac{\fvec k \cdot \fvec L_1}{2\pi} - \frac16, \frac16 - \frac{\fvec k \cdot (\fvec L_2 - \fvec L_1)}{2\pi}}\left( \frac{z}{|L_1|} e^{-i \frac{\pi}6}, e^{i \frac{2\pi}3}  \right)}{\vartheta_{-\frac16, \frac16}\left( \frac{z}{|L_1|} e^{-i \frac{\pi}6}, e^{i \frac{2\pi}3}  \right)} \label{Eqn:Etaz}
\end{align}
where $\vartheta$ is the theta function, defined as
\begin{align}
	\vartheta_{a, b}(z, \tau) = \sum_{n = - \infty}^{\infty} e^{i \pi \tau (n + a)^2} e^{2\pi i (n + a)(z + b)} \ .
\end{align}
Note that $\eta(z)$ satisfies the boundary condition $\eta(\fvec x + \fvec L_{1, 2}) = e^{i \fvec k \cdot \fvec L_{1, 2}} \eta(\fvec x)$. Then,  the Bloch wavefunction
\begin{align}
	\varPsi_{\fvec k, B}(\fvec x) =  \eta(\fvec x) \Phi_{\fvec K_m, B}(\fvec x) \label{Eqn:DxSol}
\end{align}
is a solution of Eq.~\ref{Eqn:DxEqn}. However, $\varPsi_{\fvec k, B}$ contains singular points because $\eta(\fvec z)$ is not an analytical function. The denominator in Eq.~\ref{Eqn:Etaz} vanishes at $\fvec x_0  + m \fvec L_1 + n \fvec L_2$ where $\fvec x_0 = - \frac13 (\fvec L_1 + \fvec L_2)$, and $m$ and $n$ are arbitrary integers, and in general, $\varPsi_{\fvec k, B}$ is not normalizable\cite{Grisha}. 

The solution in Eq.\ref{Eqn:DxSol} can still be physical if $\Phi_{\fvec K_m, B}(\fvec x_0) = 0$ thus canceling the zero in the denominator. Without the pseudo-vector potential $\mathcal{A}$, it can be achieved when $v_{Dirac}$ vanishes~\cite{Grisha}. The same argument also applies if $\mathcal{A}$ is present. To prove it, consider the zero mode $\Phi_{\fvec K_m, B}(\fvec x) = (\Phi_{\fvec K_m, t, B}(\fvec x), \ \Phi_{\fvec K_m, b, B}(\fvec x))$. It satisfies
\begin{align}
	0= & \Phi_{\fvec K_m, B}^T(-\fvec x) \mathcal{D}(\fvec x) \Phi_{\fvec K_m, B}(\fvec x) \nonumber \\
	= & \Phi_{\fvec K_m, B}^T(\fvec x) \mathcal{D}(- \fvec x) \Phi_{\fvec K_m, B}(- \fvec x).
\end{align}
Using $\mathcal{A}(\fvec x) = \mathcal{A}(-\fvec x)$, we obtain
\begin{align}
	\bar{\partial} \left( \Phi_{\fvec K_m, B}^T(-\fvec x) \Phi_{\fvec K_m, B}(\fvec x) \right) = 0 \ .
\end{align}
For notational convenience, we introduce $v(\fvec x) = \Phi_{\fvec K_m, B}^T(-\fvec x) \Phi_{\fvec K_m, B}(\fvec x)$. By the above formula, $v(\fvec x)$ is a constant in space.  If $v_{Dirac} = 0$, by Eq.~\ref{Eqn:vDirac}, $v(\fvec x)$ must vanish everywhere including $\fvec x = \fvec x_0$. Note that 
\begin{align} 
	v(\fvec x_0) = & \Phi_{\fvec K_m, t, B}(-\fvec x_0) \Phi_{\fvec K_m, t, B}(\fvec x_0) + \nonumber \\
	&  \Phi_{\fvec K_m, b, B}(-\fvec x_0) \Phi_{\fvec K_m, b, B}(\fvec x_0)  \ .
\end{align}
Due to the symmetry of  $C_3$ rotation around $\fvec x_0$, $\Phi_{\fvec K_m, b, B}(\pm \fvec x_0) = 0$~\cite{Grisha}. Therefore, if $v_{Dirac}=0$, then either $\Phi_{\fvec K_m, t, B}(\fvec x_0) = 0$ or $\Phi_{\fvec K_m, t, B}(-\fvec x_0) = 0$. In the former case, both components of $\Phi_{\fvec K_m, B}$ vanish at $\fvec x_0$, and therefore Eq.~\ref{Eqn:DxSol} is a normalizable solution if $v_{Dirac} = 0$. If the latter is the case, redefine
\begin{align}
	\eta(z) = \frac{\vartheta_{\frac{\fvec k \cdot \fvec L_1}{2\pi} + \frac16, -\frac16 - \frac{\fvec k \cdot (\fvec L_2 - \fvec L_1)}{2\pi}}\left( \frac{z}{|L_1|} e^{-i \frac{\pi}6}, e^{i \frac{2\pi}3}  \right)}{\vartheta_{\frac16, -\frac16}\left( \frac{z}{|L_1|} e^{-i \frac{\pi}6}, e^{i \frac{2\pi}3}  \right)} \label{Eqn:Etaz2}
\end{align}
so that the denominator vanishes at $- \fvec x_0 + m \fvec L_1 + n \fvec L_2$, and thus all the above arguments follow. 

\begin{figure}[t]
	\centering		
	\subfigure[\label{Fig:FLatBand:1}]{\includegraphics[width=0.9\columnwidth]{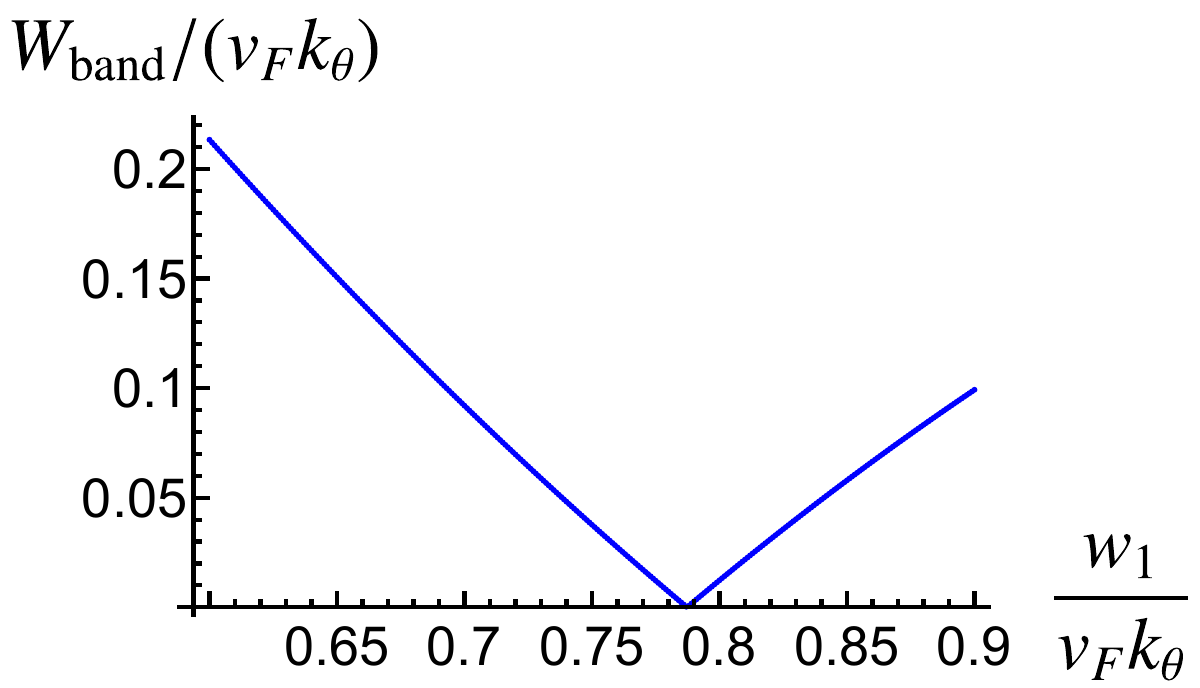}}		
	\subfigure[\label{Fig:FLatBand:2}]{\includegraphics[width=0.9\columnwidth]{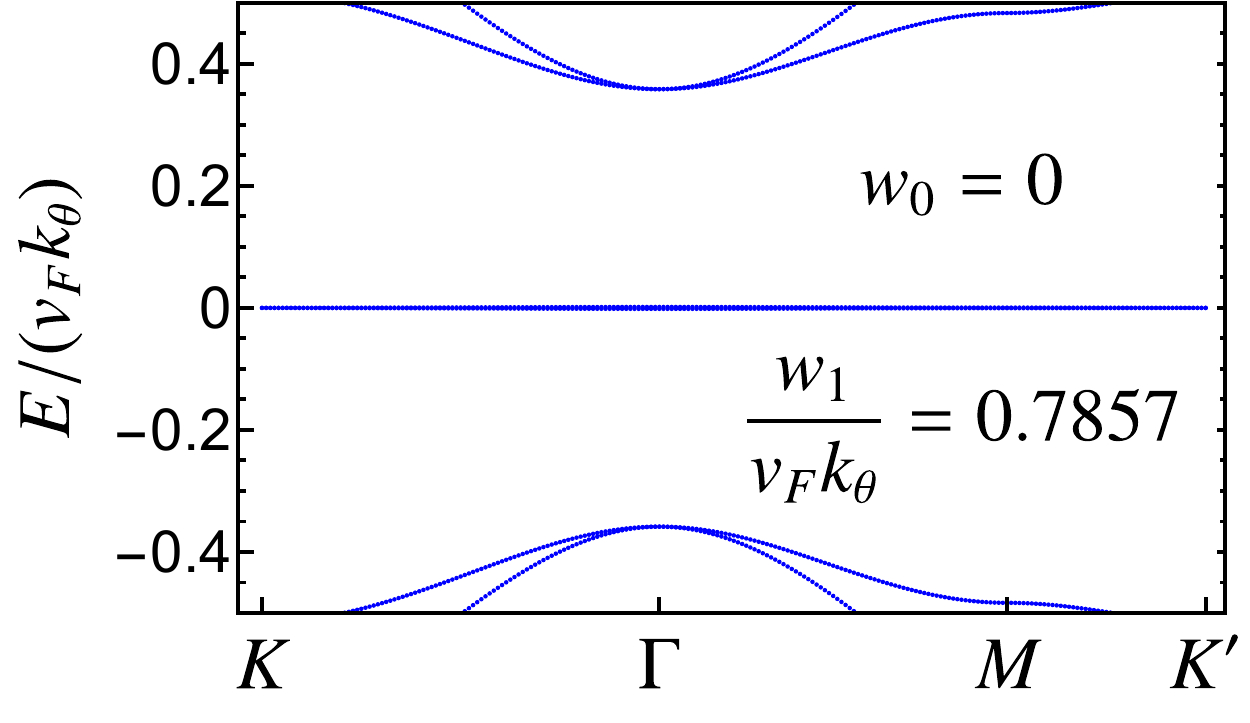}}	
	\caption{The existence of the exactly flat band for $H_{chiral}$ in the presence of the pseudo vector field $\mathcal{A}$ induced by the relaxation that is obtained from Ref.~\cite{KaxirasRelaxation}. (a): the bandwidth $W_{\text{band}}$ of the narrow bands around the CNP as a function of $w_1$. (b) The dispersion of both the narrow and remote bands when $w_1/(v_F k_{\theta}) = 0.7857$.  }
	\label{Fig:FlatBand}
\end{figure}

In the rest of this section, we consider the pseudo-vector field $\mathcal{A}$ induced by the lattice relaxation of Ref.~\cite{KaxirasRelaxation}. As illustrated in Fig.~\ref{Fig:AField}, $\varphi^{\mathcal{A}} \ll \varepsilon^{\mathcal{A}}$, suggesting that $\varphi^{\mathcal{A}}$ can be neglected, and thus $\mathcal{A} \approx \fvec \nabla \times (\hat{\fvec z}  \varepsilon^{\mathcal{A}})$. In addition, Table~\ref{Tab:ElasticParammeters} shows that the Fourier series of the lattice relaxation is dominated by the lowest six $\fvec g$s. Therefore, we can keep only these six terms and neglect others. Furthermore, because $\varepsilon_{\mathcal{A}}$ is real and odd, its lowest Fourier components are purely imaginary. By $C_3$ symmetry, they satisfy the relation
\begin{align}
	& \tilde{\varepsilon}^{\mathcal{A}}_{\fvec g_1} = \tilde{\varepsilon}^{\mathcal{A}}_{\fvec g_2} = \tilde{\varepsilon}^{\mathcal{A}}_{-(\fvec g_1 + \fvec g_2)} \nonumber \\
	= & - \tilde{\varepsilon}^{\mathcal{A}}_{-\fvec g_1} = - \tilde{\varepsilon}^{\mathcal{A}}_{-\fvec g_2} = - \tilde{\varepsilon}^{\mathcal{A}}_{\fvec g_1 + \fvec g_2} = i \tilde{\varepsilon}^{\mathcal{A}}_1.  
\end{align}
In addition, the inter-layer contact coupling field is set to be
\begin{align}
	T(\fvec x) =  w_1 \sum_{l = 1}^3 \begin{pmatrix}
		0 & e^{-i \frac{2\pi}3 (l - 1)} \\ e^{i \frac{2\pi}3 (l - 1)} & 0
	\end{pmatrix} e^{i \fvec q_{1, l} \cdot \fvec x},
\end{align}
where only the inner most $\fvec q$ shell is included. Introducing the dimensionless parameters $\alpha = w_1/(v_F k_{\theta})$ and $\gamma \tilde{\varepsilon}^{\mathcal{A}}_1$, the Fermi velocity of the Dirac cone at $\fvec K_m$ and $\fvec K_m'$ can be approximated as~\cite{appendix}
\begin{align}
	v_{Dirac} \approx v_F \frac{1 - 6(\gamma \tilde{\varepsilon}^{\mathcal{A}}_1)^2 - 3\alpha^2 + 14\sqrt{3} \alpha^2 \gamma \tilde{\varepsilon}^{\mathcal{A}}_1 }{1+3 \alpha^2 + 6(\gamma \tilde{\varepsilon}^{\mathcal{A}}_1)^2 } \ . \label{Eqn:vDiracApprox}
\end{align}
From Table.~\ref{Tab:ElasticParammeters} and \ref{Tab:ParammeterRelax}, $\gamma \tilde{\varepsilon}^{\mathcal{A}}_1 \approx 0.06$, leading to $\alpha \approx 0.79$ when $v_{Dirac}$ vanishes.

We also numerically checked the existence of the exactly flat bands in the presence of the $\mathcal{A}$ field induced by the lattice relaxation of Ref.~\cite{KaxirasRelaxation}.  As demonstrated in Fig.~\ref{Fig:FlatBand}, the bandwidth $W_{band}$ vanishes when the inter-layer coupling constant $w_1$ is tuned to be around $0.7857 v_F k_{\theta}$, very close to the value obtained from the approximate formula in Eq.~\ref{Eqn:vDiracApprox}. For the BM model where the pseudo-vector field $\mathcal{A}$ is absent, the exactly flat bands occur when $w_1/(v_F k_{\theta}) = 0.586$ (see Ref.\cite{Grisha}); if $w_1$ and $v_F$ are set to the values listed in Table~\ref{Tab:ParammeterRelax} and \ref{Tab:ParammeterInter} for the model in Ref.~\cite{KaxirasPRB16}, the corresponding twist angle is $1.07^{\circ}$.  However, this angle decreases to $0.83^{\circ}$ when the pseudo-vector field $\mathcal{A}$ induced by the lattice relaxation \cite{KaxirasPRB16} is included in the chiral limit. 

\section{Summary}

In this work, we constructed and analyzed the effective continuum theories corresponding to the microscopic tight binding models proposed in Ref.~\cite{KoshinoPRB12} and \cite{KaxirasPRB16} based on the systematic method proposed in \cite{paper1}. The nearly perfect agreement between the dispersion of the tight binding models and the dispersion of the effective continuum theories demonstrates the correctness of the constructed continuum theories and the validity of the method. We therefore envision that the experimentally measured $\bu_j(\br)$ can be plugged into our effective Hamiltonian, and the resulting energy spectra and eigenfunctions can then be used to directly compare with the scanning tunneling spectroscopy (STS) measurement of the electronic local density of states. This may pave the way for a more quantitative comparison between the theoretical predictions and the experimental results. In addition, our theory provides electron-phonon couplings as a byproduct, which are important to fully understand the role of phonons in superconductivity of TBG.

Our continuum model goes beyond the BM model in several aspects. First, the p-h symmetry of the narrow bands is only weakly broken within the BM model, while it is much more strongly broken in our continuum theory constructed for the tight binding model of Ref.~\cite{KaxirasPRB16}. While the p-h symmetry of the {\em energy} spectrum is broken in both models \cite{KoshinoPRB12,KaxirasPRB16}, we focused on the p-h asymmetry of the narrow band {\em Hilbert space}, because it is more important in determining the correlated ground states near the magic angle.  As shown in Fig.~\ref{FigS:KaxirasPH}, the p-h asymmetry is dominated by the contribution from the inter-layer contact term $w_3$ that has been overlooked in previous works. Another source of the p-h asymmetry are the inter-layer gradient terms $\fvec \Lambda$~\cite{KaxirasGradient,NickKekule}, whose numerical value listed in Table~\ref{Tab:ParammeterInter} is about two times larger than the value given in Ref.~\cite{NickKekule} and \cite{KaxirasGradient}. As a consequence, compared with the BM model and other continuum theories, our effective theory for the microscopic model of Ref.\cite{KaxirasPRB16} leads to a much larger p-h asymmetry of the wavefunctions in the narrow bands. 

Second, the inter-layer tunneling in the BM model contains terms only with the minimal momentum transfer, i.e.~the tunneling with three $\fvec q$s in the first shell and neglects all other $\fvec q$s. This approximation works quite well if the lattice relaxation is absent. In the presence of the lattice relaxation, however, Fig.~\ref{FigS:SKQShellSpec} and \ref{FigS:WannierQShellSpec} have demonstrated the necessity to include more $\fvec q$s to even qualitatively match the dispersion.

We also investigated the existence of the exactly flat bands near the CNP when the lattice induced pseudo magnetic fields are present. As long as the pseudo vector potentials respect the $C_3$ symmetry, our theoretical analysis and numerical calculations found exactly flat bands in the chiral limit, but at a smaller twist angle ($0.83^{\circ}$) then without the relaxation induced pseudo vector fields ($1.07^{\circ}$). In other words, despite the relaxation induced strain fields, the bands can be exactly flat in the chiral limit due to the compensation from lowering the twist angle. Our analysis demonstrates the importance of the $C_3$ symmetry in making this compensation possible.

\acknowledgments
J.~K.~acknowledges the support from the NSFC Grant No.~12074276, the Double First-Class Initiative Fund of ShanghaiTech University, and the start-up grant of ShanghaiTech University. O.~V.~is supported by NSF DMR-1916958 and is partially funded by the Gordon and Betty Moore Foundation's EPiQS Initiative Grant GBMF11070, National High Magnetic Field Laboratory through NSF Grant No.~DMR-1644479 and the State of Florida. Part of this work was performed at the Aspen Center for Physics, which is supported by National Science Foundation grant PHY-1607611.

\newpage

\appendix

\begin{widetext}
	\newpage

	%%%%%%%%%% Merge with supplemental materials %%%%%%%%%%
	%%%%%%%%%% Prefix a "S" to all equations, figures, tables and reset the counter %%%%%%%%%%
	\setcounter{equation}{0}
	\setcounter{figure}{0}
	\setcounter{table}{0}
	\makeatletter
	\renewcommand{\thefigure}{S\arabic{figure}}
	\renewcommand{\thetable}{S\arabic{table}}
	%\printbibliography[heading=bibempty,type=misc,prefixnumbers={S}]
	
\section{Intra-layer Dispersion}	
\label{SecS:Intra}
In this section, we derive the intra-layer part of the effective continuum Hamiltonian $H_{eff}^{\fvec K}$ from the microscopic tight binding model. The microscopic tight binding model has the general form of 
\begin{align}
	H_{tb}  = & \sum_{S S'} \sum_{j j'} \sum_{\fvec r_S, \fvec r_{S'}} t(\fvec X_{j,S} - \fvec X_{j', S'}')   c^{\dagger}_{j, S, \fvec r_S} c_{j', S', \fvec r_{S'}}, \label{EqnS:MicroH}
\end{align}
For intra-layer hoppings, the hopping displacement $\fvec X_{j,S} - \fvec X_{j', S'}'$ contains only in-plane components. In both models considered in this manuscript, the intra-layer hopping is isotropic, depends only on $|\fvec X_{j,S} - \fvec X_{j', S'}'|$, and thus the intra-layer hopping $t(\delta \fvec X) = t(|\delta \fvec X|)$. 
In addition, we only consider the lattice relaxation with which $\fvec U_{j,S} = \fvec U^{\parallel}_j(\fvec x)$ is independent of the sublattice. We also neglect the corrugation so that $\fvec U^{\perp}_{t/b}(\fvec x) = \pm \frac{d_0}2 \hat{z}$.  As a consequence,
\begin{eqnarray}
		H^{\bK}_{intra}
		& &\simeq\frac{1}{A_{mlg}} \sum_{S,S'} \sum_{j} \sum_{\fvec G} e^{i\fvec G \cdot(\fvec{\tau}_S-\fvec{\tau}_{S'})} \int \rmd^2 \fvec x
		\int \rmd^2\fvec y e^{-i(\fvec G + \fvec K) \cdot \fvec y} e^{i\frac{\fvec y}{2} \cdot \nabla_{\fvec x} 2\fvec U^\parallel_{j}(\fvec x)  \cdot (\fvec G + \fvec K)}	\nonumber\\
		& & t( \fvec y )	\left[ \Psi^\dagger_{j,S}(\fvec x) \Psi_{j,S'}(\fvec x) + \frac{\fvec y}2 \cdot\left( \left(\nabla_{\fvec x} \Psi^\dagger_{j,S}(\fvec x) \right) \Psi_{j,S'}(\fvec x) - \Psi^\dagger_{j,S}(\fvec x) \nabla_{\fvec x} \Psi_{j,S'}(\fvec x)   \right) +  \right. \nonumber \\
		& & \left. + \frac18  y^{\mu} y^{\nu} \left( (\partial_{\mu} \partial_{\nu} \Psi^{\dagger}_{j,S}(\fvec x)) \Psi_{j, S'}(\fvec x) - 2(\partial_{\mu} \Psi^{\dagger}_{j,S}(\fvec x) ) ( \partial_{\nu} \Psi_{j,S'}(\fvec x) ) + \Psi^{\dagger}_{j, S}(\fvec x) (\partial_{\mu} \partial_{\nu} \Psi_{j,S'}(\fvec x))  \right)
		  	\right].   \label{EqnS:EffContIntraH}
\end{eqnarray}
The lattice displacement 
\[ \fvec U^{\parallel}_{j, \mu} = \left(R\left( \frac{\theta_j}2 \right) - I_{2\times 2}\right)_{\mu\nu}  x_{\nu} \pm \half \delta U_{\mu}  \ .   \]
where $\theta_t = - \theta_b = \theta$, and $R(\theta/2)$ is the $2\times 2$ matrix corresponding to the counterclockwise rotation around  $z$ axis with the angle of $\theta/2$, and $+$($-$) sign is for the top (bottom) layer respectively. Notice that $\fvec \nabla_{\fvec x} \hat z \times \fvec x \cdot (\fvec G + \fvec K) = - \hat z \times (\fvec G + \fvec K)$. Thus,
\[ e^{-i (\fvec G + \fvec K ) \cdot \fvec y} e^{i \frac{\fvec y}2 \cdot \nabla_x 2 U^{\parallel}(\fvec x) \cdot (\fvec G + \fvec K)} =  e^{-i R(-\theta_j/2) \fvec y \cdot  (\fvec G + \fvec K)} e^{\pm i \fvec y/2 \cdot \fvec \nabla_{\fvec x} \delta \fvec U(\fvec x) \cdot (\fvec G + \fvec K) } \]
%\textcolor{red}{needs to rewrite this formula.}
In the main text, we consider only the lattice relaxation proposed in Ref.~\cite{KoshinoPRB17} and \cite{KaxirasRelaxation}, in which the lattice distortion is dominated by the solenoid part. However, in this section, for completeness,  we consider a more general $\fvec U_j^{\parallel}$, whose irrotational part may also be important, and only in the last step, we set $\fvec \nabla \cdot \delta \fvec U = 0$. 
As mentioned in the main text, up to the second order of the derivatives, the intra-layer part is
\begin{align}
	H_{intra} & = H^{(0)}_{intra} + \delta H_{intra} \\
	H^{(0)}_{intra} & = \int \rmd^2\fvec x\ \sum_j \sum_{S S'}\Psi_{j, S}^{\dagger}(\fvec x) \left\{ \mu \delta_{SS'} + v_F \bar{\fvec \sigma}_{S S'} \cdot \left( \fvec p^{(j)} + \gamma \mathcal{A}^{(j)} \right) + \alpha_{dp} \phi^{(j)} \delta_{SS'} + \beta_0 \fvec p^2 \delta_{S S'} -\frac{C_0}2 \left( \fvec p \cdot \mathcal{A}(\fvec x) + \mathcal{A}(\fvec x) \cdot \fvec p \right) \delta_{S S'} \right. \nonumber \\
	& \left.  + \beta_1 \left( ( p_x^2 - p_y^2) \sigma_1 + 2  p_x  p_y \sigma_2 \right)_{S S'} \pm \half\left(  p_{\mu}  \xi_{\mu,SS'}(\fvec x) +  \xi_{\mu,SS'}(\fvec x)  p_{\mu}  \right) + 2D_0 \left\{ \phi^{(j)}, \bar{\fvec \sigma}_{S S'} \cdot \fvec p \right\} \right\}  \Psi_{j, S}(\fvec x)  \\
	\delta H_{intra} & = \int \rmd^2\fvec x\ \sum_j \sum_{S S'}\Psi_{j, S}^{\dagger}(\fvec x) \left\{  \sum_{\mu\nu} \left[ C_1 \left( (\fvec \nabla \cdot \delta \fvec U)^2 +  (\partial_{\mu} \delta U_{\nu}) (\partial_{\nu} \delta U_{\mu}) \right) + C_2  (\partial_{\mu} \delta U_{\nu}) (\partial_{\mu} \delta U_{\nu}) \right]\delta_{S S'} \right. \nonumber \\
	& \left.  + \alpha \left(-\frac{\theta}2 (\fvec \nabla \times \delta \fvec U)_z - \half \left( \fvec \nabla \cdot \delta \fvec U \right)^2 \right) \delta_{S S'}  + \zeta_{S S'}(\fvec x)  \right\} \Psi_{j, S'}(\fvec x) \ . \label{EqnS:DeltaHIntra}
\end{align}
Here, we list all the expressions of the coefficients and fields that appear in the above formula.
\begin{align}
	\mu & = \sum_{\fvec a} e^{-i \fvec K \cdot \fvec a} t(|\fvec a|) \\
	v_F & = -i \sum_{\fvec a} e^{-i \fvec K \cdot (\fvec a + \delta \fvec \tau_{AB}) } (\fvec a + \delta \fvec \tau_{AB})_x t(|\fvec a + \delta \fvec \tau_{AB}|) \\
	v_F \gamma & =   \half \sum_{\fvec a} e^{-i \fvec K \cdot (\fvec a + \delta \fvec \tau_{AB})} \left[ (\fvec a + \delta \fvec \tau_{AB})_x \right]^2 \frac{t'(|\fvec a + \delta \fvec \tau_{AB}|)}{|\fvec a + \delta \fvec \tau_{AB}|} \\
	\phi^{(j)} & = \pm \half \fvec \nabla \cdot \delta \fvec U \\
		\alpha & = \frac14 \sum_{\fvec a} e^{-i \fvec K \cdot \fvec a}  |\fvec a| t'(|a|)  \quad \mbox{and} \quad  \alpha_{dp} = \frac{\sqrt{3}a}2   \frac{\partial \epsilon}{\partial |\delta_S^{\alpha}|}  + \alpha \\
	\mathcal A^{(j)}_{\mu}(\fvec x) & = \pm R\left( \frac{\theta_j}2 \right)_{\mu\nu} \mathcal A_{\nu} \approx \pm ( \mathcal A_x + \frac{\theta_j}2 \mathcal A_y\ , \ \mathcal A_y - \frac{\theta_j}2 \mathcal A_x)_{\mu}  \label{EqnS:VectorField} \\
	\mbox{with} \  \mathcal A(\fvec x) &= (\partial_x \delta  U_x - \partial_y \delta U_y\ , \ -(\partial_x \delta U_y + \partial_y \delta U_x))   \approx (2 \partial_x \partial_y \varepsilon^U(\fvec x)\ , \ (\partial_x^2 - \partial_y^2)\varepsilon^U(\fvec x) ) \\
	\beta_0 & =  - \frac14 \sum_{\fvec a} e^{- i \fvec K \cdot \fvec a} |\fvec a|^2 t(|\fvec a|) \\
	C_0 & = - \frac{i}2 \sum_{\fvec a} e^{-i \fvec K \cdot \fvec a} \left( \fvec a_x \right)^3 \frac{t'(|\fvec a|)}{|\fvec a|} \\
	\beta_1 & = - \half \sum_{\fvec a} e^{-i \fvec K \cdot (\fvec a + \delta \fvec \tau_{AB})} \left[ (\fvec a + \delta \fvec \tau_{AB})_x  \right]^2 t(|\fvec a + \delta \fvec \tau_{AB}|)  \\
	\xi_{x, SS'}(\fvec x) & =   (\frac{v_F}2 + 2D_0)(\partial_x \delta U_x ) (\sigma_1)_{S S'} - \left[ (\frac{v_F}2 + D_0)\partial_x \delta  U_x + D_0 \partial_x \delta U_y  \right] (\sigma_2)_{S S'}  \\
	\xi_{y, SS'}(\fvec x) & = \left[ (\frac{v_F}2 + D_0)\partial_x \delta  U_y + D_0 \partial_y \delta  U_x  \right] (\sigma_1)_{S S'} - (\frac{v_F}2 + 2D_0)(\partial_y \delta U_y ) (\sigma_2)_{S S'} \\
	D_0 & = - \frac{i}6 \sum_{\fvec a} e^{i \fvec K \cdot (\fvec a+ \delta \fvec \tau_{AB})} \left( (\fvec a + \delta \fvec \tau_{AB})_x \right)^3   \frac{t'(|\fvec a + \delta \fvec \tau_{AB}|)}{|\fvec a + \delta \fvec \tau_{AB}|}
\end{align}
where we have introduce the notation $\delta \fvec \tau_{S S'} = \fvec \tau_S - \fvec \tau_{S'}$, so $\delta \fvec \tau_{AB} = -\frac13(\fvec a_1 + \fvec a_2)$. $\fvec a$ is an arbitrary lattice vector. $\theta_t = - \theta_b =\theta$. The sign $+$ and $-$ in Eq.~\ref{EqnS:VectorField} are for the top and bottom layers respectively. Note that for the lattice relaxation considered in the main text, $|\fvec \nabla \cdot \delta \fvec U| \lesssim 10^{-5}$ is tiny  and thus the pseudo-scalar field $\phi^{(j)}$ can be safely neglected, as well as the term proportional to $\alpha_{dp}$. The detailed discussion on the coefficient $\alpha_{dp}$ can be found in Sec.~\ref{Sec:DeformationPotential}. 

As mentioned in the text, although the terms in $\delta H_{intra}$ are also the second order, they are numerically small compared with other second order terms in $H^{(0)}_{intra}$. Here, we express the fields and coefficients in $\delta H_{intra}$  in terms of the lattice distortion $\delta \fvec U(\fvec x)$ and the  microscopic hopping function:
\begin{align}
	C_1 & = \frac18 \sum_{\fvec a} e^{-i \fvec K \cdot \fvec a} \left(  \frac78 |\fvec a| t'(|\fvec a|) + \frac18 |\fvec a|^2 t''(|\fvec a|)  \right) \\
	C_2 & = \frac18 \sum_{\fvec a} e^{-i \fvec K \cdot \fvec a} \left( \frac38 |\fvec a| t'(|\fvec a|) + \frac18 |\fvec a|^2 t''(|\fvec a|) \right)
\end{align}
The formula of the field $\fvec \zeta_{S S'}(\fvec x)$ is listed in Eq.~\ref{EqnS:RTensor}--\ref{EqnS:ZetaField}.

\subsection{Expansion of the Jacobian Factor}
In the main text, we have expanded the the Jacobian factor $\mathcal{J}$ to the first order of $\fvec \partial \fvec U^{\parallel}$, and argues that it depends only on the divergence of $\fvec U^{\parallel}$. Since $\fvec \nabla \cdot \fvec U^{\parallel} \approx 10^{-5}$, its deviation from $1$ can be safely neglected. In this subsection, we will go to the second order of the derivatives, and derive its corresponding terms in the effective continuum $H_{eff}^{\bK}$.

Since we will expand to the second order of $\fvec \partial \fvec U^{\parallel}$, we write $\fvec U^{\parallel}$ as
\[ U^{\parallel}_{t, \mu} = \left(  I - R\left( -\frac{\theta}2 \right) \right)_{\mu\nu} x_{\nu} + \half \delta U_{\nu} = - \frac{\theta}2 \epsilon_{\mu\nu} x_{\nu} + \frac18 \theta^2 x_{\mu} + \half \delta U_{\mu} \quad \longrightarrow \quad \frac{\partial U^{\parallel}_{t,\mu}}{\partial x_{\nu}} = -\frac{\theta}2 x_{\nu} + \half \frac{\partial \delta U_{\mu}}{\partial x_{\nu}}  \]
If all the elements of the matrix $M$ has the property $|M_{\mu\nu}| \ll 1$, its determinant can be expanded as
\[  \det(I + M) = \exp\left(  \mathrm{Tr}(\ln M) \right) \approx 1 + \mathrm{Tr}(M) + \half \left( \left( \mathrm{Tr}(M) \right)^2 - \mathrm{Tr}(M^2)  \right) + O(M^3)  \]
Therefore, we obtain the expansion of the determinant up to the second order of the derivative
\begin{align}
	\mathcal{J}_{t}^2 & =  1 - \half \fvec \nabla \cdot \delta \fvec U + \frac{\theta}4 \left( \fvec \nabla \times \delta \fvec U \right)_z + \frac18 \left( \fvec \nabla \cdot \delta \fvec U \right)^2 - \frac18 \left( \partial_{\mu} \delta U_{\nu} \right) \left( \partial_{\nu} \delta U_{\mu} \right) \nonumber \\
	\mathcal{J}_{b}^2 & =  1 + \half \fvec \nabla \cdot \delta \fvec U + \frac{\theta}4 \left( \fvec \nabla \times \delta \fvec U \right)_z + \frac18 \left( \fvec \nabla \cdot \delta \fvec U \right)^2 - \frac18 \left( \partial_{\mu} \delta U_{\nu} \right) \left( \partial_{\nu} \delta U_{\mu} \right) \ . \label{EqnS:JExpansion}
\end{align}
For the lattice relaxation considered in the main text, $|\fvec \nabla \cdot \delta \fvec U| \sim 10^{-5}$ and can be safely neglected. 

\subsection{Expansion by the Order of Derivatives}
Since $\delta \fvec U \lesssim 0.3a$ varies over the moire unit cell, its gradient $|\fvec \nabla_x \delta \fvec U | \ll 1$. We can expand the exponential by the order of $\fvec \nabla_x \delta \fvec  U$, i.e.~
\[ e^{i \frac{y}2 \cdot \nabla_x \delta \fvec U \cdot (\fvec G + \fvec K)}  \approx 1 + i \frac{y_{\mu}}2  \partial_{\mu} \delta U_{\rho}  (\fvec G + \fvec K)_{\rho} - \frac18  y_{\mu}  y_{\nu} (\partial_{\mu} \delta U_{\rho} ) (\partial_{\nu} \delta U_{\sigma}) (\fvec G + \fvec K)_{\rho} (\fvec G + \fvec K)_{\sigma}   \] 
%Note that we have assumed that the intra-layer hopping function $t(\fvec x) = t(|\fvec x|)$, i.e.~it is isotropic, depending only on the magnitude of the hopping displacement.

In the rest of this section, we will derive each term in the expansion and express the coefficient in terms of the microscopic hopping function. Before doing this, we define the Fourier transformation of the hopping functions as 
\begin{align}
	\tilde{t}(\fvec q) = A_{mlg}^{-1} \int \rmd^2 \fvec y e^{-i \fvec q\cdot \fvec y} t(\fvec y)  \quad \Longrightarrow \quad  i \partial_{\fvec q_{\mu}} \tilde{t}(\fvec q) = A_{mlg}^{-1} \int \rmd^2 \fvec y e^{-i \fvec q\cdot \fvec y}   y_{\mu} t(\fvec y) 
\end{align}

\subsection{Leading term}
First, we consider the leading term that in $H_{intra}$:
\begin{align}
	\sum_{\fvec G} e^{i \fvec G \cdot \delta \fvec \tau_{SS'}} \tilde{t}(\fvec G + \fvec K) \Psi^{\dagger}_{j, S} \Psi_{j, S'} = \sum_{\fvec a} e^{-i \fvec K \cdot (\fvec a + \delta \fvec \tau_{SS'})} t(|\fvec a + \delta \fvec \tau_{SS'}|) \Psi^{\dagger}_{j, S} \Psi_{j, S'}
\end{align}
where $\fvec a$ is an arbitrary lattice vector.  In the last formula above, we have used the Poisson summation formula to transform the summation over $\fvec G$ to the summation over the lattice vectors. Due to $C_3$ symmetry, it is easy to show that the summation above vanishes when $S \neq S'$. When $S = S'$, the above summation leads to the term 
\begin{align}
	\mu \sum_{j, S} \int \rmd^2 \fvec x\ \Psi^{\dagger}_{j,S}(\fvec x) \Psi_{j, S}(\fvec x) \quad \mbox{with}\ \mu = \sum_{\fvec a} e^{-i \fvec K \cdot \fvec a} t(|\fvec a|) \ .
\end{align}
Combined with Eq.~\ref{EqnS:JExpansion}, to the second order of $\fvec \nabla \fvec U_j^{\parallel}$,  this terms leads to 
\begin{align}
	\mu \sum_{j, S} \int \rmd^2 \fvec x\  \left( 1 \mp \half \fvec \nabla \cdot \delta \fvec U + \frac{\theta}4 \left( \fvec \nabla \times \delta \fvec U \right)_z + \frac18 \left( \fvec \nabla \cdot \delta \fvec U \right)^2 - \frac18 \left( \partial_{\mu} \delta U_{\nu} \right) \left( \partial_{\nu} \delta U_{\mu} \right) \right) \Psi^{\dagger}_{j,S}(\fvec x) \Psi_{j, S}(\fvec x) \ . \label{EqnS:muJExpansion}
\end{align}

\subsection{First Order Derivative}
Next, we consider the next leading term, i.e.~the terms containing the first order derivative of either  $\delta \fvec U$ or $\Psi_{j, S}(\fvec x)$. 
\subsubsection{Fermi Velocity}
For the terms containing $\partial_{\mu} \Psi_{j,S}(\fvec x)$, we have
\begin{align}
  & \int \rmd^2 \fvec x\ v^{\mu}_{j, SS'} \Psi^{\dagger}_{j,S}(\fvec x) p_{\mu} \Psi_{j,S'}(\fvec x) \quad \mbox{with}  \\
  & v^{\mu}_{j, SS'} = \sum_{\fvec G} e^{i \fvec G \cdot \delta \fvec \tau_{SS'}} \int \rmd^2 \fvec y e^{-i (\fvec G + \fvec K)_{\rho} R(\mp \theta/2)_{\rho\nu}  y_{\nu}} (-i  y_{\mu}) t(|\fvec y|)   =  \sum_{\fvec G} e^{i \fvec G \cdot \delta \fvec \tau_{SS'}} \partial_{ q_{\nu}} \tilde{t}(\fvec q) |_{\fvec q = \fvec G + \fvec K} R\left( \pm \frac{\theta}2 \right)_{\mu\nu} \ .
\end{align}
Applying the Poisson summation formula, we obtain that
\[  \sum_{\fvec G}  e^{i \fvec G \cdot \delta \fvec \tau_{SS'}} \partial_{ q_{\nu}} \tilde{t}(\fvec q) |_{\fvec q = \fvec G + \fvec K} = -i \sum_{\fvec a} e^{- i \fvec K \cdot (\fvec a + \delta \fvec \tau_{SS'}) }   (\fvec a + \delta \fvec \tau_{SS'})_{\nu} t(|\fvec a + \delta \fvec \tau_{SS'}|) \]
For $S = S'$, it can be shown that the above summation vanishes because of $C_3$ symmetry. For $S \neq S'$, due to $C_3$ and $m_y$ (mirror reflection over the $yz$ plane), the above summation leads to $v_F (\sigma_1 \hat x - \sigma_2 \hat y)_{SS'}$, with 
\[   v_F = -i \sum_{\fvec a} e^{-i \fvec K \cdot (\fvec a + \delta \fvec \tau_{AB}) } (\fvec a + \delta \fvec \tau_{AB})_x t(|\fvec a + \delta \fvec \tau_{AB}|) \  . \]
where $\delta \fvec \tau_{AB} = - \frac13(\fvec a_1 + \fvec a_2)$.
 
Thus, making the approximation that $R(\theta)_{\mu\nu} = \delta_{\mu\nu} - \theta \epsilon_{\mu\nu}$, this gradient term can be written as
\begin{align}
	v_F \sum_j \sum_{S S'} \Psi^{\dagger}_{j, S}(\fvec x)   \left( \left( p_x + \frac{\theta_j}2   p_y \right) \sigma_1 -  \left( p_y - \frac{\theta_j}2  p_x \right) \sigma_2 \right)_{S S'}  \Psi_{j, S'}(\fvec x)
\end{align} 
For notational convenience, we can first define the layer dependent momentum operator:
\[ \fvec p^{(j)} = \left(  p_x + \frac{\theta_j}2  p_y\ ,\  p_y - \frac{\theta_j}2 p_x \right)  \ ,  \]
so that this term can be written in a simpler form:
\begin{align}
	v_F \sum_{j} \sum_{S S'} \Psi^{\dagger}_{j, S}(\fvec x)  \bar{\sigma}_{S S'} \cdot \fvec p^{(j)} \Psi_{j, S'}(\fvec x)
\end{align}
Now, combined with the expansion of the Jacobi factor in Eq.~\ref{EqnS:JExpansion} and expanded to the second order of the derivatives, we obtain
\begin{align}
	v_F \sum_{j} \sum_{S S'} \Psi^{\dagger}_{j, S}(\fvec x)  \left( 1 \mp \half \fvec \nabla \cdot \delta \fvec U \right) \bar{\sigma}_{S S'} \cdot \fvec p^{(j)} \Psi_{j, S'}(\fvec x) 
\end{align}

\subsubsection{Pseudo-vector and Pseudo-scalar Fields}
To the first order derivative of $\delta \fvec U$, $H_{intra}$ also contains term that couples to $\Psi^{\dagger}_{j,S} \Psi_{j, S'}$
\begin{align}
	 & \pm \sum_{\fvec G} e^{i \fvec G \cdot \delta \fvec \tau_{S S'}} \int \rmd^2 \fvec y\ e^{-i (\fvec G + \fvec K) \cdot R(-\theta_j/2) \fvec y} t(|\fvec y|) \frac{i}2 y_{\mu} (\partial_{\mu}  \delta  U_{\rho} ) (\fvec G + \fvec K)_{\rho}  \nonumber \\
	= & \pm \frac{i}2 \sum_{\fvec G} e^{i \fvec G \cdot \delta \fvec \tau_{S S'}} \int \rmd^2 \fvec z \ e^{-i (\fvec G + \fvec K) \cdot \fvec z} t(|\fvec z|) R( \frac{\theta_j}2)_{\mu\nu} z_{\nu} (\fvec G + \fvec K)_{\rho} (\partial_{\mu}  \delta  U_{\rho} ) \label{EqnS:PseudoCoupling}
\end{align}
where by Poisson summation formula, 
\begin{align}
	& \frac{i}2 \sum_{\fvec G} e^{i \fvec G \cdot \delta \fvec \tau_{S S'}} \int \rmd^2 \fvec z \ e^{-i (\fvec G + \fvec K) \cdot \fvec z} t(\fvec z)  z_{\nu} (\fvec G + \fvec K)_{\rho} = \half \sum_{\fvec a} e^{-i \fvec K \cdot (\fvec a + \delta \fvec \tau_{S S'})} \partial_{\rho} \left[  (\fvec a + \delta \fvec \tau_{S S'})_{\nu} t(|\fvec a + \delta \fvec \tau_{S S'}|) \right] \nonumber \\
	= & \half \sum_{\fvec a} e^{-i \fvec K \cdot (\fvec a + \delta \fvec \tau_{S S'})}  \left[   \delta_{\rho \nu} t(|\fvec a + \delta \fvec \tau_{S S'}|) + (\fvec a + \delta \fvec \tau_{S S'})_{\nu}  (\fvec a + \delta \fvec \tau_{S S'})_{\rho} \frac{t'(|\fvec a + \delta \fvec \tau_{S S'}|)}{|\fvec a + \delta \fvec \tau_{S S'}|} \right]   \label{EqnS:PseudoFields}
\end{align}
This summation, when $S = S'$, leading to
\begin{align}  
	\half \sum_{\fvec a} e^{-i \fvec K \cdot \fvec a} \left[ t(|a|) + \half |\fvec a| t'(|a|) \right] \delta_{\nu\rho} = \left( \frac{\mu}2 + \frac14 \sum_{\fvec a} |\fvec a| t'(|\fvec a|) \right) \delta_{\nu\rho} \ .  \label{EqnS:AlphaPFormula}
\end{align}
For notational convenience, we introduce $\alpha = \frac14 \sum_{\fvec a} |\fvec a| t'(|\fvec a|)$. When $S = S'$, Eq.~\ref{EqnS:PseudoCoupling} leads to
\begin{align}
	\pm \left( \frac{\mu}2 + \alpha \right) \delta_{\rho \nu} R\left(\frac{\theta_j}2 \right)_{\mu\nu} \partial_{\mu} \delta \fvec U_{\rho} \approx  \left( \frac{\mu}2 + \alpha \right)  \left( \pm \fvec \nabla \cdot \delta \fvec U - \frac{\theta}2 \left( \fvec \nabla \times \delta \fvec U \right)_z \right)
\end{align} 
Now, combined with the expansion of the Jacobi determinant, we see that  the terms that couple to $\Psi^{\dagger}_{j,S} \Psi_{j, S}$ are
\begin{align}
	& \left( \frac{\mu}2 + \alpha \right)  \left( \pm \fvec \nabla \cdot \delta \fvec U - \frac{\theta}2 \left( \fvec \nabla \times \delta \fvec U \right)_z \right) + \left( \frac{\mu}2 + \alpha \right)  \left( \pm \fvec \nabla \cdot \delta \fvec U  \right) \left(  \mp \half \fvec \nabla \cdot \delta \fvec U \right)  \nonumber \\
	= & \left( \frac{\mu}2 + \alpha \right)  \left( \pm \fvec \nabla \cdot \delta \fvec U - \frac{\theta}2 \left( \fvec \nabla \times \delta \fvec U \right)_z - \half (\fvec \nabla \cdot \delta \fvec U)^2 \right) 
\end{align}
Note that in the main text where we consider only the lattice relaxation proposed in Ref.~\cite{KoshinoPRB17} and \cite{KaxirasRelaxation}. The divergence of the atomic displacement field $|\fvec \nabla \cdot \delta \fvec U| \lesssim 10^{-5}$ is tiny and can be safely neglected.

When $S \neq S'$, by $C_3$ and $m_y$ mirror symmetries, the summation in Eq.~\ref{EqnS:PseudoFields} produces 
\begin{align}
   &  \half \sum_{\fvec a} e^{-i \fvec K \cdot (\fvec a + \delta \fvec \tau_{S S'})}  \left[   \delta_{\rho \nu} t(|\fvec a + \delta \fvec \tau_{S S'}|) + (\fvec a + \delta \fvec \tau_{S S'})_{\nu}  (\fvec a + \delta \fvec \tau_{S S'})_{\rho} \frac{t'(|\fvec a + \delta \fvec \tau_{S S'}|)}{|\fvec a + \delta \fvec \tau_{S S'}|} \right]  \nonumber \\
   = & \half \sum_{\fvec a} e^{-i \fvec K \cdot (\fvec a + \delta \fvec \tau_{S S'})} (\fvec a + \delta \fvec \tau_{S S'})_{\nu}  (\fvec a + \delta \fvec \tau_{S S'})_{\rho} \frac{t'(|\fvec a + \delta \fvec \tau_{S S'}|)}{|\fvec a + \delta \fvec \tau_{S S'}|}     \nonumber \\
	 = & v_F\gamma \left( \left(\tau_3\right)_{\rho\nu} (\sigma_1)_{SS'}  +  \left(\tau_1\right)_{\rho\nu} (\sigma_2)_{SS'}  \right) \\
	 \mbox{with}\quad v_F \gamma = &  \half \sum_{\fvec a} e^{-i \fvec K \cdot (\fvec a + \delta \fvec \tau_{AB})} \left[ (\fvec a + \delta \fvec \tau_{AB})_x \right]^2 \frac{t'(|\fvec a + \delta \fvec \tau_{AB}|)}{|\fvec a + \delta \fvec \tau_{AB}|}
\end{align}
To further simplify the notation, we introduce the layer dependent pseudo-vector field $\mathcal{A}^{(j)}$ as
\begin{align}
	& \mathcal A^{(j)}_{\mu}(\fvec x) = \pm R\left(\frac{\theta_j}2 \right)_{\mu\nu} \mathcal A_{\nu} \approx \pm ( \mathcal A_x - \frac{\theta_j}2 \mathcal A_y\ , \ \mathcal A_y + \frac{\theta_j}2 \mathcal A_x)_{\mu}  \nonumber \\
\mbox{with} \quad & \mathcal A = (\partial_x \delta  U_x - \partial_y \delta  U_y\ , \ -(\partial_x \delta  U_y + \partial_y \delta  U_x))   \approx (2 \partial_x \partial_y \varepsilon^U(\fvec x)\ , \ (\partial_x^2 - \partial_y^2)\varepsilon^U(\fvec x) ) 
\end{align}
where the sign $+$ and $-$ are for the top and bottom layers respectively. 

Now, combining with the expansion of the Jacobi factor and keep the terms up to the second order of derivatives, 
\begin{align}
	 v_F \gamma \int \rmd^2 \fvec x\ \Psi_{j,S}^{\dagger}(\fvec x) \left( \bar{\fvec \sigma}_{SS'} \cdot \mathcal{A}^{(j)} - \left( \bar{\fvec \sigma}_{SS'} \cdot \mathcal{A} \right) \half \fvec \nabla \cdot \delta \fvec U  \right) \Psi_{j, S'}(\fvec x)
\end{align}
%It is clear that the first two terms can be written as a gauge-independent form. The last term, although seems to be the first order derivative of $\delta \fvec U$ by naive counting, actually turns out to be the second order, since it is proportional to the twist angle $\theta$ and the  first order derivative of $\delta \fvec U$. 

\subsection{second order derivative terms}
In this section, we considered the next order terms, i.e.~the 2nd order of combined derivative of $\delta \fvec U$ and  gradient of fermion fields. 
\subsubsection{second order gradient of fermion field}
First, consider the term
\[  \sum_{\fvec G} e^{i \fvec G \cdot \delta \fvec \tau_{S S'}} \int \rmd^2 \fvec y e^{-i (\fvec G + \fvec K)\cdot \fvec y} t(|\fvec y|) \half (-i y_{\mu})  (-i y_{\nu})  \Psi^{\dagger}_{j, S}(\fvec x)  p_{\mu}  p_{\nu} \Psi_{j, S'}(\fvec x) \]
Applying the Poisson summation formula,
\begin{align}
	& -\half \sum_{\fvec G} e^{i \fvec G \cdot \delta \fvec \tau_{S S'}} \int\rmd^2 \fvec y\ e^{-i (\fvec G + \fvec K)\cdot \fvec y} t(|\fvec y|)  y_{\mu}  y_{\nu} = - \half \sum_{\fvec G} e^{i\fvec G \cdot \delta \fvec \tau_{S S'}} \left. \frac{\partial^2}{\partial q_{\mu} \partial q_{\nu}} \tilde{t}(\fvec q) \right|_{\fvec q = \fvec G + \fvec K} \nonumber \\
	= & -\half \sum_{\fvec a} e^{-i \fvec K \cdot (\fvec a + \delta \fvec \tau_{S S'})} (\fvec a + \delta \fvec \tau_{S S'})_{\mu}  (\fvec a + \delta \fvec \tau_{S S'})_{\nu} t(|\fvec a + \delta \fvec \tau_{S S'}|) 
\end{align} 
For $S = S'$, $\delta \fvec \tau = 0$. We obtain
\[ \beta_0 = -\half \sum_{\fvec a} e^{-i \fvec K \cdot \fvec a} a_{\mu}  a_{\nu} t(|\fvec a|) = - \frac14 \sum_{\fvec a} e^{- i \fvec K \cdot \fvec a} |\fvec a|^2 t(|\fvec a|) \ . \] 
For $S \neq S'$, by $C_3$ and $m_y$ symmetry, we obtain
\begin{align}
 & -\half \sum_{\fvec a} e^{-i \fvec K \cdot (\fvec a + \delta \fvec \tau_{S S'})} (\fvec a + \delta \fvec \tau_{S S'})_{\mu}  (\fvec a + \delta \fvec \tau_{S S'})_{\nu} t(|\fvec a + \delta \fvec \tau_{S S'}|) = \beta_1 \left[ (\tau_3)_{\mu\nu} \sigma_1 + (\tau_1)_{\mu\nu} \sigma_2 \right]_{S S'}   \\
 \mbox{with} \quad & \beta_1 = - \half \sum_{\fvec a} e^{-i \fvec K \cdot (\fvec a + \delta \fvec \tau_{AB})} \left[ (\fvec a + \delta \fvec \tau_{AB})_x  \right]^2 t(|\fvec a + \delta \fvec \tau_{AB}|) \ .
\end{align}
Thus, combining together, we obtain
\begin{align}
   \int \rmd^2\fvec x\ \sum_j \sum_{S S'}\Psi_{j, S}^{\dagger}(\fvec x) \left[  \beta_0 \fvec p^2 \delta_{S S'}  + \beta_1 \left( ( p_x^2 - p_y^2) \sigma_1 + 2  p_x p_y \sigma_2 \right)_{S S'}   \right] \Psi_{j, S}(\fvec x)   \ . 
\end{align}

\subsubsection{Cross terms between atomic displacement gradients and gradients of the fermion field}
Next, consider the cross term between the first order derivative of $\delta \fvec U$ and the  gradient of the fermion field.
\begin{align}
	A_{mlg}^{-1} \sum_{\fvec G}e^{i \fvec G \cdot  \delta \fvec \tau_{S S'}} \int\rmd^2 \fvec y\ e^{-i (\fvec G+ \fvec K)\cdot \fvec y} t(|\fvec y|) \frac{i}2   y_{\mu}  \partial_{\mu} \delta  U_{\rho} (\fvec G + \fvec K)_{\rho} (-i) y_{\nu} \half \left( (  p_{\nu} \Psi_{j, S}(\fvec x))^{\dagger} \Psi_{j, S'} + \Psi_{j, S}^{\dagger}(\fvec x)  p_{\nu} \Psi_{j, S'}(\fvec x) \right) \label{EqnS:Cross2nd}
\end{align}
Applying the Poisson summation formula, we can obtain:
\begin{align}
	& \half \sum_{\fvec G} e^{i \fvec G \cdot  \delta \fvec \tau_{S S'}} \int\rmd^2 \fvec y\ e^{-i (\fvec G+ \fvec K)\cdot \fvec y} t(|\fvec y|) y_{\mu} y_{\nu} (\fvec G + \fvec K)_{\rho}	=  -\half \sum_{\fvec G} e^{i \fvec G \cdot  \delta \fvec \tau_{S S'}} \left. \frac{\partial^2}{\partial q_{\mu} \partial q_{\nu}} \right|_{\fvec q = \fvec G + \fvec K} (\fvec G + \fvec K)_{\rho}  \nonumber \\
	= & - \frac{i}2 \sum_{\fvec a} e^{-i \fvec K \cdot (\fvec a + \delta \fvec \tau_{SS'})} \left[ \delta_{\mu \rho} (\fvec a + \delta \fvec \tau_{S S'})_{\nu} t(|\fvec a + \delta \fvec \tau_{S S'}|) + \delta_{\nu \rho} (\fvec a + \delta \fvec \tau_{S S'})_{\mu} t(|\fvec a + \delta \fvec \tau_{S S'}|) \right. \nonumber \\ 
	& \left. + (\fvec a + \delta \fvec \tau_{S S'})_{\mu} (\fvec a + \delta \fvec \tau_{S S'})_{\nu} (\fvec a + \delta \fvec \tau_{S S'})_{\rho} \frac{t'(|\fvec a + \delta \fvec \tau_{S S'}|)}{|\fvec a + \delta \fvec \tau_{S S'}|} \right]  \label{EqnS:CrossCoeff}
\end{align}	
For $S = S'$, the above formula leads to
\begin{align}
	C_{\mu\nu\rho} = - \frac{i}2 \sum_{\fvec a} e^{-i \fvec K \cdot \fvec a} a_{\mu} a_{\nu}  a_{\rho} \frac{t'(|\fvec a|)}{|\fvec a|} 
\end{align}
By $C_3$ and $m_y$ symmetries, we can prove that the only non-zero components of the tensor $C_{\mu\nu\rho}$ are
\[  C_{xxx} = - C_{xyy} = - C_{yyx} = - C_{yxy} = C_0 \quad \mbox{with} \quad C_0 = - \frac{i}2 \sum_{\fvec a} e^{-i \fvec K \cdot \fvec a} \left(  a_x \right)^3 \frac{t'(|\fvec a|)}{|\fvec a|}  \]
Then Eq.~\ref{EqnS:Cross2nd}, when $S = S'$,  can be written as
\begin{align}
	C_0 \int \rmd^2 \fvec x\ \Psi_{j, S}^{\dagger}(\fvec x) \half \left\{ \fvec p\ , \mathcal{A}(\fvec x) \right\} \Psi_{j, S}(\fvec x) 
\end{align}
Now, for $S \neq S'$,   the summation in Eq.~\ref{EqnS:CrossCoeff} gives
\begin{align}
	& \frac{v_F}2 \left[ \delta_{\mu\rho} (\bar{\fvec \sigma}_{S S'})_{\nu} + \delta_{\nu\rho} \left( \bar{\fvec \sigma}_{S S'} \right)_{\mu} \right] + D_0 \left[ \delta_{\mu\rho} (\bar{\fvec \sigma}_{S S'})_{\nu} + \delta_{\nu\rho} \left( \bar{\fvec \sigma}_{S S'} \right)_{\mu} + \delta_{\mu\nu} (\bar{\fvec \sigma}_{S S'})_{\rho}  \right] \nonumber \\
	= &  \left( \frac{v_F}2 + D_0 \right)  \delta_{\mu \rho} \bar{\sigma}_{\nu, S S'} +  \left( \frac{v_F}2 + D_0 \right)  \delta_{\nu \rho} \bar{\sigma}_{\mu, S S'} +   D_0   \delta_{\mu \nu} \bar{\sigma}_{\rho, S S'} \\
	\mbox{with} \quad D_0 & = - \frac{i}6 \sum_{\fvec a} e^{i \fvec K \cdot (\fvec a+ \delta \fvec \tau_{AB})} \left( (\fvec a + \delta \fvec \tau_{AB})_x \right)^3   \frac{t'(|\fvec a + \delta \fvec \tau_{S S'}|)}{|\fvec a + \delta \fvec \tau_{S S'}|} \ .
\end{align}
This leads to the term
\begin{align}
	& \pm\Psi_{j, S}^{\dagger}(\fvec x) \half\left(  p_{\mu}    \xi_{\mu,SS'}(\fvec x) +  \xi_{\mu,SS'}(\fvec x)   p_{\mu} +  \left( \frac{v_F}2 + D_0 \right) \half \left\{   \fvec \nabla \cdot \delta \fvec U, \bar{\fvec \sigma}_{SS'} \cdot \fvec p \right\} \right) \Psi_{j, S'}(\fvec x)  \\
	\mbox{with}\quad &   \xi_{x, SS'}(\fvec x) =   \left[ (\frac{v_F}2 + 2D_0)\partial_x \delta   U_x  \sigma_1 - \left[ (\frac{v_F}2 + D_0)\partial_y \delta  U_x + D_0  \partial_x \delta  U_y  \right] \sigma_2 \right]_{S S'} \ , \nonumber \\
	& \fvec \xi_{y, SS'}(\fvec x) =   \left( \left[ (\frac{v_F}2 + D_0)\partial_x \delta   U_y + D_0 \partial_y \delta   U_x  \right] \sigma_1 - (\frac{v_F}2 + 2D_0)(\partial_y \delta U_y ) \sigma_2  \right)_{S S'}  \ .
\end{align}

\subsubsection{$(\fvec \nabla \delta \fvec U)^2$ terms}
Lastly, we consider the terms containing the square of the gradient of $\delta \fvec U$:
\begin{align}
		A_{mlg}^{-1} \half \sum_{\fvec G}e^{i \fvec G \cdot  \delta \fvec \tau_{S S'}} \int\rmd^2 \fvec y\ e^{-i (\fvec G+ \fvec K)\cdot \fvec y} t(|\fvec y|) \frac{i}2   y_{\mu}   \frac{i}2  y_{\nu} \partial_{\mu} \delta  U_{\rho} (\fvec G + \fvec K)_{\rho} \partial_{\nu} \delta  U_{\sigma} (\fvec G + \fvec K)_{\sigma} \Psi_{j, S}(\fvec x)^{\dagger} \Psi_{j, S'}(\fvec x) 
\end{align}
Again, we consider the coefficient
\begin{align}
	& - \frac18 A_{mlg}^{-1} \sum_{\fvec G} e^{i \fvec G \cdot \delta \fvec \tau_{SS'}}  \int\rmd^2 \fvec y\ e^{-i (\fvec G+ \fvec K)\cdot \fvec y} t(|\fvec y|)  y_{\mu} y_{\nu} (\fvec G + \fvec K)_{\rho} (\fvec G + \fvec K)_{\sigma} \nonumber \\
	= & \frac18 \sum_{\fvec a} e^{-i \fvec K \cdot (\fvec  a+ \delta \fvec \tau_{SS'})} \left. \frac{\partial^2}{\partial x_{\rho} \partial x_{\sigma}} \left( (\fvec x + \delta \fvec \tau_{S S'})_{\mu}  (\fvec x + \delta \fvec \tau_{S S'})_{\nu} t(|\fvec x + \delta \fvec \tau_{S S'}|) \right)  \right|_{\fvec  x= \fvec a}  \nonumber \\
	= &  \frac18 \sum_{\fvec a} e^{-i \fvec K \cdot (\fvec  a+ \delta \fvec \tau_{SS'})} \left[  ( \delta_{\mu\sigma} \delta_{\rho\nu} + \delta_{\nu\sigma} \delta_{\rho\mu} ) t(|\fvec a + \delta \fvec \tau_{S S'}|) + \right.\nonumber \\
	&   \left( \delta_{\nu\sigma}  (\fvec a + \delta \fvec \tau_{SS'})_{\mu} (\fvec a + \delta \fvec \tau_{SS'})_{\rho}  + \delta_{\mu\sigma}  (\fvec a + \delta \fvec \tau_{SS'})_{\nu} (\fvec a + \delta \fvec \tau_{SS'})_{\rho} + \delta_{\rho\sigma}  (\fvec a + \delta \fvec \tau_{SS'})_{\mu} (\fvec a + \delta \fvec \tau_{SS'})_{\nu} +  \right. \nonumber \\  
	& \left. + \delta_{\nu\rho}  (\fvec a + \delta \fvec \tau_{SS'})_{\mu} (\fvec a + \delta \fvec \tau_{SS'})_{\sigma} +  \delta_{\rho\mu}  (\fvec a + \delta \fvec \tau_{SS'})_{\nu} (\fvec a + \delta \fvec \tau_{SS'})_{\sigma}  \right) \frac{t'(|\fvec a + \delta \fvec \tau_{S S'}|)}{|\fvec a + \delta \fvec \tau_{S S'}|} + \nonumber \\
	&  \left. (\fvec a + \delta \fvec \tau_{SS'})_{\mu}  (\fvec a + \delta \fvec \tau_{SS'})_{\nu}  (\fvec a + \delta \fvec \tau_{SS'})_{\rho}  (\fvec a + \delta \fvec \tau_{SS'})_{\sigma}  \left( \frac{t''(|\fvec a + \delta \fvec \tau_{S S'}|)}{|\fvec a + \delta \fvec \tau_{S S'}|^2} - \frac{t'(|\fvec a + \delta \fvec \tau_{S S'}|)}{|\fvec a + \delta \fvec \tau_{S S'}|^3}  \right)    \right] \ . \label{EqnS:SecondTerm}
\end{align}
For $S = S'$, the above summation can be simplified as
\begin{align}
	& \frac18 \sum_{\fvec a} e^{-i \fvec K \cdot \fvec a} \left[ \left( \delta_{\mu\sigma} \delta_{\rho \nu} + \delta_{\nu\sigma} \delta_{\rho\mu} \right)  \left( t(|\fvec a|) + \frac78 |\fvec a| t'(|\fvec a|) + \frac18 |\fvec a|^2 t''(|\fvec a|)  \right) + \delta_{\rho\sigma} \delta_{\mu \nu} \left( \frac38 |\fvec a| t'(|\fvec a|) + \frac18 |\fvec a|^2 t''(|\fvec a|) \right) \right] \nonumber \\
	= &\left(  \frac{\mu}8 + C_1  \right) \left( \delta_{\mu\sigma} \delta_{\rho \nu} + \delta_{\nu\sigma} \delta_{\rho\mu} \right) + C_2 \delta_{\rho\sigma} \delta_{\mu \nu} \ ,
\end{align}
with the expression for the two coefficients $C_1'$ and $C_2$ listed below
\begin{align}
	C_1 & = \frac18 \sum_{\fvec a} e^{-i \fvec K \cdot \fvec a} \left( \frac78 |\fvec a| t'(|\fvec a|) + \frac18 |\fvec a|^2 t''(|\fvec a|)  \right) \\
	C_2 & = \frac18 \sum_{\fvec a} e^{-i \fvec K \cdot \fvec a} \left( \frac38 |\fvec a| t'(|\fvec a|) + \frac18 |\fvec a|^2 t''(|\fvec a|) \right)
\end{align}

This leads to the term 
\begin{align}
	\left[ \left(  \frac{\mu}8 + C_1 \right) \left(  (\fvec \nabla \cdot \delta  U)^2 + (\partial_{\mu} \delta  U_{\nu}) (\partial_{\nu} \delta  U_{\mu}) \right) + C_2  (\partial_{\mu} \delta   U_{\nu}) (\partial_{\mu} \delta   U_{\nu}) \right] \Psi^{\dagger}_{j, S}(\fvec x) \Psi_{j, S}(\fvec x)
\end{align}	
%where we have made the approximation $\partial_{\mu} \delta \fvec U_{\mu} \approx 0$. 	

For $S \neq S'$,  the first term in the square bracket of Eq.~\ref{EqnS:SecondTerm} vanishes by $C_3$ symmetry.  Introducing the notation
\begin{align}
	R_{\mu\nu\rho\sigma, S S'} & =  \frac14 v_F\gamma \left[   \delta_{\mu\sigma} \left( (\tau_3)_{\rho\nu} \sigma_1 + (\tau_1)_{\rho\nu} \sigma_2 \right) + (\mu \leftrightarrow \rho) + (\mu\sigma \leftrightarrow \rho\nu) \right]_{S S'} \label{EqnS:RTensor} \\
	V_{\mu\nu\rho\sigma, S S'} & = \frac18 \sum_{\fvec a}  e^{-i \fvec K \cdot (\fvec  a+ \delta \fvec \tau_{SS'})}  (\fvec a + \delta \fvec \tau_{SS'})_{\mu}  (\fvec a + \delta \fvec \tau_{SS'})_{\nu}  (\fvec a + \delta \fvec \tau_{SS'})_{\rho}  (\fvec a + \delta \fvec \tau_{SS'})_{\sigma}  \left( \frac{t''(|\fvec a + \delta \fvec \tau_{S S'}|)}{|\fvec a + \delta \fvec \tau_{S S'}|^2} - \frac{t'(|\fvec a + \delta \fvec \tau_{S S'}|)}{|\fvec a + \delta \fvec \tau_{S S'}|^3}  \right)  \nonumber \\
	& = \frac{D_1}6 \left[   \delta_{\nu\mu} \left( (\tau_3)_{\rho\sigma} \sigma_1 + (\tau_1)_{\rho\sigma} \sigma_2 \right) + (\mu \leftrightarrow \rho) + (\mu \leftrightarrow \sigma) + (\rho \leftrightarrow \nu) + (\nu \leftrightarrow \sigma) + (\rho\mu \leftrightarrow \nu\sigma) \right]_{S S'} \nonumber \\
	& \ + D_2 \left( \cos(\frac{\pi}2 n_2) \sigma_1 - \sin(\frac{\pi}2 n_2) \sigma_2 \right)_{S S'} \label{EqnS:VTensor}  \\
	\zeta_{S S'}(\fvec x) & = \left(R_{\mu\nu\rho\sigma, S S'} + V_{\mu\nu\rho\sigma, S S'} \right) \partial_{\mu} \delta   U_{\rho} \partial_{\nu} \delta  U_{\sigma} \label{EqnS:ZetaField}
\end{align}
where $n_2$ in  Eq.~\ref{EqnS:VTensor} is the number of times that the index $y$ appears in the subscripts $\mu$, $\nu$, $\rho$, and $\sigma$, i.e.
\[ n_2 = \delta_{\mu y} + \delta_{\nu y} + \delta_{\rho y} + \delta_{\sigma y} \  .  \]
Eq.~\ref{EqnS:VTensor} is derived by $C_3$ symmetry, and the constants $D_1$ and $D_2$ must be real as constrained by $m_y$ symmetry. To be more specific,
\[ \left\{   \begin{array}{l}
	D_1 = \dfrac1{16} \sum\limits_{\fvec a} e^{-i \fvec K \cdot (\fvec a + \delta \fvec \tau_{AB})} \left( \dfrac{t''(|\fvec a + \delta \fvec \tau_{AB}|)}{|\fvec a + \delta \fvec \tau_{AB}|^2} - \dfrac{t'(|\fvec a + \delta \fvec \tau_{AB}|)}{|\fvec a + \delta \fvec \tau_{AB}|^3}  \right) \left( \left[ (\fvec a + \delta \fvec \tau_{AB})_x \right]^4 + \left[ (\fvec a + \delta \fvec \tau_{AB})_y \right]^4 \right) \\
	D_2 = \dfrac1{16} \sum\limits_{\fvec a} e^{-i \fvec K \cdot (\fvec a + \delta \fvec \tau_{AB})} \left( \dfrac{t''(|\fvec a + \delta \fvec \tau_{AB}|)}{|\fvec a + \delta \fvec \tau_{AB}|^2} - \dfrac{t'(|\fvec a + \delta \fvec \tau_{AB}|)}{|\fvec a + \delta \fvec \tau_{AB}|^3}  \right) \left( \left[ (\fvec a + \delta \fvec \tau_{AB})_x \right]^4 - \left[ (\fvec a + \delta \fvec \tau_{AB})_y \right]^4 \right)
\end{array}    \right.    \]
Thus, we have the terms 
\begin{align}
	\sum_{j} \sum_{SS'} \Psi^{\dagger}_{j, S} \left( \zeta_{S S'} + \half v_F \gamma \left( \fvec \nabla \cdot \delta \fvec U \right) \bar{\fvec \sigma}_{SS'} \cdot \mathcal{A}  \right) \Psi_{j, S'}
\end{align}
with the numerical values of the parameters listed in Table.~\ref{TabS:DeltaParammeterRelax}.
%	\centering
\begin{table}[htb]
	\centering
	\begin{tabular}{|c|c|c|c|c|c|} \hline
		intra-layer & $\alpha$ (eV) & $C_1$ (eV)  & $C_2$ (eV) &  $D_1$ (eV)  &  $D_2$ (eV)  \\ \hline
		Ref.~\cite{KoshinoPRB12}  & $-1.0542$ & $-0.1165$ & $0.1448$ & $-0.9829$ & $3.5837$  \\
		Ref.~\cite{KaxirasPRB16}  & $2.5472$  & $0.4822$ & $-0.1546$ & $-1.0862$   &  $-2.1330$  \\ \hline 
	\end{tabular}
	\caption{Numerical values of the parameters in $\delta H_{intra}$ in Eq.(\ref{Eqn:HIntra}) and (\ref{EqnS:DeltaHIntra})  for two different microscopic tight binding models.}
	\label{TabS:DeltaParammeterRelax}
\end{table}

\subsection{Deformation Potential}
\label{Sec:DeformationPotential}
In this subsection, we consider how the strain can couple with the charge density of the electrons on a single layer. In the previous subsections, we have calculated the effective continuum Hamiltonian $H_{eff}^{\bK}$, that contains the term
\begin{align}
  H_{dp, 1} = \alpha \int \rmd^2 \fvec x\ \sum_{j, S}  \pm \left( \fvec \nabla \cdot \delta \fvec U \right) \Psi_{j,S}^{\dagger} \Psi_{j,S} = 2\alpha \int \rmd^2 \fvec x\ \sum_{j, S}   \left( \fvec \nabla \cdot \fvec U_j^{\parallel} \right) \Psi_{j,S}^{\dagger} \Psi_{j,S}  \label{EqnS:DP1}
\end{align}
with $\alpha = \frac14 \sum_{\fvec a} e^{-i \fvec K \cdot \fvec a} |\fvec a| t'(|a|)$. In the above formula,  $+$ and $-$ sign are form the top and bottom layers respectively, and we have used the fact that $\fvec \nabla \cdot \fvec U_t^{\parallel} = - \fvec \nabla \cdot \fvec U_b^{\parallel} = \half \fvec \nabla \cdot \delta \fvec U$. However,  we have assumed that the all the hopping constants depend only on the displacement, and neglect the possible dependence of the onsite energy on the nearby atomic configurations. If the onsite energy $\epsilon$ depends on the lengths of three nearest bonds, it is expected that this dependence should also induce the coupling between the strain and the fermion density. In this case, $\epsilon$ is a function of $\left\{ |\fvec n_{j, S}^{(i)}(\fvec x) | \right\})$ ($i = 1, \cdots, 3$) where $\fvec n^{(i)}_{j, S}(\fvec x)$ is the nearest neighbor bond on layer $j$, sublattice $S$, and at the position of $\fvec x$~\cite{paper1}. As a result,  the onsite energy can be expanded to the linear order of the strain~\cite{paper1}:
\begin{align}
	\epsilon\left( \left\{ |\fvec  n_{j, S}^{(\alpha)}(\fvec x) |\right\} \right) \approx \epsilon_0 + \kappa \fvec \nabla \cdot \fvec U_j^{\parallel} \quad \mbox{with} \quad \kappa = \frac{\sqrt{3}}2 a  \frac{\partial \epsilon}{\partial |\delta_S^{\alpha}|} 
\end{align}
This induces another term that couples the strain with the fermion density,
\begin{align}
	H_{dp, 2}^{\bK} =   \sum_{j, S} \int \rmd^2 \fvec x\ \left( \epsilon_0 + \kappa \fvec \nabla \cdot \fvec U_j^{\parallel} \right) \Psi_{j,S}^{\dagger} \Psi_{j,S}  \ .  \label{EqnS:DP2}
\end{align}
Now, combining Eq.~\ref{EqnS:DP1} and \ref{EqnS:DP2}, we obtain
\begin{align}
	H_{dp}^{\bK} =  \sum_{j, S} \int \rmd^2 \fvec x\ \left( \epsilon_0 + \alpha_{dp} \fvec \nabla \cdot \fvec U_j^{\parallel} \right) \Psi_{j,S}^{\dagger} \Psi_{j,S}  \ . 
\end{align}
with
\begin{align}
	\alpha_{dp} = \kappa + 2\alpha = \frac{\sqrt{3}}2 a  \frac{\partial \epsilon}{\partial |\delta_S^{\alpha}|}  + \half \sum_{\fvec a} e^{-i \fvec K \cdot \fvec a} |\fvec a| t'(|\fvec a|)
\end{align}
Since Ref.~\cite{KoshinoPRB12} and \cite{KaxirasPRB16} does not provide enough information on how the onsite energy depends on the nearby atomic configuration, we can calculate only the value of $\alpha$, but not $\kappa$. The numerical value of $\kappa$ can be obtained either by  {\em ab initio}  method, such as DFT, or calibrated by the experimental measurements on $\alpha_{dp}$ with the formula $\kappa = \alpha_{dp} -  2\alpha$.  

%Combined with Eq.~\ref{EqnS:DP}, we obtain 

%The values of the second term in the expression of $\alpha_{dp}$ for two microscopic hopping  models are listed in Table~\ref{TabS:DeltaParammeterRelax}.  

% We will leave this form and do not simplify it. 

\section{Inter-layer Tunneling}	
In this section, we will present the detailed derivation and the symmetry analysis of the inter-layer tunneling terms. Starting from the master formula in Eq.~\ref{Eqn:EffContH}, and keep only the first order gradient of the fermion fields, we obtain
\begin{align}
	H_{inter} & = A_{mlg}^{-1} \sum_{\fvec G} \sum_{S S'} e^{i \fvec G \cdot \fvec \tau_{S S'}} e^{i (\fvec G + \fvec K)\cdot 2\fvec U_{t, S}^{\parallel}(\fvec x) } \int \rmd^2 \fvec y\ e^{-i(\fvec G + \fvec K)\cdot \fvec y} t(\fvec y + d_0 \hat z, \{\fvec n^{(\alpha)}_{t, S}(\fvec x + \fvec y/2) \},  \{\fvec n_{b, S'}^{(\alpha)}( \fvec x - \fvec y/2)\} ) \times \nonumber \\
	& \ \left[ \Psi^{\dagger}_{t,S}(\fvec x) \Psi_{b, S'}(\fvec x) + \frac{\fvec y}2 \cdot \left( \fvec \nabla \Psi^{\dagger}_{t,S}(\fvec x)  \Psi_{b, S'}(\fvec x) - \Psi^{\dagger}_{t,S}(\fvec x)  \fvec \nabla \Psi_{b, S'}(\fvec x) \right)  \right]  + h.c.
\end{align}
where we have applied the formula that the lattice distortion $\fvec U^{\parallel}_{t} = - \fvec U^{\parallel}_b$ and are independent of the sublattice. In addition, the microscopic inter-layer hopping function are assumed to depend not only on the hopping displacement $\fvec y + d_0 \hat z$, but also on the direction of the nearest neighbor vectors $\fvec n^{(\alpha)}_{t, S}$ and $\fvec n^{(\alpha)}_{b, S'}$ on each layer~\cite{KaxirasPRB16}.  Note that the terms in Eq.~\ref{Eqn:EffContH} are expanded only to the first order of the gradients of fermion fields, and therefore, the inter-layer tunneling can be written as
\begin{align}
	H_{inter} = \int\rmd^2 \fvec x\ \Psi_{t,S}(\fvec x) \left[ T_{SS'}(\fvec x) + \half \left\{ \fvec p, \fvec \Lambda_{SS'}(\fvec x)  \right\}  \right] \Psi_{b, S'}(\fvec x) + h.c.
\end{align}
In the rest of this section, we will study the properties of these two tunneling fields $T_{SS'}(\fvec x)$  and $\fvec \Lambda_{S S'}(\fvec x)$ and express them in terms of the microscopic hopping functions.

Different from the intra-layer terms that depends only on the gradient of $\fvec U^{\parallel}$, the inter-layer tunnelings are the function of $\fvec U^{\parallel}$ that is not tiny compared with $|\fvec a|$. As a consequence, we do not expand the inter-layer tunneling terms in the powers of $\delta \fvec U$. 

\begin{figure}[htb]
	\centering
	\includegraphics[width=0.5\columnwidth]{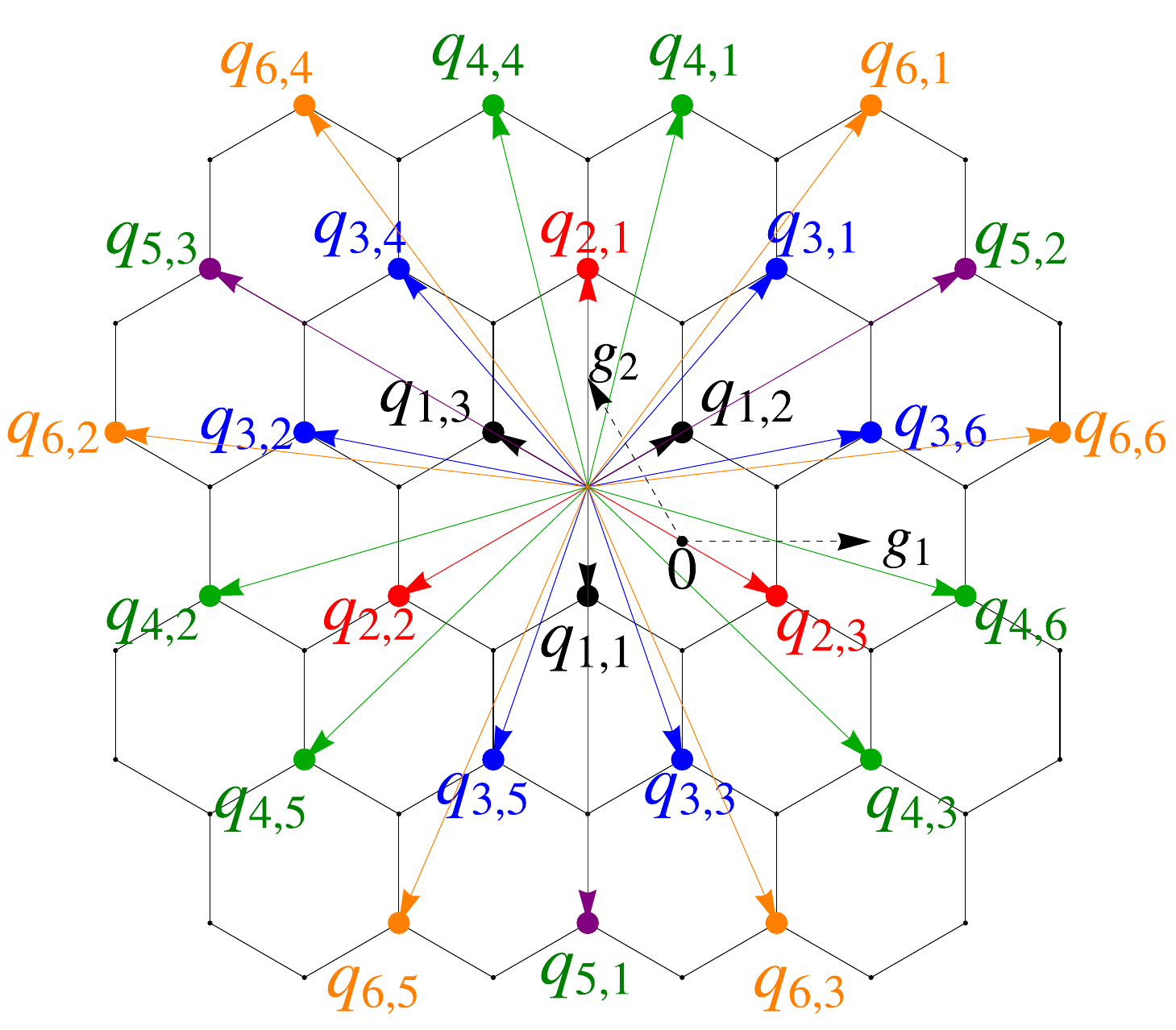}
	%	\subfigure[\label{Fig:AField:2}]{\includegraphics[width=0.95\columnwidth]{PseudoMagField.pdf}}					
	\caption{The vectors $\fvec q_{\mu,l}$ in the first six shells.}
	\label{FigS:ExtraQShell}
\end{figure}

As mentioned in the main text, in general, $\fvec U^{\parallel}_t(\fvec x) = \frac{\theta}2 \hat z \times \fvec x + \half \delta \fvec U(\fvec x)$, leading to
\[ e^{i(\fvec G + \fvec K)\cdot 2  \fvec U^{\parallel}(\fvec x)} = e^{i \theta (\fvec G + \fvec K) \cdot (\hat z \times \fvec x)} e^{i (\fvec G+ \fvec K) \cdot \delta \fvec U} = e^{i \fvec x \cdot (\fvec q_{1,1} + \fvec g_{\fvec G})} e^{i (\fvec G + \fvec K) \cdot \delta \fvec U(\fvec x)}  \]
where $\fvec q_{1, 1} = - \theta z\times \fvec K$ and $\fvec g_{\fvec G} = -\theta \hat z \times \fvec G$. For small twist angle $\theta \ll 1$, it can be shown that $\fvec g_{\fvec G}$ is a reciprocal lattice vector of the moire superlattice. For $\fvec G_i$ and $\fvec g_i$ defined in Fig.~\ref{Fig:Schematic}, it is easy to see that 
\[ \fvec g_{\fvec G_1} = -(\fvec g_1 + \fvec g_2) \quad \mbox{and}\quad \fvec g_{\fvec G_2} = \fvec g_1 \quad \Longrightarrow \quad   \fvec g_{m_1 \fvec G_1 + m_2 \fvec G_2} = (m_2 - m_1) \fvec g_1 - m_2 \fvec g_2 \]
In addition, the lattice relaxation $\delta \fvec U(\fvec x) = \delta \fvec U(\fvec x + \fvec L_i)$ ($i = 1$, $2$) is a periodic function, and so is $e^{i (\fvec G + \fvec K) \cdot \delta \fvec U}$. This suggests that 
\[  e^{i (\fvec G + \fvec K) \cdot \delta \fvec U} = \sum_{\fvec g} u_{\fvec g}(\fvec G) e^{i \fvec g \cdot \fvec x} \quad \Longrightarrow \quad e^{i (\fvec G + \fvec K)\cdot 2U^{\parallel}_t(\fvec x)} = e^{i \fvec x \cdot \fvec q_{(1,1)}} \sum_{\fvec g} e^{i (\fvec g_{\fvec G} + \fvec g) \cdot \fvec x} u_{\fvec g}(\fvec G) \]
where $\sum_{\fvec g}$ sums over all the reciprocal lattice vector of the moire superlattice. This gives the Fourier transformation of $e^{i(\fvec G + \fvec K) \cdot 2 \fvec U^{\parallel}_t(\fvec x)}$. 
 
Now, we obtain the expression of the inter-layer tunneling terms:
\begin{align}
	T_{SS'}(\fvec x) & = \sum_{\fvec G} e^{i \fvec G \cdot \fvec \tau_{S S'}} e^{i \fvec q_{(1,1)} \cdot \fvec x} \sum_{\fvec g} e^{i (\fvec g_{\fvec G} + \fvec g) \cdot \fvec x} u_{\fvec g}(\fvec G) \int \rmd^2 \fvec y\ e^{-i (\fvec G + \fvec K) \cdot \fvec y} t(\fvec y + d_0 \hat z, \{\fvec n^{(\alpha)}_{t, S}(\fvec x + \fvec y/2) \},  \{\fvec n_{b, S'}^{(\alpha)}( \fvec x - \fvec y/2)\} ) \label{EqnS:TComp} \\
	\fvec \Lambda_{SS'}(\fvec x) & = \sum_{\fvec G} e^{i \fvec G \cdot \fvec \tau_{S S'}} e^{i \fvec q_{(1,1)} \cdot \fvec x} \sum_{\fvec g} e^{i (\fvec g_{\fvec G} + \fvec g) \cdot \fvec x} u_{\fvec g}(\fvec G) \int \rmd^2 \fvec y\ (-i \fvec y) e^{-i (\fvec G + \fvec K) \cdot \fvec y} t(\fvec y + d_0 \hat z, \{\fvec n^{(\alpha)}_{t, S}(\fvec x + \fvec y/2) \},  \{\fvec n_{b, S'}^{(\alpha)}( \fvec x - \fvec y/2)\} ) \label{EqnS:LambdaComp}
\end{align} 
As shown in the next section, the Fourier transformation of the inter-layer tunneling in both microscopic tight binding models decay fast as a function of the momentum. Therefore, it is convenient to express and calculate the Fourier transformation of the tunneling terms. Since the direction of the nearest bond $\fvec n_{j, S}$ is also periodic with the period of $\fvec L_i$, its' Fourier transformation can be written as
\begin{align}
	T_{SS'}(\fvec x) = \sum_{\mu, l} e^{i \fvec q_{\mu, l} \cdot \fvec x} T_{SS'}^{(\mu, l)} \ , \qquad	\fvec \Lambda_{SS'}(\fvec x) = \sum_{\mu, l} e^{i \fvec q_{\mu, l} \cdot \fvec x} \fvec \Lambda_{SS'}^{(\mu, l)} 
\end{align}
where the vectors $\fvec q_{(\mu, l)}$ are plotted in Fig.~\ref{FigS:ExtraQShell} for the first six $\fvec q$ shells ($\mu \leq 6$). The subscripts $\mu$ refers to the shell of the vectors, and $l$ is used to distinguish different vectors in the same shell. It is easy to see that $\fvec q_{\mu, l} - \fvec q_{(1, 1)}$ is a reciprocal lattice vector of the moire superlattice. 
 
Here, we consider the microscopic tight binding models with the inter-layer hopping depending only on the hopping displacement, as in Ref.~\cite{KoshinoPRB12}. We introduce the notation for the Fourier transformation of the hopping 
\[ \tilde{t}_{d_0}(\fvec q) = A_{mlg}^{-1} \int \rmd^2 \fvec y\ e^{- i \fvec q \cdot \fvec y} t(\fvec y + d_0 \hat z)  \quad \Longrightarrow \quad A_{mlg}^{-1} \int \rmd^2 \fvec y\ e^{- i \fvec q \cdot \fvec y} \fvec y t(\fvec y + d_0 \hat z) = i \fvec \nabla_{\fvec q} \tilde{t}_{d_0}(\fvec q)  \]   
Therefore, 
\begin{align}
	T_{SS'}(\fvec x) & = e^{i \fvec q_{(1,1)} \cdot \fvec x}  \sum_{\fvec G} e^{i \fvec G \cdot \fvec \tau_{S S'}} \tilde{t}_{d_0}(\fvec G + \fvec K)  \sum_{\fvec g} e^{i (\fvec g_{\fvec G} + \fvec g) \cdot \fvec x} u_{\fvec g}(\fvec G) \\
	\fvec \Lambda_{SS'}(\fvec x) & = e^{i \fvec q_{(1,1)} \cdot \fvec x}  \sum_{\fvec G} e^{i \fvec G \cdot \fvec \tau_{S S'}}  \left. \fvec \nabla_{\fvec q} \tilde{t}_{d_0}(\fvec q) \right|_{\fvec q = \fvec G + \fvec K} \sum_{\fvec g} e^{i (\fvec g_{\fvec G} + \fvec g) \cdot \fvec x} u_{\fvec g}(\fvec G) 
\end{align}

With the rigid twist only, $\delta \fvec U = 0$, and thus $u_{\fvec g}(\fvec G) = \delta_{\fvec g, 0}$. If we only focus on the first $\fvec q$ shell ($\mu =1$), we see that
\begin{align}
	T_{S S'}(\fvec x) & = e^{i \fvec q_{(1,1)} \cdot x} \sum_{\fvec G} e^{i \fvec G \cdot \delta \fvec \tau_{S S'}} \tilde{t}_{d_0}(\fvec G + \fvec K) e^{i \fvec g_{\fvec G} \cdot \fvec x} \nonumber \\
	& \approx e^{i \fvec q_{(1,1)} \cdot \fvec x} \tilde{t}(\fvec K) + e^{i \fvec q_{1,2} \cdot \fvec x} e^{- i \fvec G_1 \cdot  \delta \fvec \tau_{S S'}} \tilde{t}_{d_0}(-\fvec G_1 + \fvec K) + e^{i \fvec q_{1,2} \cdot \fvec x} e^{-i (\fvec G_1 + \fvec G_2 ) \cdot  \delta \fvec \tau_{S S'}} \tilde{t}_{d_0}(-\fvec G_1 - \fvec G_2 + \fvec K)
\end{align}
Due to $C_3$ symmetry, $\tilde{t}_{d_0}(\fvec K) = \tilde{t}_{d_0}(-\fvec G_1 + \fvec K) = \tilde{t}_{d_0}(-\fvec G_1 - \fvec G_2 + \fvec K)$ and are real by $C_{2x}$ symmetry. Thus, by calculating the component of $w_i^{(1, l)}$, we see that
\[  w_0^{(1,l)} = w_0 \ , \quad w_2^{(1,1)} = 0 \ , \quad w_3^{(1, l)} = 0 \ ,  \quad w_1^{(1, 2)} = w_1^{(1, 3)} = - \frac12 w_1^{(1,1)} \ , \quad  \quad w_2^{(1, 2)} = - w_2^{(1, 3)} = -\frac{\sqrt{3}}2 w_0  \ . \]
where $w_0 = \tilde{t}_{d_0}(\fvec K)$.  Thus, if we further neglect the gradient coupling fields $\fvec \Lambda_{SS'}$, we fully recover the BM model~\cite{BMModel} in which $w_0 = w_1$.

The values of $w_i^{(\mu, l)}$ and $\fvec \lambda_i^{(\mu, l)}$ have been listed in Table.~\ref{Tab:ParammeterInter} and \ref{TabS:ExtraParammeterInter} for the two microscopic tight binding models proposed in Ref.~\cite{KoshinoPRB12} and \cite{KaxirasPRB16}. Notice that we only listed the values of $w$s and $\fvec \lambda$s only for the first several $\fvec g$ shells. The contributions from other shells are negligible. 

In Sec.~\ref{Sec:TBMKaxiras}, we will discuss the inter-layer tunnelings terms  for another more complicated microscopic model in which the inter-layer hopping depends not only on the hopping displacement, but also on the direction of the nearest neighbor bonds on the two layers.

\begin{table}[htb]
	\centering
	\begin{tabular}{|c|c|c|c|c|c|c|c|c|c|c|c|c|} \hline
		&  $w_0^{(4,1)}$ & $w_1^{(4,1)}$  & $w_2^{(4,1)}$  & $w_3^{(4,1)}$  &   $w_0^{(5,1)}$  & $w_1^{(5,1)}$  & $w_2^{(5,1)}$  & $w_3^{(5,1)}$ &  $w_0^{(6,1)}$  & $w_1^{(6,1)}$ & $w_2^{(6,1)}$  & $w_3^{(6,1)}$    \\ \hline
		\begin{tabular}{@{}c@{}}Ref.~\cite{KoshinoPRB12} \\ relaxed \end{tabular} & $-0.39$ & $1.13$ & $-0.20$ & $0$ & $5.79$ & $6.36$ & $0$ & $0$ & $5.18$ & $-2.78$ & $-4.98$ & $0$ \\ \hline 
		\begin{tabular}{@{}c@{}}Ref.~\cite{KaxirasPRB16} \\ relaxed \end{tabular} & $0.16$ & $0.34$ & $0.10$ & $-0.07$  & \multicolumn{8}{|c|}{negligible}  \\ \hline
		& $\fvec \lambda_{0}^{(4,1)}/a$   & $\fvec \lambda_{1}^{(4,1)}/a$   & $\fvec \lambda_2^{(4,1)}/a$ & $\fvec \lambda_3^{(4,1)}/a$  & $\fvec \lambda_{0}^{(5,1)}/a$   & $\fvec \lambda_{1}^{(5,1)}/a$   & $\fvec \lambda_2^{(5,1)}/a$ & $\fvec \lambda_3^{(5,1)}/a$ & $\fvec \lambda_{0}^{(6,1)}/a$   & $\fvec \lambda_{1}^{(6,1)}/a$   & $\fvec \lambda_2^{(6,1)}/a$ & $\fvec \lambda_3^{(6,1)}/a$ \\ \hline
		\begin{tabular}{@{}c@{}}Ref.~\cite{KoshinoPRB12} \\ relaxed \end{tabular} & $(-0.1, -0.2)$ & $(0.5, 0.1)$ & $(-0.1, -0.7)$ & $(0, 0)$ & $(-5.2, 0)$ & $(-5.1, 0)$ & $(0, -0.2)$ & $(0, 0)$ & $(2.2, -4.1)$ & $(-1.1, 2.1)$ & $(-2.1, 3.3)$ & $(0, 0)$ \\ \hline 
		\begin{tabular}{@{}c@{}}Ref.~\cite{KaxirasPRB16} \\ relaxed \end{tabular} & $(0.3, 0.1)$ & $(0.3, -0.1)$ & $(-0.1, 0)$  & $(0, 0)$ & \multicolumn{8}{|c|}{negligible} \\ \hline
	\end{tabular}
	\caption{Parameters of the inter-layer tunneling terms for two microscopic tight binding models, in the presence of the lattice relaxation. $a$ is the magnitude of the primitive lattice vector, and all numbers are in the unit of meV. }
	\label{TabS:ExtraParammeterInter}
\end{table}

\section{Symmetry Analysis}
\label{SecS:Sym}
In this section, we consider the constraints on them by various symmetries. Here, we focus on three different symmetry transformations, $C_2\mathcal{T}$, $C_3$, and $C_{2x}$, and the fermion fields transform as 
\begin{align}
	C_2\mathcal{T}: \quad & \Psi_{j, S}(\fvec x) \longrightarrow \mathcal{K} \Psi_{j, \bar{S}}(-\fvec x)    \\
	C_3: \quad  & \Psi_{j, S}(\fvec x) \longrightarrow  e^{-i \frac{2\pi}3 (\sigma_z)_{SS}} \Psi_{j, S}\left( R\left( - \frac{2\pi}3 \right) \fvec x \right) \\
	C_{2x}: \quad & \Psi_{j, S}(\fvec x) \longrightarrow  \Psi_{\bar{j}, \bar{S}}(m_y \fvec x)  \ , 
\end{align}
where $\bar{j}$ and $\bar{S}$ are the layer and sublattice index different from $j$ and $S$ respectively. $R\left(-\frac{2\pi}3 \right) \fvec x$ is the vector $\fvec x$ rotated clockwisely by the angle of $2\pi/3$, i.e. 
\[ R\left(-\frac{2\pi}3 \right) (x, y)^T = \left( - \frac{x}2 + \frac{\sqrt{3}}2 y \ ,  - \frac{y}2 - \frac{\sqrt{3}}2 x \right)^T  \]
and $m_y$ is the reflection symmetry through $xz$ plane, i.e.~$m_y (x, y)^T = (x, -y)^T$.
 
The inter-layer tunneling terms are invariant under $C_2\mathcal{T}$ transformation. It leads to the constraints
\begin{align}  
	& \left( \sigma_1  T(\fvec x) \sigma_1 \right)_{SS'} = \left(T_{SS'}(-\fvec x) \right)^* \quad  \left( \sigma_1  \fvec \Lambda(\fvec x) \sigma_1 \right)_{SS'} = \left( \fvec \Lambda_{SS'}(-\fvec x) \right)^*   \\ 
\Longrightarrow & \left( \sigma_1 T^{(\mu, l)} \sigma_1 \right)_{SS'} = \left( T_{SS'}^{(\mu, l)}\right)^* \ , \quad \left( \sigma_1 \fvec \Lambda^{(\mu, l)} \sigma_1 \right)_{SS'} = \left( \fvec \Lambda_{SS'}^{(\mu, l)}\right)^* 
\end{align}
This suggests that we can write 
\begin{align}
	T_{SS'}^{(\mu, l)} = \left( w_0^{(\mu, l)} \sigma_0 + w_1^{(\mu, l)} \sigma_1 + w_2^{(\mu, l)} \sigma_2 + i w_3^{(\mu, l)} \sigma_3 \right)_{SS'} \\ 
	\fvec \Lambda_{SS'}^{(\mu, l)} = \left( \fvec \lambda_0^{(\mu, l)} \sigma_0 + \fvec \lambda_1^{(\mu, l)} \sigma_1 + \fvec \lambda_2^{(\mu, l)} \sigma_2 + i \fvec \lambda_3^{(\mu, l)} \sigma_3 \right)_{SS'}
\end{align}
where all $w_i^{(\mu, l)}$ ($i = 0$, $\cdots$, $4$) are real numbers and $\fvec \lambda^{(\mu, l)}_i$ are two-component real vectors. 

The invariance under $C_3$ transformation leads to the following constraints on $T_{SS'}$ and $\fvec \Lambda_{SS'}$:
\begin{align}
	 T_{SS'}(\fvec x)  & = \left( e^{i \frac{2\pi}3 \sigma_z} T\left( R\left( \frac{2\pi}3 \right) \fvec x \right) e^{-i \frac{2\pi}3 \sigma_z} \right)_{S S'} \nonumber \\
	 \Lambda_{\alpha, SS'}(\fvec x) & = \left( e^{i \frac{2\pi}3 \sigma_z} R_{\alpha\beta}\left(- \frac{2\pi}3 \right) \Lambda_{\beta}\left( R\left( \frac{2\pi}3 \right) \fvec x \right) e^{-i \frac{2\pi}3 \sigma_z} \right)_{S S'} \ .
\end{align}
For notational convenience, we introduce $\fvec q_{(\mu, l')} = R(2\pi/3)\fvec q_{(\mu, l)}$. Then,
\begin{align}
	T^{(\mu, l)}_{SS'} = \left( e^{i\frac{2\pi}3 \sigma_z} T^{(\mu, l')} e^{-i\frac{2\pi}3 \sigma_z}  \right)_{S S'} \quad , \quad \Lambda_{\alpha, S S'}^{(\mu, l)} = R_{\alpha\beta}\left(- \frac{2\pi}3 \right) \left( e^{i\frac{2\pi}3 \sigma_z}  \Lambda_{\beta}^{(\mu, l')} e^{-i\frac{2\pi}3 \sigma_z}  \right)_{SS'} \ .
\end{align}
Eventually, the contact and gradient coupling constants should satisfy the following relations:
\begin{align}
	& w_0^{(\mu, l')} = w_0^{(\mu, l)} \ , \qquad w_3^{(\mu, l')} = w_3^{(\mu, l)} \ , \qquad w_1^{(\mu, l')} - i w_2^{(\mu, l')} = e^{i\frac{2\pi}3} 
	\left( w_1^{(\mu, l)} - i w_2^{(\mu, l)} \right)  \nonumber \\
	& ( \fvec \lambda_0^{(\mu, l')} )_{\alpha} =  \left( R\left(\frac{2\pi}3\right) \right)_{\alpha \beta} \left( \fvec \lambda_0^{(\mu, l)} \right)_{\beta} \ , \qquad 	( \fvec \lambda_3^{(\mu, l')} )_{\alpha} =  \left( R\left(\frac{2\pi}3\right) \right)_{\alpha \beta} \left( \fvec \lambda_3^{(\mu, l)} \right)_{\beta} \nonumber \\
	& ( \fvec \lambda_1^{(\mu, l')} - i \fvec \lambda_2^{(\mu, l')} )_{\alpha} = e^{i \frac{2\pi}3} \left( R\left( \frac{2\pi}3 \right) \right)_{\alpha \beta} \left( \fvec \lambda_1^{(\mu, l)} - i \fvec \lambda_2^{(\mu, l')} \right)_{\beta} \ . \label{EqnS:C3Constraints}
\end{align}
%As shown in Fig.~\ref{FigS:ExtraQShell}, $l' = l+1$ for $l \neq 3$ or $6$. Otherwise, $l' = 1$ if $l = 3$, and $l' = 4$ if $l = 6$. 

Lastly, we consider $C_{2x}$ symmetry. It impose the constraints:
\begin{align}
	     T_{SS'}(\fvec x)  & = \left( \sigma_x T( m_y \fvec x) \sigma_x \right)_{S' S}^* \ ,  \qquad
	\fvec \Lambda_{\alpha, SS'}(\fvec x)  = ( \tau_3 )_{\alpha\beta} \left( \sigma_x \fvec \Lambda_{\beta}(m_y \fvec x) \sigma_x \right)_{S' S}^*  \ .
\end{align}
Again, we introduce the notation $\fvec q_{(\mu, n)} = - m_y \fvec q_{(\mu, l)}$. It leads to
\begin{align}
	& w_0^{(\mu, n)} =  w_0^{(\mu, l)} \ , \qquad w_1^{(\mu, n)} =  w_1^{(\mu, l)} \ ,  \qquad w_2^{(\mu, n)} = - w_2^{(\mu, l)} \ , \qquad w_3^{(\mu, n)} = w_3^{(\mu, l)}  \nonumber \\
	& \left( \fvec \lambda_{0}^{(\mu, n)} \right)_{\alpha}= \left( \tau_3 \right)_{\alpha \beta} \left( \fvec \lambda_{0}^{(\mu, l)}  \right)_{\beta} \ , \quad \left( \fvec \lambda_{1}^{(\mu, n)} \right)_{\alpha}= \left( \tau_3 \right)_{\alpha \beta} \left( \fvec \lambda_{1}^{(\mu, l)}  \right)_{\beta} \ , \quad \left( \fvec \lambda_2^{(\mu, n)} \right)_{\alpha}= - \left( \tau_3 \right)_{\alpha \beta} \left( \fvec \lambda_2^{(\mu, l)}  \right)_{\beta}   \nonumber  \\
	& \left( \fvec \lambda_3^{(\mu, n)} \right)_{\alpha}= \left( \tau_3 \right)_{\alpha \beta} \left( \fvec \lambda_3^{(\mu, l)}  \right)_{\beta} \ . \label{EqnS:C2xConstraints}
\end{align}
The formulas above are the general constraints for  the inter-layer tunneling fields $T$ and $\fvec \Lambda$. As an example, we explicitly write down the formula here  for $\fvec q$s in the innermost shell $\mu = 1$. It is obvious that $n = 1$ for $l = 1$. Combining Eq.~\ref{EqnS:C3Constraints} and \ref{EqnS:C2xConstraints}, 
\begin{align}  
	& w_0^{(1, l)} = w_0 \ , &  w_3^{(1, l)} & = w_3 \ , & w_2^{(1, 1)} & = 0 \\
	& w_1 = w_1^{(1,1)} \ , & w_1^{(1,2)} & = w_1^{(1,3)} = - \frac{w_1}2 \ , & w_2^{(1,2)} & = - w_2^{(1,3)} = - \frac{\sqrt{3}}2 w_1 \\
	& \fvec \lambda_0^{(1,1)} = (\lambda_{0, x}, 0) \ , & \fvec \lambda_1^{(1,1)} &= (\lambda_{1, x}, 0) \ , &  \fvec \lambda_3^{(1,1)} & = (\lambda_{3, x}, 0) \ , & \fvec \lambda_2^{(1,1)} = (0, \lambda_{2, y}) \ .
\end{align}

\section{Microscopic Hopping Function}
\label{Sec:microscopicTBM}

\subsection{Slater-Koster like hopping parameterization}  
\label{Sec:TBMKoshino}
First, we consider the microscopic tight binding model proposed in Ref.~\cite{KoshinoPRB12}, in which
\begin{align}
		t(\fvec d) & = V_{pp\pi}^0 e^{- \frac{|\fvec d| - a_0}{\delta}} \left[  1 - \left( \frac{\fvec d \cdot \hat z}{|\fvec d|} \right)^2 \right] + V_{pp\sigma}^0 e^{- \frac{|\fvec d| - d_0}{\delta}}  \left( \frac{\fvec d \cdot \hat z}{|\fvec d|} \right)^2
		\label{Eqn:SlaterKoster}
\end{align}
where $V^0_{pp\pi} = -2.7$eV, $V^0_{pp\sigma} = 0.48$eV. $a_0 = 0.142$nm is the distance between the two nearest-neighbor carbon atoms on the same layer; $d_0 =0.335$nm is the inter-layer distance. The decay length for the hopping is $\delta = 0.319a_0$. The intra-layer hopping thus can be expressed as 
\begin{align}
	t_{intra}(\fvec d) = V_{pp\pi}^0 e^{- \frac{|\fvec d| - a_0}{\delta}} \label{Eqn:KIntra}
\end{align}
where $\fvec d$ is the in-plane hopping displacement.It is easy to show that the Fourier transformation of the intra-layer hopping is
\begin{align}
	\tilde{t}_{intra}(\fvec q) = A_{mlg}^{-1} \int \rmd^2 \fvec y\ e^{-i \fvec y \cdot \fvec q} t(\fvec y ) = V_{pp\pi}^0 \frac{2\pi \delta^2}{A_{mlg}} e^{a_0/\delta} \left[ 1 + (q \delta)^2\right]^{-\frac32}
\end{align}
It is now clear that the $\tilde{t}_{intra}$ decays in $q^{-3}$ in the momentum space, but the intra-layer hopping $t$ exponentially decays in real space. 
The Fourier transformations $\tilde{t}$ of the inter-layer hoppings are  
\begin{align}
		%	& \tilde{t}(|\fvec q|, 0)  = \frac1{A_{uc}} \int \rmd^2 \fvec y\ e^{-i \fvec y \cdot \fvec q} t(\fvec y, 0) \nonumber \\
		%	 = & \frac{V_{pp\pi}^0}{A_{uc}} \frac{2\pi}{(1 + q^2 \delta^2)^{\frac32}} \delta^2 e^{a_0/\delta}  \\
		& \tilde{t}_{d_0}(|\fvec q|)  = \frac1{\mathcal{A}_{mlg}} \int \rmd^2 \fvec y\ e^{-i \fvec y \cdot \fvec q} t(\fvec y + d_0 \hat z) \nonumber \\
		= & \frac{2\pi d_0^2}{\mathcal{A}_{mlg}} \left[ V_{pp\pi}^0 \int_0^{\infty} \rmd y\ y J_0(q d_0 y) e^{-\frac{d_0}{\delta} \left( \sqrt{y^2+1} - a_0/d_0 \right)} \frac{y^2}{y^2 + 1}   +  V_{pp\sigma}^0 \int_0^{\infty} \rmd y\ y J_0(q d_0 y) e^{-\frac{d_0}{\delta} \left( \sqrt{y^2+1} - 1 \right)} \frac1{y^2 + 1}  \right] \label{EqnS:KInter}
\end{align}
With the Fourier transformation in Eq.~\ref{EqnS:KInter}, we are able to calculate all the components $w_i^{(\mu, l)}$ and $\fvec \lambda_i^{(\mu, l)}$ of the inter-layer tunneling fields $T_{SS'}(\fvec x)$ and $\fvec \Lambda_{S S'}(\fvec x)$, with their values listed in Tab.~\ref{TabS:ExtraParammeterInter}.

\subsection{Wannier based hopping parameterization} 
\label{Sec:TBMKaxiras}
In this subsection, we consider the model proposed in Ref.~\cite{KaxirasPRB16}. Note that Ref.~\cite{KaxirasPRB16} does not provide a general formula for the intra-layer hoppings, but lists its magnitude for a set of discrete hopping distance.  Here, we fit the intra-layer hopping with the following formula:
\begin{align}
	t_{intra}(\fvec r) = t_0 e^{- \alpha_0 \bar{r}^2} \cos(\beta_0 \bar{r}) + t_1 \bar{r}^2 e^{- \alpha_1 (\bar{r} - r_1)^2} \label{Eqn:KaxirasIntra}
\end{align} 
The values of the fitted parameters are listed in Table.~\ref{TabS:KaxirasPar}. It is obvious that the intra-layer hopping decays exponentially as a function of $|\fvec r|$.

According to Ref.~\cite{KaxirasPRB16}, the general form of the inter-layer hopping can be written as
\begin{align}
	t_{inter}(\fvec r) = & V_0(r) + V_3(r) \left( \cos(3\theta_{12}) + \cos(3\theta_{21}) \right) + V_6(r)\left( \cos(6\theta_{12}) + \cos(6\theta_{21})\right) \label{EqnS:KaxirasInter} \\
	V(r) = & \lambda_0 e^{- \xi_0 \bar{r}^2} \cos(\kappa_0 \bar{r}) \\
	V_3(r) = & \lambda_3 \bar{r}^2  e^{- \xi_3 (\bar{r} - x_3)^2} \\
	V_6(r) = & \lambda_6 e^{- \xi_6 (\bar{r} - x_3)^2}\sin(\kappa_6 \bar{r})  
\end{align}
where the vector $\fvec r$ is  the in-plane projected vector of the hopping displacement and $\bar{r} = r/a$. The variables $\theta_{12}$ and $\theta_{21}$ are the angles between $\fvec r$  and the nearest neighbor bond vectors $\fvec n_{j,S}$ on two layers, i.e.
\begin{align}
	&\theta_{12} 	= \cos^{-1}\left( -\frac{\fvec r \cdot \fvec n_{j,S}}{r |\fvec n_{j,S}|} \right) = \theta_{\fvec r} - \theta_{j,S} + \pi, \\
	&\theta_{21}  = \cos^{-1} \left( \frac{\fvec r \cdot \fvec n_{j',S'}}{r |\fvec n_{j',S'}|}  \right) = \theta_{\fvec r} - \theta_{j',S'} \ .
\end{align}
In the above formula, we define $\theta_{\fvec r}$ to be the angle between the  vector $\fvec r$ and the $x$ axis, and $\theta_{j,S}$ ($\theta_{j',S'}$) to be the angle between the bond vector $\fvec n_{j, S}$ ($\fvec n_{j', S'}$) and the $x$ axis. Note that each carbon atom has three different bond vectors. We label the angles of these three bonds by $\theta_{j,S}^{(\alpha)}$ ($\theta_{j',S'}^{(\alpha)}$), where the superscript $\alpha$ is the index of the bond vectors. Without the lattice distortion (e.g. as for a rigid twist), the three in-plane nearest neighbors of a carbon atom are $C_3$ symmetric about the carbon atom, and $\theta^{(\alpha)}_{j, S} = \theta_{j, S}^{(1)} + 2\pi(\alpha - 1)/3$. Therefore, the angles $\theta_{12}$ and $\theta_{21}$ could differ by $2\pi/3$ if choosing a different nearest neighbor bond, leading to the same $\cos(3 m \theta_{12})$ and $\cos(3 m \theta_{21})$ with $m$ being an integer. As a consequence, Eq.~\ref{EqnS:KaxirasInter} is independent of the choice of the bond vectors. The values of all microscopic hopping parameters are listed in Table.~\ref{TabS:KaxirasPar}.

\begin{table}[htb]
	\centering
	\begin{tabular}{|c|c|c|c|c|c|c|} \hline
		Inter &  $\lambda_i$ (eV) & $\xi_i$  & $x_i$ & $\kappa_i$ & & \\		\hline
		$V_0$ & $0.3155$ & $1.7543$ &  & $2.0010$  & & \\ \hline 
		$V_3$   & $-0.0688$ & $3.4692$  & $0.5212$ & & & \\ \hline
		$V_6$   & $-0.0083$ & $2.8764$  & $1.5206$ & $1.5731$ & & \\ \hline
		Intra & $t_0$ (eV) & $\alpha_0$ & $\beta_0$ & $t_1$ (eV) & $\alpha_1$ & $r_1$ \\ \hline
		& $-18.4295$ & $1.2771$ & $2.3934$ & $-3.7183$ & $6.2194$ & $0.9071$ \\ \hline
	\end{tabular}
	\caption{Parameters for the inter-layer and intra-layer hoppings in the model proposed in Ref.~\cite{KaxirasPRB16}. }
	\label{TabS:KaxirasPar}
\end{table}
%Table.~\ref{TabS:KaxirasPar} lists all the values of  that will be used to construct its continuum model.

In the presence of the lattice distortion, however, the local $C_3$ symmetry is broken, and thus Eq.~\ref{EqnS:KaxirasInter} depends on the choice of the bond vectors. In this case, we set the inter-layer hopping as
\begin{align}
		t_{inter}(\fvec r) = & V_0(r) + V_3(r) \left( \frac13 \sum_{\alpha = 1}^3 \cos(3\theta_{12}^{(\alpha)}) + \frac13 \sum_{\alpha = 1}^3  \cos(3\theta_{21}^{(\alpha)}) \right) + V_6(r)\left( \frac13 \sum_{\alpha = 1}^3  \cos(6\theta_{12}^{(\alpha)}) + \frac13 \sum_{\alpha = 1}^3  \cos(6\theta_{21}^{(\alpha)})\right)
\end{align}
where $\theta_{12}^{(\alpha)} = \theta_{\fvec r} - \theta^{(\alpha)}_{j,S} + \pi$ and $\theta_{21}^{(\alpha)} = \theta_{\fvec r} - \theta^{(\alpha)}_{j',S'}$.

To obtain the Fourier transformation of the hopping function, consider
\begin{align}
	& \int\rmd^2\fvec y\ e^{-i \fvec q \cdot \fvec y} t_n(|\fvec y|) e^{i n(\theta_{\fvec y} - \theta_j)} = \int_0^{\infty} \rmd y\ y t_n(y) \int_0^{2\pi}\rmd \theta e^{- i q y \cos(\theta - \theta_q)}  e^{i n(\theta - \theta_j)} \nonumber \\
	= & \int_0^{\infty} \rmd y\ y t_n(y) \int_0^{2\pi}\rmd \theta\ \sum_{m=-\infty}^{\infty} ( - i e^{i(\theta - \theta_q)})^m  J_m(q y) e^{i n (\theta - \theta_j)} = 2\pi \int_0^{\infty} \rmd y\ y t_n(y) (-i)^{-n} J_{-n}(q y) e^{i n(\theta_q -  \theta_j)} \nonumber \\
	= & 2\pi (-i)^n e^{i n (\theta_q - \theta_j)} \int_0^{\infty}\rmd y\ y t_n(y) J_n(q y)  \label{EqnS:BesselJ}
\end{align}
where $\theta_q$ is the angle between the vector $\fvec q$ and the $\hat x$ axis. In the derivation above, we have applied the formula $e^{\frac{z}2(t - t^{-1})} = \sum_{m = -\infty}^{\infty} t^n J_n(z)$ and $J_{-n}(z) = (-1)^n J_n(z)$. Starting from Eq.~\ref{EqnS:BesselJ}, it is easy to obtain 
\begin{align}
	& \int\rmd^2\fvec y\ e^{-i \fvec q \cdot \fvec y} t_n(|\fvec y|) \cos( n(\theta_{\fvec y} - \theta_j) ) = 2\pi (- i)^n \cos(n (\theta_q - \theta_j)) \int_0^{\infty} \rmd y\ y J_n(q y) t_n(y)
\end{align}
For notation convenience, we introduce 
\begin{align}
	\tilde{V}_i(q) = \frac{2\pi}{A_{mlg}}  \int_0^{\infty} \rmd y\ y V_i(y) J_i(q y) 
\end{align}
where $i = 0$, $3$, and $6$.

We first consider the Fourier transformation when the lattice is locally $C_3$ symmetric, i.e.~the lattice relaxation is absent. In this case, 
\begin{align}
	& \tilde{t}_{C_3}(\fvec q, \theta_{j,S}, \theta_{j', S'}) =  \tilde{t}(\fvec q, \theta_{j,S}, \theta_{j', S'}) = A_{mlg}^{-1} \int\rmd^2\fvec y\ e^{-i \fvec q \cdot \fvec y} t(\fvec y + d_0 \hat z, \theta_{j,S}, \theta_{j', S'})  \nonumber \\
	 = & \tilde{V}_0(q) +  i \left(-\cos(3(\theta_q - \theta_{j, S})) + \cos(3(\theta_q - \theta^{(\alpha)}_{j', S'})) \right)  \tilde{V}_3(q)  - \left(\cos(6(\theta_q - \theta_{j, S})) + \cos(6(\theta_q - \theta_{j', S'})) \right) \tilde{V}_6(q) \label{EqnS:FTHoppingC3Sym}
%	& \frac1{A_{mlg}}\int\rmd^2\fvec y\ y^{\mu} e^{- i \fvec q \cdot \fvec y} t(\fvec y, \{ \theta^{(\alpha)}_{j,S} \}, \{\theta^{(\alpha)}_{j', S'}\})  = i \frac{\partial}{\partial q^{\mu}} \tilde{t}(\fvec q, \{ \theta^{(\alpha)}_{j,S} \}, \{\theta^{(\alpha)}_{j', S'}\})
\end{align}
where $\tilde{t}_{C_3}$ is introduced to refer to the Fourier transformation of the inter-layer hopping function when the lattice is locally $C_3$ symmetric.

In the presence of the lattice distortion that varies slowly in the real space, the angle $\theta_{j, S}^{(\alpha)}$ can be approximated as
\begin{align}
	& \theta_{j,S}^{(\alpha)} = \theta_{\fvec \delta_S^{(\alpha)}} + \delta \theta^{(\alpha)}_{j,S},  \\
	& \delta \theta_{j,S}^{(\alpha)} = \frac{(\hat z \times \fvec \delta_S^{(\alpha)})\cdot \delta \fvec n_{j,S}^{(\alpha)}}{|\fvec \delta_S^{(\alpha)}|^2} =  \frac{\epsilon_{\mu\nu}}{|\fvec \delta_S^{(\alpha)}|^2}  \delta_{S,\mu}^{(\alpha)} \frac{\partial U_{j,\nu}^{\parallel}}{\partial x_{\rho}} \delta_{S,\rho}^{(\alpha)}. 
\end{align}
where $\theta_{\fvec \delta_B^{(\alpha)}}=\theta_{\fvec \delta_A^{(\alpha)}}+\pi$, and for our choice of the coordinate system, $\theta_{\fvec \delta_A^{(1)}}=\pi/6$, $\theta_{\fvec {\delta}_A^{(2)}}=\pi/6+2\pi/3$, $\theta_{\fvec \delta_A^{(3)}}=\pi/6-2\pi/3$. Since $\fvec U^{\parallel}_{t, S} = - \fvec U^{\parallel}_{b, S} = \half (\theta  \hat z \times \fvec y + \delta \fvec U)$, we obtain 
\[  \delta \theta_{t,S}^{(\alpha)} = - \delta \theta_{b,S}^{(\alpha)} = \frac{\theta}2 + \frac{\epsilon_{\mu\nu}}{2|\fvec \delta_S^{(\alpha)}|^2}  \delta_{S,\mu}^{(\alpha)} \frac{\partial \delta \fvec U_{\nu}}{\partial x_{\rho}} \delta_{S,\rho}^{(\alpha)} \ .  \]
Consequently, the hopping function is approximated as
\begin{align}
	 & t(\fvec y + d_0 \hat z, \{\fvec n^{(\alpha)}_{j, S}\},  \{\fvec n^{(\alpha)}_{j', S'}\}) \approx t(\fvec y + d_0 \hat z, \{ \theta_{\fvec \delta_S^{(\alpha)}} \},  \{ \theta_{\fvec \delta_{S'}^{(\alpha)}} \}) + \sum_{\alpha} \frac{\partial t}{\partial \theta^{(\alpha)}_{j, S}} \delta\theta^{(\alpha)}_{j, S} + \sum_{\alpha} \frac{\partial t}{\partial \theta^{(\alpha)}_{j', S'}} \delta\theta^{(\alpha)}_{j', S'} \nonumber \\
	 = & V_0(y) + V_3(y) \left( -\cos3(\theta_{\fvec y} - \theta_{\fvec \delta_S}) + \cos3(\theta_{\fvec y} - \theta_{\fvec \delta_{S'}}) \right) + V_6(y) \left( \cos6(\theta_{\fvec y} - \theta_{\fvec \delta_S}) + \cos6(\theta_{\fvec y} - \theta_{\fvec \delta_{S'}}) \right)   \nonumber \\
	 & + V_3(y) \left( \left( \sum_{\alpha} \delta \theta_{j, S}^{(\alpha)} \right) \sin3(\theta_{\fvec y} - \theta_{\fvec \delta_S} )  - \left( \sum_{\alpha} \delta \theta_{j', S'}^{(\alpha)} \right) \sin3(\theta_{\fvec y} - \theta_{\fvec \delta_{S'}}  )  \right)  \nonumber \\
	 & - 2 V_6(y) \left( \left( \sum_{\alpha} \delta \theta_{j, S}^{(\alpha)} \right) \sin6(\theta_{\fvec y} - \theta_{\fvec \delta_S} )  + \left( \sum_{\alpha} \delta \theta_{j', S'}^{(\alpha)} \right) \sin6(\theta_{\fvec y} - \theta_{\fvec \delta_{S'}}  )  \right)
\end{align}
Correspondingly, its Fourier transformation can be written as
\begin{align}
	\tilde{t}(\fvec q, \{\theta_{j, S}^{(\alpha)}\}, \{\theta_{j', S'}^{(\alpha)}\}) \approx \tilde{t}_{C_3}(\fvec q, \theta_{\fvec \delta_S}, \theta_{\fvec \delta_{S'}}) + \frac13 \left( \sum_{\alpha} \delta \theta_{j,S}^{(\alpha)} \right) \frac{\partial \tilde{t}_{C_3}(\fvec q, \theta_{\fvec \delta_S}, \theta_{\fvec \delta_{S'}})}{\partial \theta_{\fvec \delta_S}} + \frac13 \left( \sum_{\alpha} \delta \theta_{j', S'}^{(\alpha)} \right) \frac{\partial \tilde{t}_{C_3}(\fvec q, \theta_{\fvec \delta_S}, \theta_{\fvec \delta_{S'}})}{\partial \theta_{\fvec \delta_{S'}}}
\end{align}
where $\tilde{t}_{C_3}$, defined in Eq.~\ref{EqnS:FTHoppingC3Sym}, is the Fourier transformation of the inter-layer hopping for locally $C_3$ symmetric lattice. 	
	
%In the following, we will focus on $W_{SS'}(\fvec x)$ and derive its formula  in the absence/presence of the lattice relaxation. 
%\section{Impact of $\fvec q$ shells}   \label{SecS:QShells}

\section{Impact of Sub-leading terms}
We also consider the impact of the subleading terms in the constructed continuum model. Fig.~\ref{FigS:SKQShellSpec} and \ref{FigS:WannierQShellSpec} demonstrates the spectrum after truncating to a different number of $\fvec q$ shells in the inter-layer tunneling terms, for two microscopic models in  Ref.~\cite{KoshinoPRB12} and in Ref.~\cite{KaxirasPRB16}. The impact of the higher order terms on the energy spectrum is illustrated in Fig.~\ref{FigS:SKGradient} and \ref{FigS:WannierGradient}. Fig.~\ref{FigS:KaxirasPH} demonstrates the p-h asymmetry induced by $w_3$ and momentum dependent $\Lambda$ in the interlayer tunneling.  

\begin{figure}[htb]
	\centering		
    \subfigure[\label{FigS:SKQShellSpec:NoRelax}]{\includegraphics[width=0.7\columnwidth]{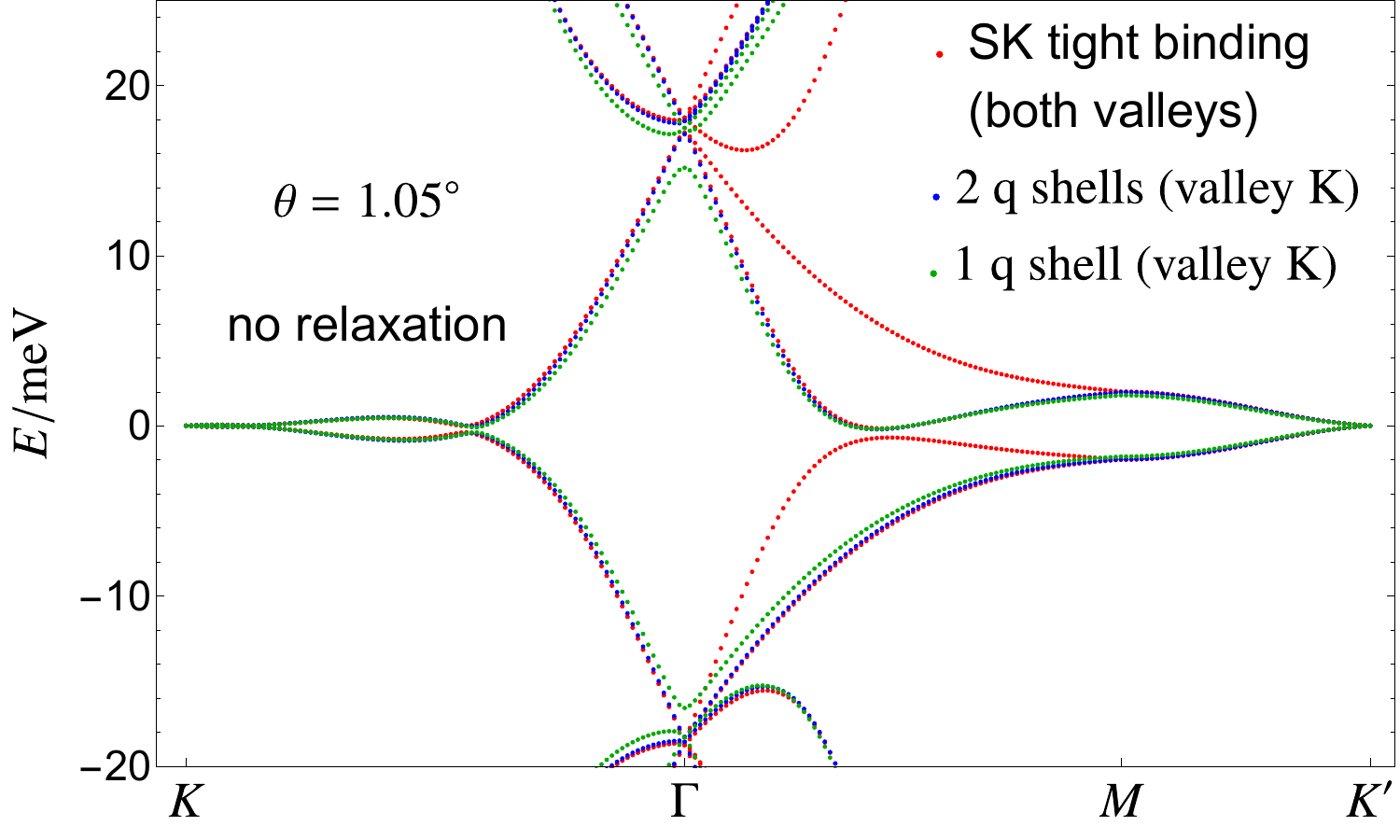}}	
	\subfigure[\label{FigS:SKQShellSpec:Relax}]{\includegraphics[width=0.7\columnwidth]{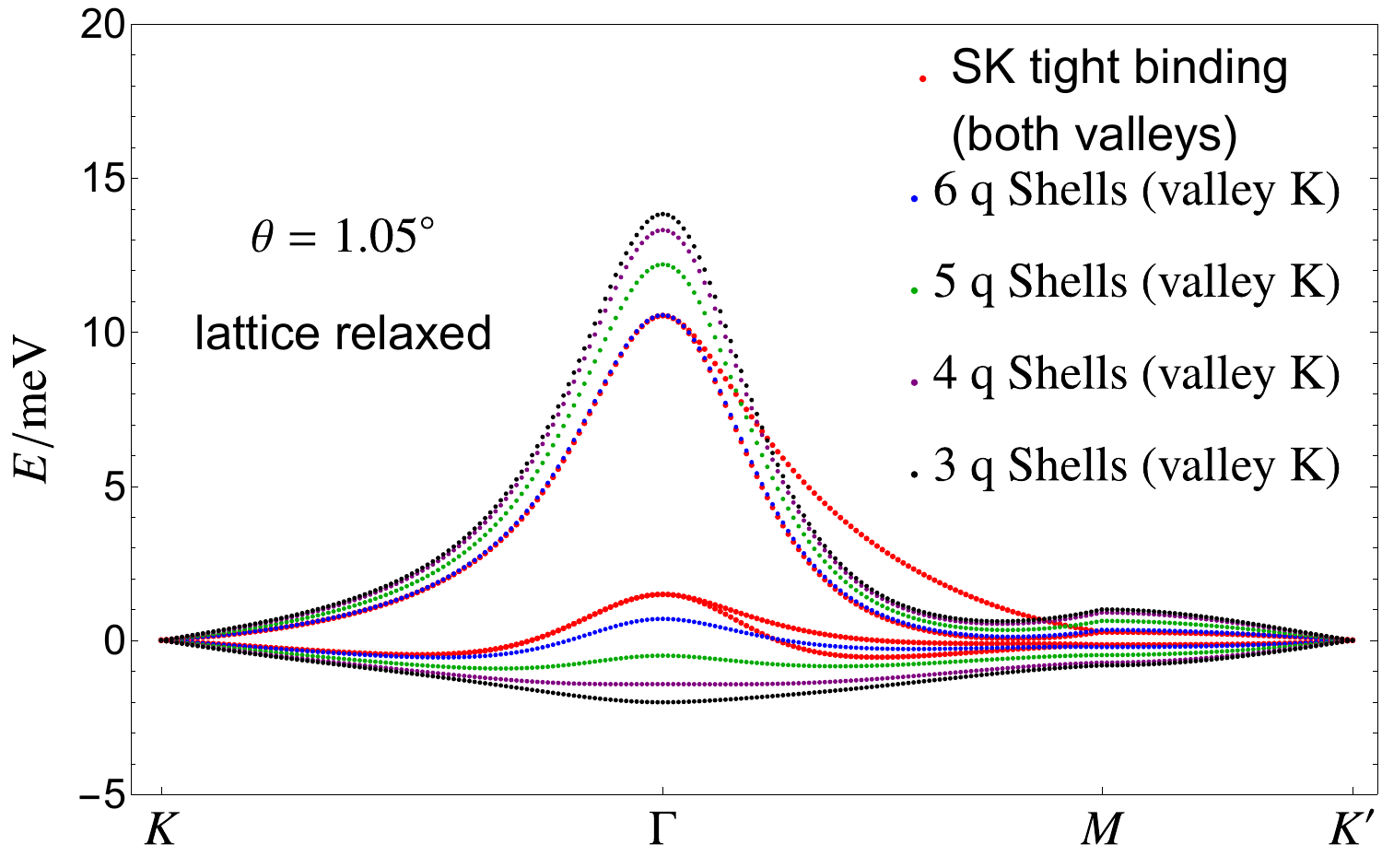}}		
%	\subfigure[\label{FigS:QShellSpec:KaxirasNoRelax}]{\includegraphics[width=0.6\columnwidth]{ShiangFangNoRelaxQShells.pdf}}	
%	\subfigure[\label{FigS:QShellSpec:KaxirasRelax}]{\includegraphics[width=0.6\columnwidth]{ShiangFangRelaxQShells.pdf}}		
	\caption{Spectrum after truncating to a different number of ${\fvec q}$ shells in the inter-layer tunneling terms. We have considered the effective continuum Hamiltonian $H_{eff}^{\bK}$ constructed for the microscopic models in  Ref.~\cite{KoshinoPRB12}. With the rigid twist only (above), the inclusion of two ${\fvec q}$ shells leads to an almost perfect agreement between the spectrum produced by $H_{eff}^{\bK}$ and the microscopic model. In the presence of the lattice relaxation (below), however, more ${\fvec q}$ shells are needed to achieve comparable accuracy. Note that while the tight binding spectra automatically contain both valleys, for the continuum model we show the spectra only for one valley in order to avoid clutter.}
	\label{FigS:SKQShellSpec}
\end{figure}

\begin{figure}[htb]
	\centering		
%	\subfigure[\label{FigS:QShellSpec:KoshinoNoRelax}]{\includegraphics[width=0.6\columnwidth]{KoshinoNoRelaxQShellComp.pdf}}	
%	\subfigure[\label{FigS:QShellSpec:KoshinoRelax}]{\includegraphics[width=0.6\columnwidth]{KoshinoRelaxQShellComp.pdf}}		
	\subfigure[\label{FigS:WannierQShellSpec:NoRelax}]{\includegraphics[width=0.7\columnwidth]{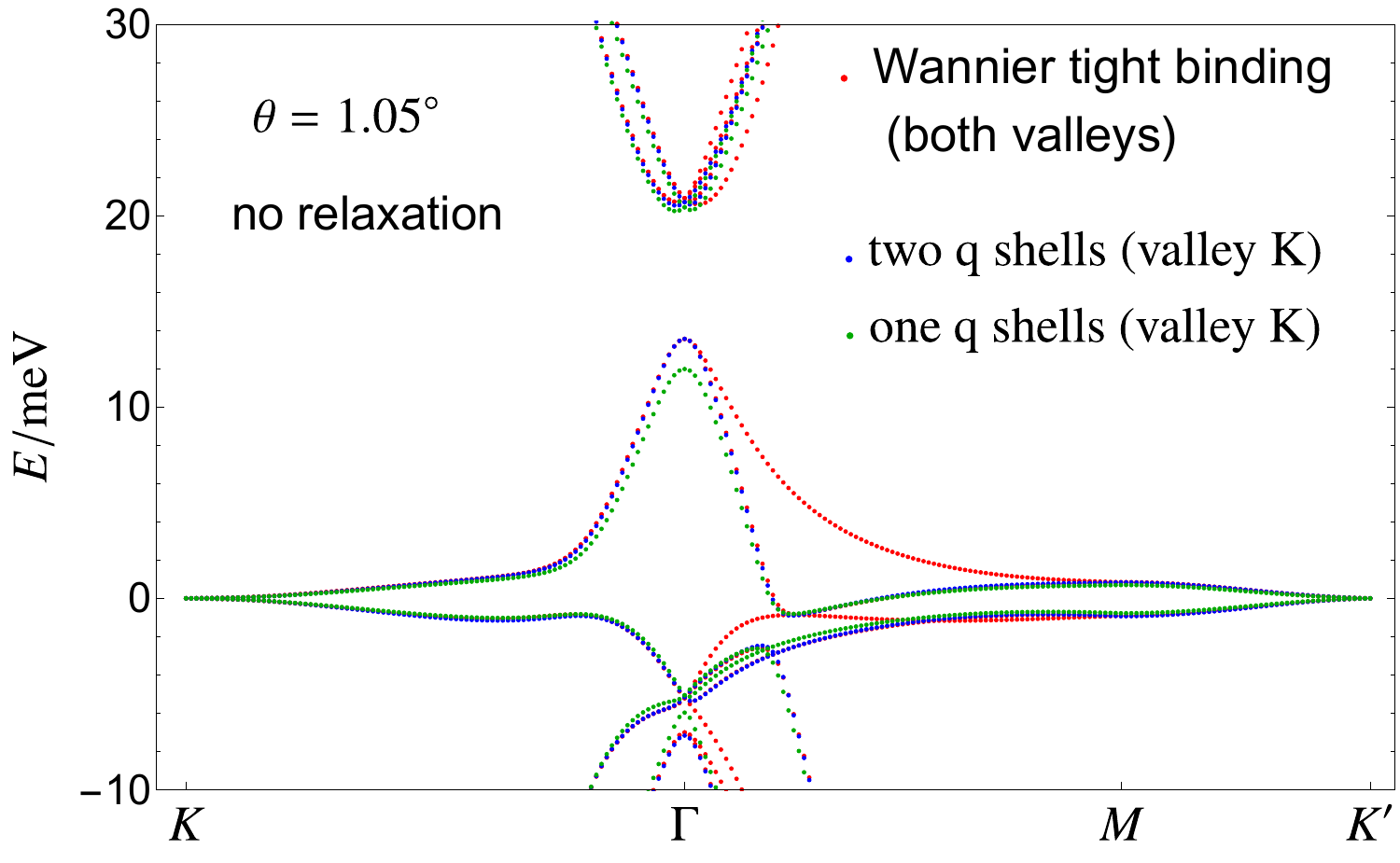}}	
	\subfigure[\label{FigS:WannierQShellSpec:Relax}]{\includegraphics[width=0.7\columnwidth]{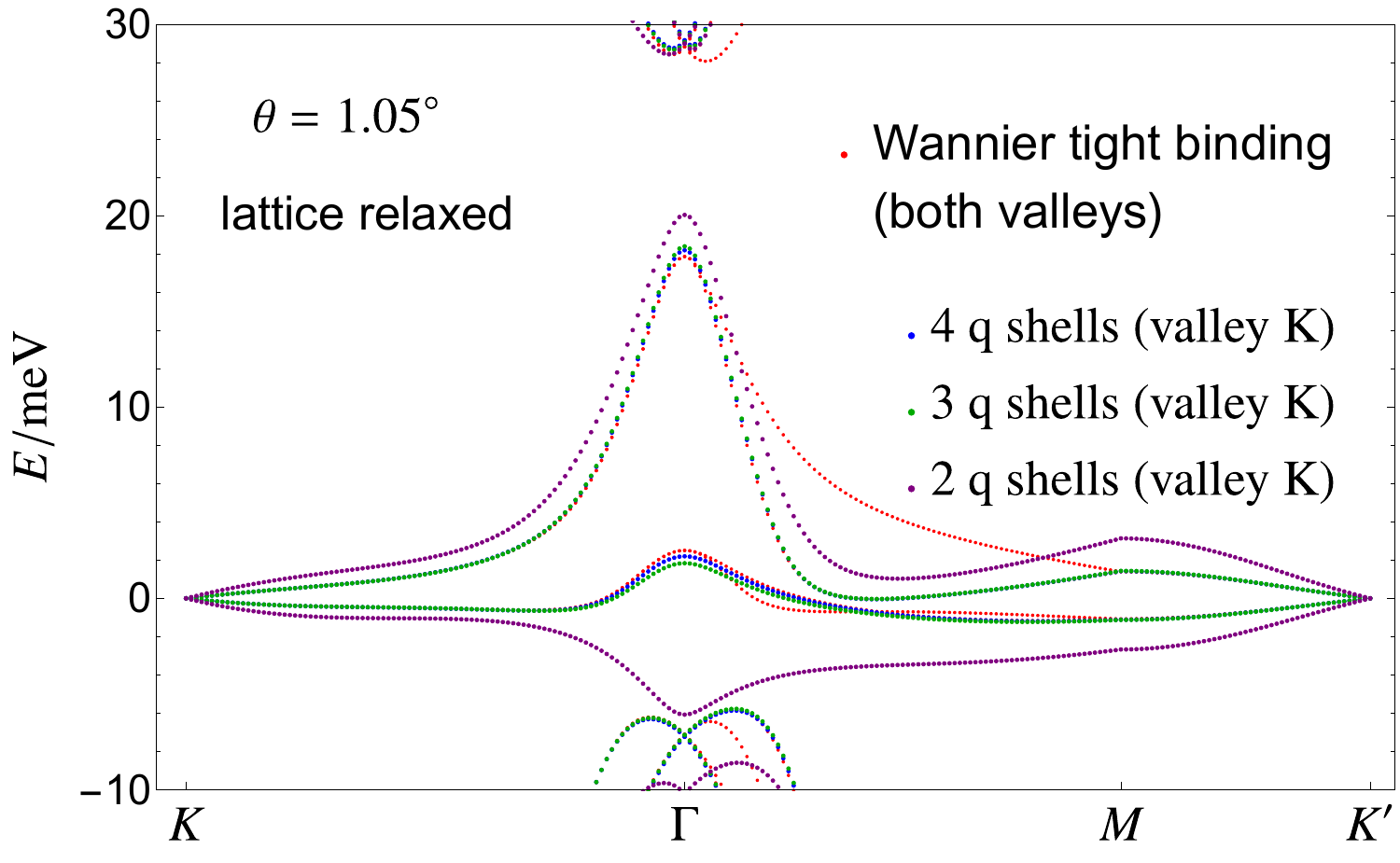}}		
	\caption{Same as Fig.~\ref{FigS:SKQShellSpec} but for another microscopic model in Ref.~\cite{KaxirasPRB16}.}
	\label{FigS:WannierQShellSpec}
\end{figure}

%In this section, we  compare the spectrum produced by the microscopic tight binding model and the effective continuum theory. Fig.~\ref{FigS:QShellSpec} have demonstrated the spectrum by including different $\fvec q$ shells with and without the lattice relaxation. In the absence of the lattice relaxation, Fig.~\ref{FigS:QShellSpec}(a) and \ref{FigS:QShellSpec}(c) have demonstrated that the spectrum of the continuum theory almost exactly match the one of the tight binding model with two $\fvec q$ shells only in the inter-layer tunnelings of $H_{eff}^{\bK}$. However, in the presence of the lattice relaxation, four and six $\fvec q$ shells are needed to match the microscopic tight binding models in Ref.~\cite{KoshinoPRB12} and \cite{KaxirasPRB16} respectively.

%\section{Higher order derivatives}
\begin{figure}[htb]
	\centering		
	\subfigure[\label{FigS:SKGradientc:NoRelax}]{\includegraphics[width=0.9\columnwidth]{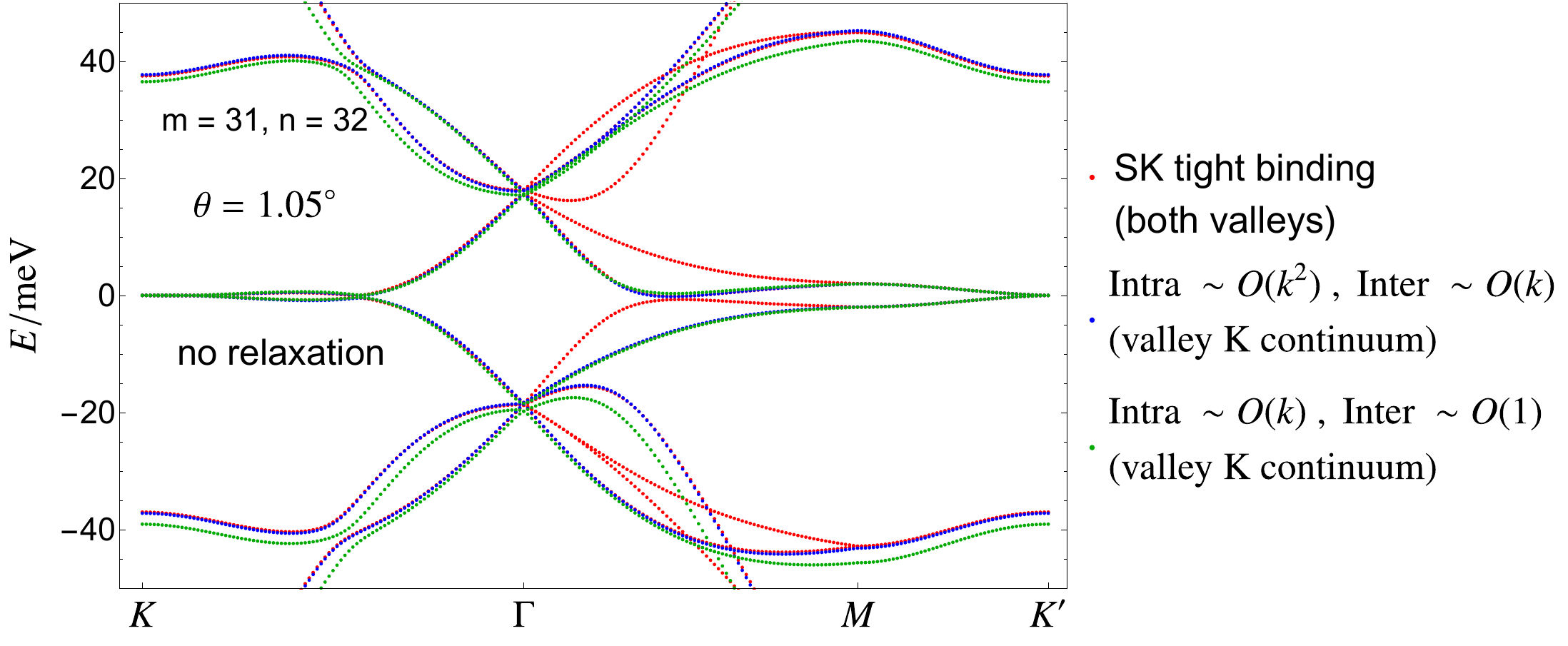}}	
	\subfigure[\label{FigS:SKGradient:Relax}]{\includegraphics[width=0.9\columnwidth]{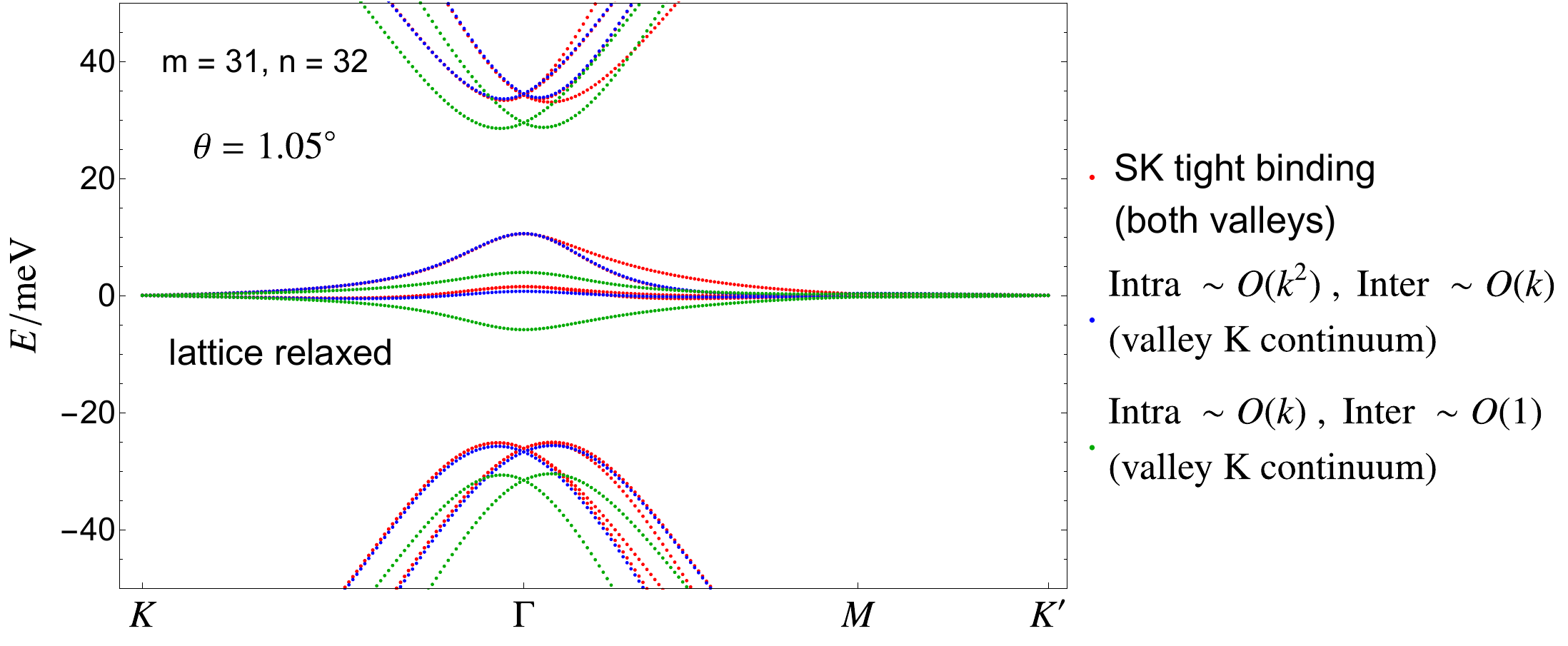}}		
%	\subfigure[\label{FigS:Gradient:KaxirasNoRelax}]{\includegraphics[width=0.7\columnwidth]{KaxirasGradientNoRelax.pdf}}	
%	\subfigure[\label{FigS:Gradient:KaxirasRelax}]{\includegraphics[width=0.7\columnwidth]{KaxirasGradientRelax.pdf}}		
	\caption{Impact of the higher order terms in the effective Hamiltonian on the energy spectrum near the CNP. The spectra of $H_{eff}^{\bK}$ constructed for the microscopic models in  Ref.~\cite{KoshinoPRB12} produced by  (blue) keeping or (green) dropping the second order gradient terms in the intra-layer continuum Hamiltonian and gradient couplings in the inter-layer tunneling terms. For comparison, the spectra of the microscopic tight binding model are also plotted and marked as red. Dropping the higher order derivative terms leads to a mismatch of $\sim5$-$10$meV, that is consistent with the estimate of this energy scale in the main text. Note that while the tight binding spectra automatically contain both valleys, for the continuum model we show the spectra only for one valley in order to avoid clutter.}
	\label{FigS:SKGradient}
\end{figure}

\begin{figure}[htb]
	\centering		
%	\subfigure[\label{FigS:Gradientc:KoshinoNoRelax}]{\includegraphics[width=0.7\columnwidth]{KoshinoGradientNoRelax.pdf}}	
%	\subfigure[\label{FigS:Gradient:KoshinoRelax}]{\includegraphics[width=0.7\columnwidth]{KoshinoGradientRelax.pdf}}		
	\subfigure[\label{FigS:WannierGradient:NoRelax}]{\includegraphics[width=0.9\columnwidth]{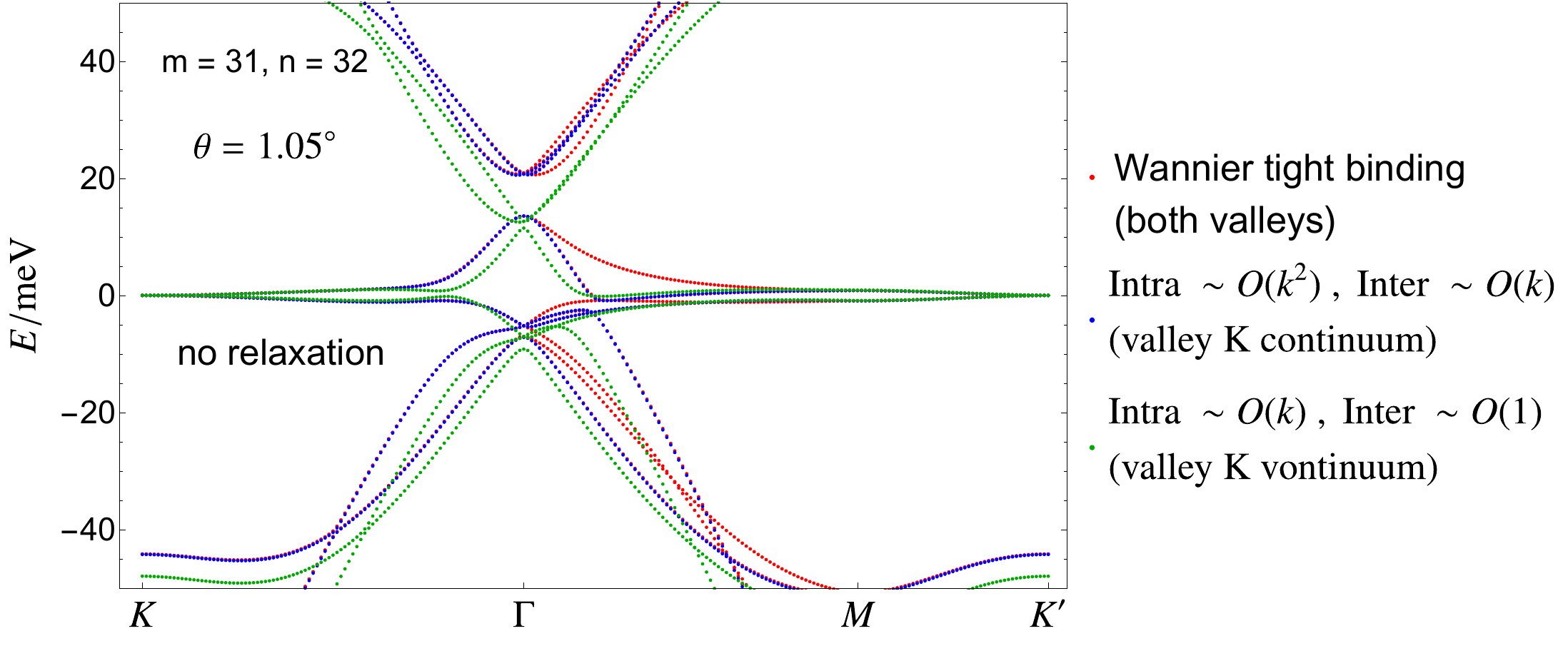}}	
	\subfigure[\label{FigS:WannierGradient:Relax}]{\includegraphics[width=0.9\columnwidth]{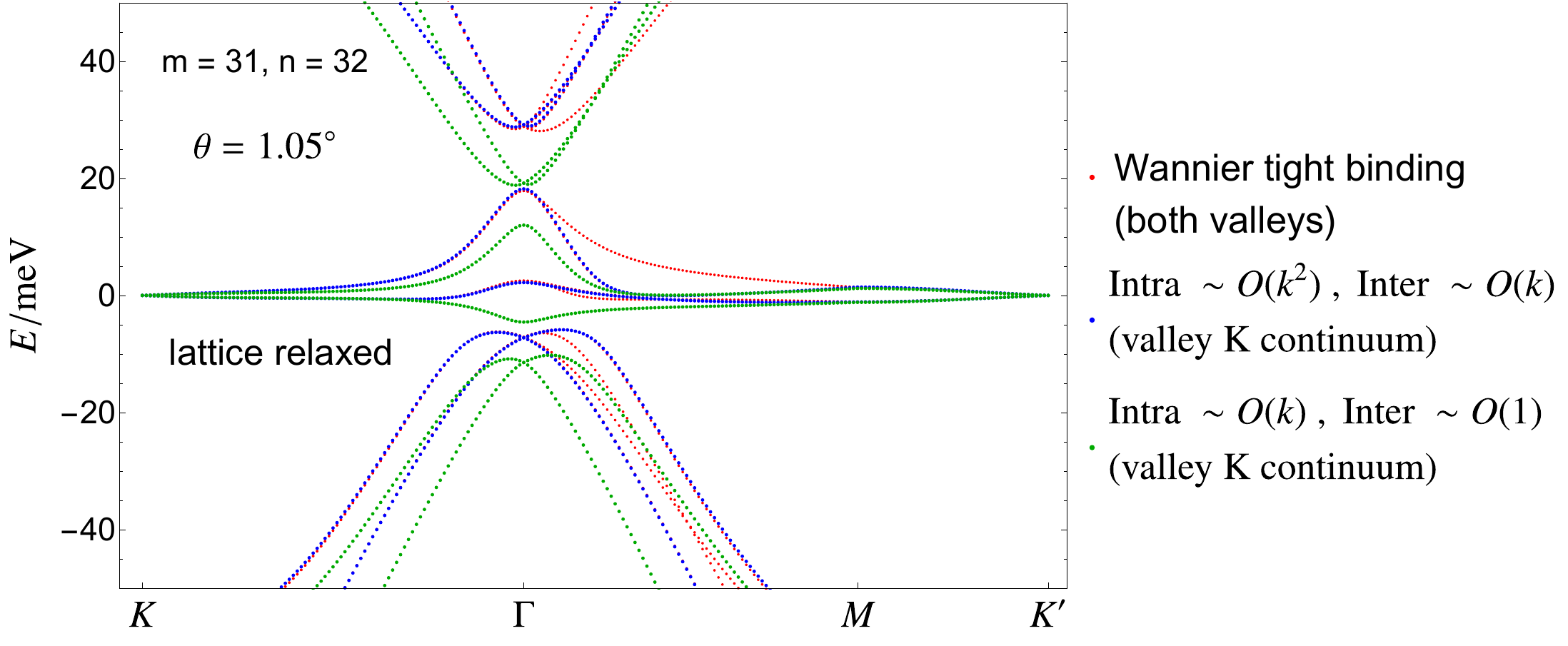}}		
	\caption{Same as Fig.~\ref{FigS:SKGradient} but for another microscopic model in Ref.~\cite{KaxirasPRB16}.}
	\label{FigS:WannierGradient}
\end{figure}

%As mentioned in the main text, we go beyond the BM model by expanding  to the second order of derivatives in the intra-layer Hamiltonian and the first order derivatives in the inter-layer tunneling. To obtain the effect of these higher order terms, we plot the spectrum of the continuum theories with and without these higher order terms in Fig.~\ref{FigS:Gradient}. For both microscopic models in Ref.~\cite{KoshinoPRB12} and \cite{KaxirasPRB16}, the absence of the higher order terms leads to the difference of  $\sim10$meV, compared with spectra of the tight binding models. 

%\section{Origin of p-h Asymmetry}

\begin{figure}[htb]
	\centering		
	\subfigure[\label{FigS:KaxirasPH:All}]{\includegraphics[width=0.45\columnwidth]{KaxirasPHSVD.pdf}}	
	\subfigure[\label{FigS:KaxirasPH:W30}]{\includegraphics[width=0.45\columnwidth]{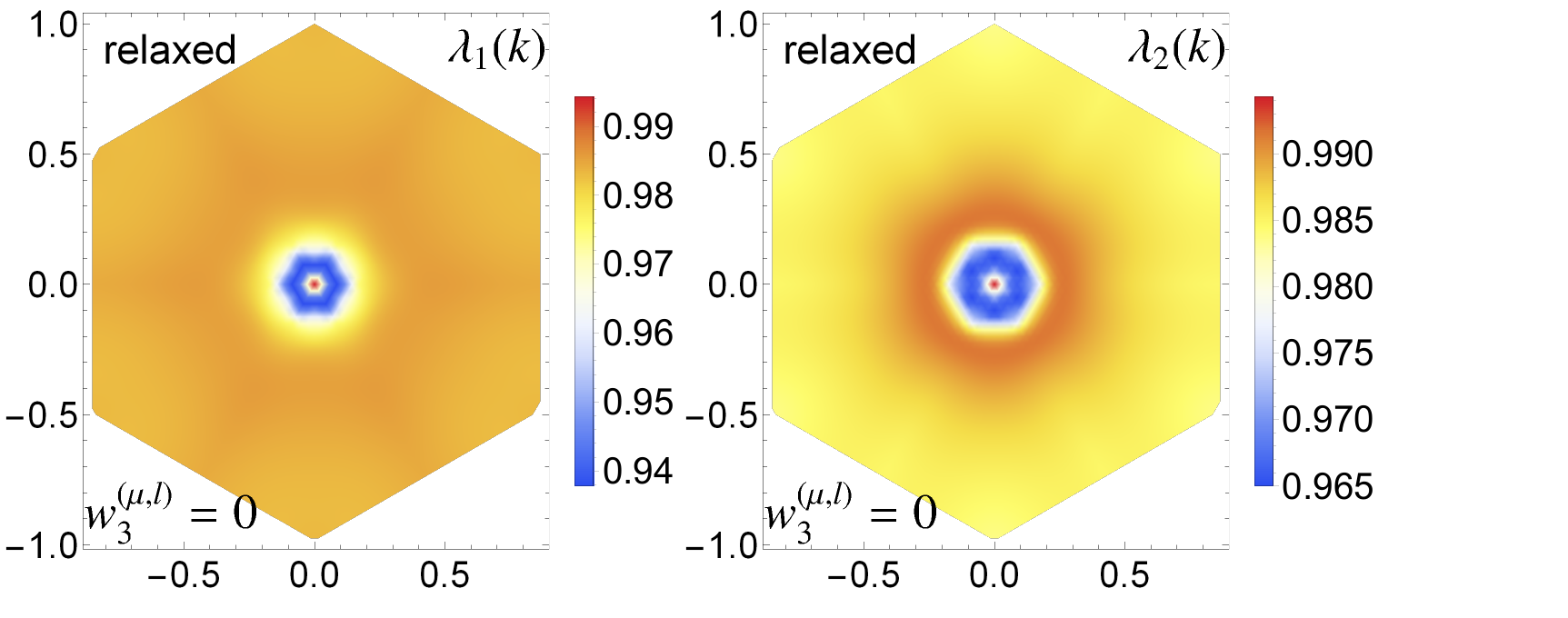}}	
	\subfigure[\label{FigS:KaxirasPH:W30Lambda0}]{\includegraphics[width=0.6\columnwidth]{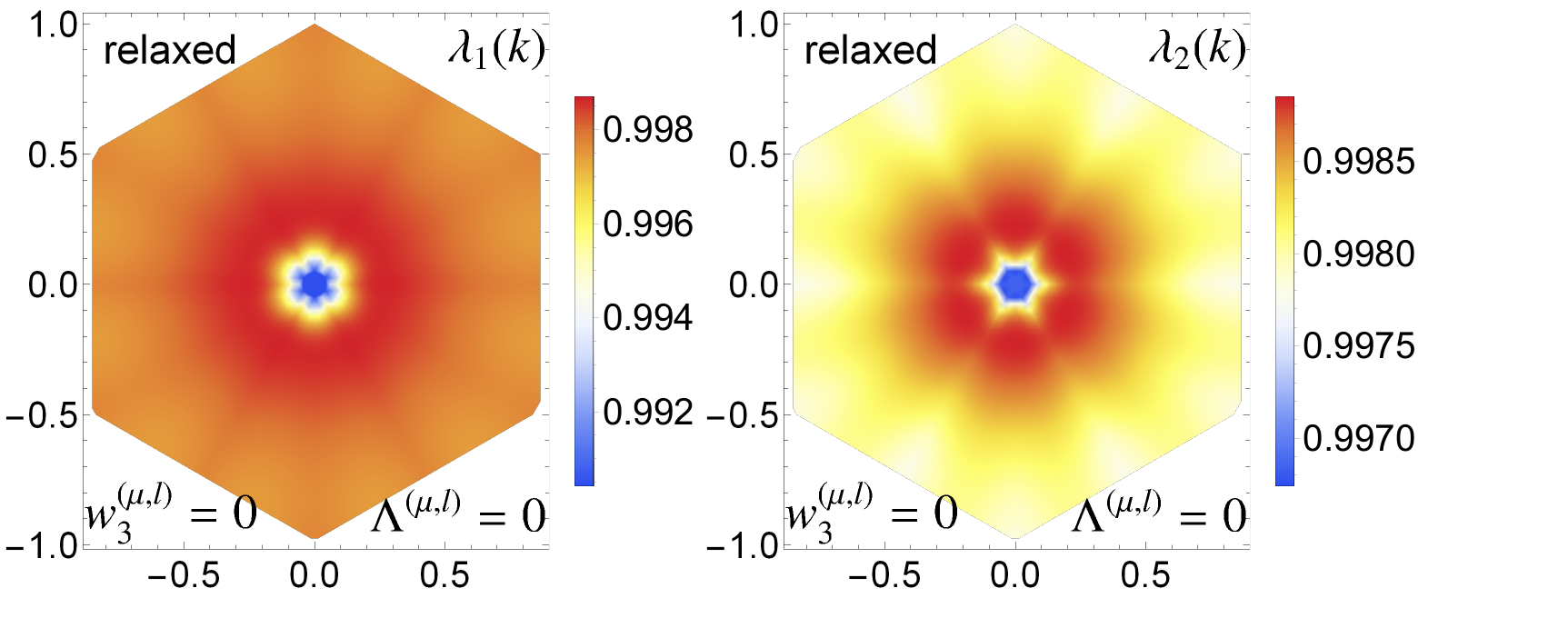}}	
	\caption{The two singular values, $\lambda_1(\fvec k)$ (left column) and $\lambda_2(\fvec k)$ (right column), of the projected p-h operator $\hat{\mathcal{P}}$ defined in Sec.~\ref{Sec:Analysis} by turning off different inter-layer terms in $H_{eff}^{\bK}$ for the microscopic tight binding model in Ref.~\cite{KaxirasPRB16}, and keeping all terms in the intra-layer part of $H^\bK_{eff}$ in Eq.\ref{Eqn:H0Intra}. (a) all the terms in Eq.~\ref{Eqn:HamInter} are kept in the inter-layer tunnelings, (b) the contact couplings $w_3^{(\mu, l)}$ for all the shells are set to $0$, and (c) both the contact $w_3^{(\mu, l)}$ and the gradient couplings $\fvec \Lambda_{SS'}^{(\mu, l)}$ are set to $0$. It is clear that the p-h asymmetry is dominated by the contribution from $w_3^{(\mu, l)}$. }
	\label{FigS:KaxirasPH}
\end{figure}

%In this section, we consider the source of the p-h asymmetry found in  the  model of Ref.~\cite{KaxirasPRB16}. As mentioned in the text, the p-h symmetry could be broken by various terms, including the $O(k^2)$ and $O(k \partial U)$ in the intra-layer coupling, as well as $w_3$ and $\fvec \lambda_{i \neq 3}$ in the inter-layer tunnelings. As shown in Fig.~\ref{FigS:KaxirasPH}, the dominant source of the particle-asymmetry is $w_3$, while the subdominant is the vector couplings $\fvec \Lambda$. The $O(k^2)$ and $O(k \partial \fvec U )$ has little contributions to the p-h asymmetry. 

\section{Approximate Formula of $v_{Dirac}$}
In this section, we derive the approximate formula of the $v_{Dirac}$ in the presence of the pseudo-vector field that is induced by the lattice relaxation proposed in Ref.~\cite{KaxirasRelaxation}. Motivated by Fig.~\ref{Fig:AField}, we neglect $\fvec \nabla \varphi^{\mathcal{A}}$, the irrotational part of the pseudo-vector field $\mathcal{A}$, and thus $\mathcal{A} \approx \fvec \nabla \times (\hat z \varepsilon^{\mathcal{A}})$. Furthermore, as demonstrated in Table~\ref{Tab:ElasticParammeters}, the lattice relaxation $\delta \fvec U$ in Ref.~\cite{KaxirasRelaxation} is dominated by the lowest harmonics. Because $\varepsilon_{\mathcal{A}}$ is even and real, its Fourier components must be pure imaginary and also odd. Considering $C_3$ symmetry, we can assume that
\begin{align}
	\varepsilon^{\mathcal{A}}(\fvec  x) \approx i \tilde{\varepsilon}^{\mathcal{A}}_1  \sum_{j = 1}^3 \left( e^{i \fvec g_j \cdot \fvec x} - e^{-i \fvec g_j \cdot \fvec x} \right)
\end{align}
where $\fvec g_3 = - (\fvec g_1 + \fvec g_2)$.  Under these approximation, we can obtain
\begin{align}
	\mathcal{A}_+ = \frac1{\tilde{\varepsilon}^{\mathcal{A}}_1} \left( \mathcal{A}_1 + i \mathcal{A}_2 \right) \approx i |\fvec g_1| \sum_{j = 1}^3  \omega^{j - 1} \left( e^{i \fvec g_j \cdot \fvec x} + e^{-i \fvec g_j \cdot \fvec x}  \right)
\end{align}
where $\fvec g_3 = - (\fvec g_1 + \fvec g_2)$ and $\omega = e^{i 2\pi/3}$. Now, for the equation
\begin{align}
	v_F \begin{pmatrix}
		p_+ + \gamma \tilde{\varepsilon}^{\mathcal{A}}_1 \mathcal{A}_+ & \alpha U(\fvec x) \\ \alpha U(- \fvec x) & p_+ - \gamma \tilde{\varepsilon}^{\mathcal{A}}_1 \mathcal{A}_+
	\end{pmatrix} \begin{pmatrix}
	\Phi_1(\fvec x) \\ \Phi_2(\fvec x)
\end{pmatrix} = 0  \label{EqnS:ZeroModeEqn}
\end{align}
where $\alpha = w_1 /(v_F k_{\theta})$, and $U(\fvec x) = \sum_j \omega^{j - 1} e^{i \fvec q_j \cdot \fvec x}$. We further assume that $\alpha$ and $\gamma \tilde{\varepsilon}^{\mathcal{A}}_1$ are small and expand the wavefunction in terms of the powers of $\alpha$ and $\gamma \tilde{\varepsilon}^{\mathcal{A}}_1$:
\begin{align}
	\Phi_1(\fvec x) & = 1 + \gamma \tilde{\varepsilon}^{\mathcal{A}}_1 \Phi_1^{(0, 1)}(\fvec x) + \alpha^2 \Phi_1^{(2., 0)}(\fvec x) + (\gamma \tilde{\varepsilon}^{\mathcal{A}}_1)^2 \Phi_1^{(0, 2)}(\fvec x) + \cdots \label{EqnS:Phi1Expansion} \\
	\Phi_2(\fvec x) & = \alpha \Phi_2^{(1, 0)}(\fvec x) + \alpha \gamma \tilde{\varepsilon}^{\mathcal{A}}_1 \Phi_2^{(1, 1)}(\fvec x) + \cdots \label{EqnS:Phi2Expansion} 
\end{align}
where we have used the fact that $\Phi_1(\fvec x)$ is even in $\alpha$ and $\Phi_2(\fvec x)$ is odd in $\alpha$. Substituting Eq.~\ref{EqnS:Phi1Expansion} and \ref{EqnS:Phi2Expansion} into Eq.~\ref{EqnS:ZeroModeEqn} and  comparing the powers of $\alpha$ and $\gamma \tilde{\varepsilon}^{\mathcal{A}}_1$, we obtain the equations 
\begin{align}
  & p_+ \Phi_1^{(0, 1)} + \mathcal{A}_+ = 0  \ , & & p_+ \Phi_1^{(2, 0)} + U(\fvec x) \Phi_2^{(1, 0)} = 0 \ , & p_+ \Phi_1^{(0, 2)} + \mathcal{A}_+ \Phi_1^{(0, 1)} = 0 \\
  & p_+ \Phi_2^{(1, 0)} + U(-\fvec x) = 0 \ , && p_+ \Phi_2^{(1, 1)} - \mathcal{A}_+ \Phi_2^{(1, 0)} + U(- \fvec x) \Phi_1^{(0, 1)} = 0 \ .
\end{align}
We consider the correction of $v_{Dirac}$ by the inclusion of $\mathcal{A}$ by  expanding it to the powers of $O(\alpha^2 \gamma \tilde{\varepsilon}^{\mathcal{A}}_1)$. This allows us to focus only on the lowest harmonics of the Fourier expansion of $\Phi_1$ and $\Phi_2$.  After some calculations, we found
\begin{align}
	\Phi_1^{(0, 1)} & = -i \sum_{j=1}^3 \left( e^{i \fvec g_j \cdot \fvec x} - e^{-i \fvec g_j \cdot \fvec x} \right) \ , &
	\Phi_2^{(1, 0)} & = i \sum_j e^{-i \fvec q_j \cdot \fvec x} \\
	\Phi_1^{(2, 0)} & = \frac{-i}{\sqrt{3}} \sum_j \left( \omega e^{i \fvec g_j \cdot \fvec x}  - \omega^* e^{-i \fvec g_j \cdot \fvec x} \right) \ ,  & \Phi_2^{(1, 1)} &\approx -2i \sqrt{3} \sum_j e^{-i \fvec q_j \cdot \fvec x} \\
	\Phi_1^{(0, 2)} & = - \sum_j \left( e^{i \fvec g_j \cdot \fvec x} + e^{- i\fvec g_j \cdot \fvec x} \right) \ .
\end{align}
Now, Eq.~\ref{Eqn:vDirac} gives
\begin{align}
	v_{Dirac} \approx \frac{1 - 6(\gamma \tilde{\varepsilon}^{\mathcal{A}}_1)^2 - 3\alpha^2 (1 - \frac{14}3\sqrt{3} \gamma \tilde{\varepsilon}^{\mathcal{A}}_1)}{1+3 \alpha^2 +  6(\gamma \tilde{\varepsilon}^{\mathcal{A}}_1)^2 } \ . \label{EqnS:VDiracAppr}
\end{align}
Based on Table.~\ref{Tab:ElasticParammeters} and \ref{Tab:ParammeterRelax}, $\gamma \tilde{\varepsilon}^{\mathcal{A}}_1 \approx 0.06$. Eq.~\ref{EqnS:VDiracAppr} gives $\alpha \approx 0.79$ when $v_{Dirac} = 0$. This value is very close to $\alpha = 0.7857$, the numerical result obtained in Sec.~\ref{Sec:ChiralLimit}.

\end{widetext}


\begin{thebibliography}{90}
\bibitem{appendix} See Appendix for more detailed derivation of the interlayer Hamiltonian $H_{intra}$, the interlayer tunneling, and the approximate formula of $v_{Dirac}$ when the relaxation induced pseudo-magnetic field is included. Note that the numerical values of the coupling constants of $\delta H_{intra}$, the contact and gradient couplings of the interlayer tunnelings with $\fvec q$s in the fourth and fifth shells are also listed in the Appendix.
\end{thebibliography}
\end{document}